\documentclass[twocolumn]{aastex631}

\usepackage{xcolor}
\newcommand {\msun}{\,M$_\odot \ $}
\newcommand {\dg}{$^\circ$}
\newcommand {\kms}{$\rm{km\ s^{-1}}$}

\begin{document}

\title{The compact object and innermost ejecta of SN 1987A}

\correspondingauthor{J. Larsson}
\email{josla@kth.se}

\author[0000-0003-0065-2933]{J. Larsson}
\affiliation{Department of Physics, KTH Royal Institute of Technology, The Oskar Klein Centre, AlbaNova, SE-106 91 Stockholm, Sweden}

\author[0000-0001-8532-3594]{C.\ Fransson}
\affiliation{Department of Astronomy, Stockholm University, The Oskar Klein Centre, AlbaNova, SE-106 91 Stockholm, Sweden}

\author[0000-0001-6872-2358]{P.\ J.\ Kavanagh}
\affil{Department of Physics, Maynooth University, Maynooth, Co. Kildare, Ireland}

\author[0000-0001-9855-8261]{B.\ Sargent}
\affiliation{Space Telescope Science Institute, 3700 San Martin Drive, Baltimore, MD 21218, USA}
\affiliation{Center for Astrophysical Sciences, The William H. Miller III Department of Physics and Astronomy, Johns Hopkins University, Baltimore, MD 21218, USA}

\author[0000-0002-3875-1171]{M.\ J.\ Barlow}
\affiliation{Department of Physics and Astronomy, University College London (UCL), Gower Street, London WC1E 6BT, UK}

\author[0000-0002-5529-5593]{M.\ Matsuura}
\affiliation{Cardiff Hub for Astrophysical Research and Technology (CHART), School of Physics and Astronomy, Cardiff University, The Parade, Cardiff CF24 3AA, UK}

\author[0000-0002-8526-3963]{C.\ Gall}
\affiliation{DARK, Niels Bohr Institute, University of Copenhagen, Jagtvej 155, DK-2200 Copenhagen N, Denmark}

\author[0000-0003-1319-4089]{R.~D.\ Gehrz}
\affiliation{Minnesota Institute for Astrophysics, University of Minnesota, 116 Church Street, S.E., Minneapolis, MN 55455, USA}

\author[0000-0002-2667-1676]{N.\ Habel}
\affil{Jet Propulsion Laboratory, California Institute of Technology, 4800 Oak Grove Dr., Pasadena, CA 91109, USA}

\author[0000-0002-2954-8622]{A.~S.\ Hirschauer}
\affil{Department of Physics \& Engineering Physics, Morgan State University, 1700 East Cold Spring Lane, Baltimore, MD 21251, USA}

\author[0000-0003-4870-5547]{O.~C.\ Jones}
\affil{UK Astronomy Technology Centre, Royal Observatory, Blackford Hill, Edinburgh, EH9 3HJ, UK}

\author[0000-0002-1966-3942]{R.~P.\ Kirshner}
\affil{TMT International Observatory, 100 W Walnut Street, Pasadena, CA 91124, USA}

\author[0000-0002-0522-3743]{M.\  Meixner}
\affil{Jet Propulsion Laboratory, California Institute of Technology, 4800 Oak Grove Dr., Pasadena, CA 91109, USA}

\author[0000-0002-2461-6913]{S.\ Rosu}
\affiliation{Department of Astronomy, University of Geneva, Chemin Pegasi 51, 1290 Versoix, Switzerland}
 
\author[0000-0001-7380-3144]{T.\ Temim}
\affil{Department of Astrophysical Sciences, Princeton University, Princeton, NJ 08544, USA} 

\begin{abstract}

The first JWST observations of SN 1987A provided clear evidence that a compact object is ionizing the innermost ejecta. Here we analyze a second epoch of JWST NIRSpec and MIRI/MRS observations to better characterize the properties of this region, aided by a higher spectral resolving power for the new NIRSpec data. We confirm the presence of the previously identified narrow lines from the central region;  [\ion{Ar}{6}]~4.5292\ $\mu$m, [\ion{Ar}{2}]~$6.9853\ \mu$m, [\ion{S}{4}]\ $10.5105\ \mu$m, and [\ion{S}{3}]\ $18.7130\ \mu$m, and also identify similar components in [\ion{Ca}{5}]\ $4.1585\ \mu$m, [\ion{Cl}{2}]\ $14.3678\ \mu$m, and possibly [\ion{Fe}{2}]\ $1.6440\ \mu$m. These lines are blueshifted by $\sim -250$~\kms, while the emission region is spatially unresolved and located south-east of the center. The offset and blueshift could imply a kick velocity of $510 \pm 55$~\kms\  for the neutron star. We also identify [\ion{Ca}{4}]~$3.2068\ \mu$m near the center, but it is displaced to the north and has a redshift of $\sim 700$~\kms. We find that scattering by dust in the ejecta with a typical grain size $\sim 0.3\ \mu$m can explain the  [\ion{Ca}{4}] properties and the absence of other narrow lines at shorter wavelengths, while dust absorption is important at $\lambda \gtrsim 8\ \mu$m. Photoionization models for a pulsar wind nebula and a cooling neutron star are both compatible with the observations, with the exception of the [\ion{Fe}{2}] feature. The two models primarily differ at short wavelengths, where new lines are expected to emerge over time as the optical depth of dust in the expanding ejecta decreases.  

\end{abstract}



\section{Introduction}

The explosion of a massive star as a core-collapse supernova (SN) leaves behind a compact object, which can be a neutron star (NS) or a black hole (BH). A NS is thought to be created in the vast majority of successful explosions (e.g., \citealt{Sukhbold2016}), with the population of NSs exhibiting diverse properties. Some NSs are only detected through the thermal X-ray emission from the hot surface (e.g., \citealt{Ho2009}), while others are pulsars that can be surrounded by pulsar wind nebulae (PWNe; e.g., \citealt{Buhler2014}). A major open question is how the properties of the progenitor stars and explosions are related to the properties of compact objects left behind. 

The nearby SN~1987A, located in the Large Magellanic Cloud (LMC), is an interesting case study in this context. Its progenitor was caught in pre-explosion images \citep{Walborn1987} and the properties of the explosion have been studied in detail throughout the electromagnetic spectrum (see \citealt{McCray2016} for a review). The progenitor is believed to be the result of a binary merger, which created a star with a mass of 15--20$~M_{\odot}$ that exploded as a blue supergiant (BSG; e.g., \citealt{Podsiadlowski1990,Menon2017,Utrobin2021}).  The binary merger model also offers an explanation for the complex circumstellar medium (CSM), which is composed of an equatorial ring (ER) and two larger outer rings (ORs) located above and below its plane \citep{Morris2007,Morris2009}. 

The explosion of SN~1987A was accompanied by a burst of neutrinos \citep{Alekseev1987,Bionta1987,Hirata1987}, which signaled the formation of a NS. A clear electromagnetic signal from the compact object remained elusive for decades, however, leaving open questions about whether the NS is also a pulsar with a PWN, or if it has collapsed further into a BH. \cite{Alp2018} summarized the situation up until 2018 and presented multiwavelength upper limits on a central point source.  

More recently, possible signs of a compact object were reported on the basis of observations at submillimeter wavelengths with the Atacama Large Millimeter Array (ALMA; \citealt{Cigan2019}) and hard X-rays with  the Nuclear Spectroscopic Telescope Array (NuSTAR; \citealt{Greco2021,Greco2022}). The ALMA observations revealed a peak in the dust emission in the form of an irregularly shaped blob located close to the center, which may be due to local heating of the dust by a compact object \citep{Cigan2019}. However, alternative interpretations in terms of heating by radioactive $^{44}$Ti or a higher density of dust (rather than a higher temperature) could not be ruled out. In the case of the NuSTAR observations, \cite{Greco2021} and \cite{Greco2022} reported a non-thermal component in their spectral analysis, which they attribute to emission from a PWN. However, NuSTAR does not spatially resolve the system and the hard X-rays may alternatively be explained as originating from the shock interaction with the CSM \citep{Alp2021}.  The presence of a PWN was also previously suggested based on radio observations with the Australia Telescope Compact Array (ATCA; \citealt{Zanardo2014}), though the evidence was inconclusive also  in this case, with free-free emission and cold dust offering alternative explanations for the observed excess.  

The first JWST \citep{Gardner2023} observations provided new insights regarding the compact object in SN~1987A. The observations were performed with NIRSpec \citep{Jakobsen2022} and the MIRI Medium-Resolution Spectrometer (MRS; \citealt{Wells2015}), which together provide spatially resolved spectroscopy over the 1--28~$\mu$m wavelength range. The data revealed narrow (FWHM $\sim 150$~\kms) emission lines from [\ion{Ar}{6}]~4.5292~$\mu$m, [\ion{Ar}{2}]~6.9853~$\mu$m, [\ion{S}{4}]~10.5105~$\mu$m, and [\ion{S}{3}]~18.7130~$\mu$m originating from the center of the ejecta (\citealp{Fransson2024}; F24 from hereon). This is in contrast to all previously observed lines from the ejecta, which are broad (FWHM $\sim 3500$~\kms) and primarily due to  \ion{H}{1},  H$_2$, \ion{He}{1},  [\ion{Si}{1}], and [\ion{Fe}{1-II}]   (e.g., \citealt{Jerkstrand2011, Kjaer2010, Kangas2022, Larsson2023}). 

As demonstrated by F24, the narrow Ar and S lines provide strong evidence for the presence of a compact object, which is ionizing the innermost ejecta. This is clear from multiple pieces of evidence: (i) the emission region is small (narrow lines from a spatially unresolved region) and located close to the center of the ejecta; (ii) the ionization state of the region is higher than for the surrounding ejecta; (iii) Ar and S are created deep inside the progenitor star and are expected to reside close to the compact object; (iv) the line luminosities and ratios can be reproduced in models of photoionization by a compact object; and (v) alternative explanations can be ruled out, including other energy sources and contamination by narrow lines from the ER. 

There are, however, open questions regarding the properties of the compact object. Most importantly, F24 could not discriminate between the scenarios where the ionizing radiation is dominated by the thermal emission from a hot NS surface or the non-thermal emission from a PWN.  Models where the ionization is caused by PWN shocks were found to match the observations less well, but could  
not be conclusively ruled out. Finally, there is a remote possibility that the ionizing radiation is due to fallback accretion onto a BH. This scenario was not modeled in  F24, but is very unlikely considering the relatively long duration of the neutrino signal, the small Fe core mass expected for the progenitor of SN~1987A, as well as the small amount of fallback accretion expected at these late epochs (F24). 

Our understanding of the properties of the compact object is limited by considerable uncertainties regarding the properties of the innermost ejecta. These include the elemental abundances,  physical conditions, the details of the asymmetries/clumping, as well as the properties of dust. It is well-known that there is a large reservoir of dust in the ejecta of SN~1987A \citep{Matsuura2015,Cigan2019}, and F24 concluded that dust has a major effect on the observed line luminosities. Specifically, the low luminosities of the [\ion{S}{4}]~10.5105\ $\mu$m and [\ion{S}{3}]~18.7130\ $\mu$m lines compared to the predictions from the photoionization models can be explained by absorption by silicate dust, which is strong at wavelengths $\gtrsim 8\  \mu$m. At shorter wavelengths, in the NIR and optical, scattering by dust is expected to be more important than absorption, with the magnitude and wavelength dependence of the optical depth being strongly dependent on the unknown grain size.  Scattering can broaden the line profiles and spread the light over a larger spatial region,  which may explain why no central point source has yet been observed at optical wavelengths \citep{Rosu2024}. 

Another important result from the JWST observations is that all the narrow lines are observed to be blueshifted by $\sim -250$~\kms\  with respect to the systemic velocity of SN~1987A, and that the centroid of the emission is located slightly south-east of the geometric center of the ER.  Assuming that the emission site is located close to the compact object, the Doppler shift and offset from the center imply a total kick velocity of $\sim 500$~\kms\ for the compact object, which provides an  important diagnostic of the asymmetric explosion. The interpretation in terms of kick velocity depends on the nature of the compact object though, as the line-emitting ejecta may be located further away from it in the PWN scenario, in which case the blueshift may be attributed to dust absorption within the PWN.   

Here, we aim to improve our understanding of the innermost ejecta and compact object in SN~1987A by analyzing new  JWST Cycle 2 observations with NIRSpec and MIRI/MRS. Importantly, the new NIRSpec observations were obtained with the high-resolution gratings, which have a factor $\sim 3$ higher spectral resolving power than the medium-resolution gratings used for the Cycle 1 observations. This is key for characterizing the properties of the [\ion{Ar}{6}] line and for searching for  additional narrow lines from the central region, which can constrain the properties of the ejecta and compact object. Many such lines are predicted in the NIR range (including lines from Si, Ca and S), but they are expected to be blended with strong broad emission lines from the surrounding faster ejecta, so a high spectral resolution is needed to place meaningful constraints on them. The new NIRSpec and MRS observations also enable us to study the time evolution of the narrow lines.  

We describe the observations and data reduction in Section~\ref{sec:obs}, followed by a presentation of the analysis and results in Section~\ref{sec:analysis}, where a summary of the most important observational results can be found in Section~\ref{sec:analyis:summary}. We present models for the line emission and discuss the results in Section~\ref{sec:disc} and summarize our conclusions in Section \ref{sec:conclusions}. 

\section{Observations and data reduction}
\label{sec:obs}

\subsection{NIRSpec Cycle 2 observations}

JWST Cycle 2 observations with the NIRSpec IFU were carried out on 2024 February 20--21, 13,511--13,512 days after the explosion (PID 3131, PI J. Larsson). The details of the observations are summarized in Table~\ref{tab:obs}. All three high-resolution gratings were used, i.e., G140H, G235H, and G395H, where the latter was used together with both the F070LP and F100LP filters. The total wavelength range included in these observations is 0.92--5.27~$\mu$m, though there are gaps in the wavelength coverage caused by the physical gaps between the NIRSpec detectors (see Table~\ref{tab:obs}). In the wavelength gaps, the spatial region with missing data moves across the field-of-view (FOV). In Table~\ref{tab:obs}, we also list the smaller wavelength gaps where there are missing data at the center of the system, which is the focus of our analysis.  

\begin{figure}[t]
\centering
\includegraphics[width=\hsize]{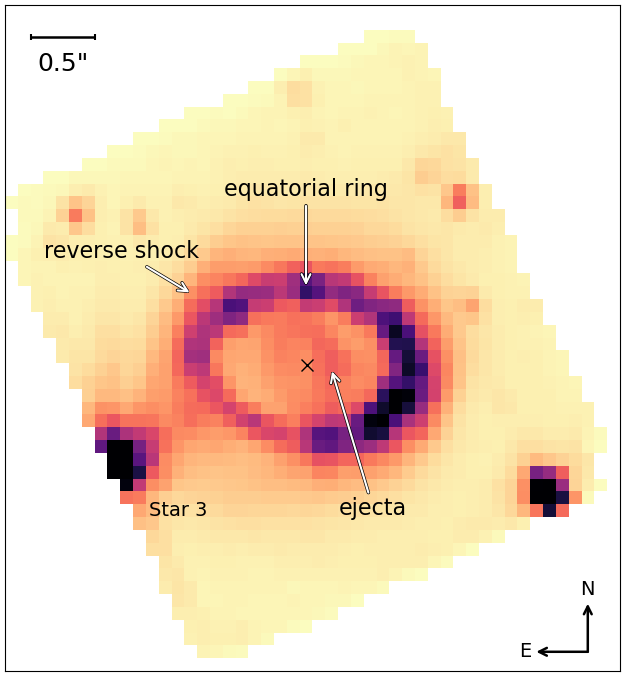}
\caption{Image integrated over G140H/F070LP, illustrating the FOV captured by the NIRSpec observations of SN~1987A. The emission in this wavelength range is dominated by H and He lines, the strongest one being \ion{He}{1}~1.0833~$\mu$m. The center is marked with a black x-symbol, and the main emission components are labeled. Projection effects cause diffuse emission from the RS to appear both inside and outside the ER.}
\label{fig:g140im}
\end{figure}

The spectral resolving power ($R=\lambda / \Delta \lambda$) in these observations ranges from $R \sim 1900$--$3600$ from short to long wavelengths in each of G140H/F100LP, G235H/F100LP, and G395/F290LP \citep{Jakobsen2022}. For the smaller wavelength range covered by G140H/F070LP, the spectral resolving power is $R \sim 1700$--$2400$. The IFU covers a $3^{\prime \prime} \times 3^{\prime \prime}$ FOV with a sampling of $0\farcs{1} \times 0\farcs{1}$. The FOV captures the ejecta and ER of SN~1987A, as illustrated in Figure~\ref{fig:g140im}. 

The observations were carried out using a small four-point dither cycling pattern and the NRSIRS2RAPID readout mode, which is the same set-up as in the Cycle 1 observations. So-called leakcal observations were carried out for all grating/filter combinations using the same exposure parameters as the science observations. Leakcal observations are obtained with the IFU aperture closed and are used to remove contaminating emission that enters through open shutters in the micro-shutter array (MSA) and/or leaks through closed MSA shutters. 

The data were downloaded from the Mikulski Archive for Space Telescopes (MAST) and processed with version 1.13.4 of the JWST Calibration Pipeline \citep{Bushouse2023}. We used versions 11.17.19 and “jwst\_1181.pmap” of the Calibration Reference Data System (CRDS) and CRDS context, respectively. We used the same procedure and input parameters for the pipeline as for the Cycle~1 data \citep{Larsson2023}, with the only difference being that the outlier detection step was turned on. This part of the pipeline has improved significantly since the time of the Cycle 1 analysis and serves to remove bad pixels and a large fraction of the artifacts induced by cosmic rays. We experimented with varying the outlier detection threshold between 50--99.8\% and found that 70\% gave the best performance in terms of removing outliers.  

We used an HST image of SN~1987A to improve the absolute astrometry of the NIRSpec data. The HST image was obtained on 2024 February 27, just a week after the NIRSpec observations, using the WFC3/F657N filter (PID 16996, PI: J. Larsson). The image was processed with DrizzlePac \citep{Hoffmann2021} and aligned with Gaia DR3 \citep{Gaia2023}. For each NIRSpec grating/filter combination, we created images integrated over the full bandpass, as well as continuum-subtracted images of bright H lines. The latter trace bright clumps of emission in the ER (“hotspots"), which are expected to coincide with the H$\alpha$ emission that dominates the HST/F657N image. We determined the offset between NIRSpec and HST by fitting the positions of the 3--5 brightest hotspots in the H-dominated images, as well as the star to the north-west in the full-band images (see Figure~\ref{fig:g140im}). We corrected the NIRSpec world coordinate solution (WCS) by the resulting offset and estimated an uncertainty of 0\farcs{01} in the astrometry based on the sample standard deviation from these fits.  

We do not subtract a background from the NIRSpec data cubes due to the lack of sufficiently large clean background regions within the FOV. The background is low and flat \citep{Larsson2023} and gets removed as part of the continuum subtraction when analyzing spectral lines, which is the main focus of this work. We do, however, subtract a background spectrum when analyzing the continuum. The background spectrum was extracted from three $0\farcs{3} \times 0\farcs{3}$ source-free regions located outside the ER and reverse shock (RS).

\begin{deluxetable*}{lccrcc}[t]
\tablecaption{Summary of NIRSpec Cycle 2 observations. \label{tab:obs}}
\tablecolumns{6}
\tablenum{1}
\tablewidth{0pt}
\tablehead{
\colhead{Grating/filter} &
\colhead{Date} & 
\colhead{Epoch\tablenotemark{a}} & 
\colhead{$t_{\rm exp}$\tablenotemark{b}} &
\colhead{$\lambda$ range} &
\colhead{$\lambda$ gap (center)\tablenotemark{c}} \\
\colhead{} &
\colhead{(YYYY-mm-dd)} &
\colhead{(d)} &
\colhead{(s)} &
\colhead{($\mu$m)} &
\colhead{($\mu$m)} 
}
\startdata
G140H/F070LP & 2024-02-20 & 13512 & 6127 & 0.92--1.27 & $<0.96$ ($<0.93$) \\
G140H/F100LP & 2024-02-19 & 13511 & 6127 & 0.97--1.89 & 1.41--1.49 (1.44--1.46) \\
G235H/F170LP & 2024-02-19 & 13511 & 4493 & 1.66--3.17 & 2.36--2.49 (2.40--2.45) \\
G395H/F290LP & 2024-02-19 & 13511 & 7353 & 2.87--5.27 & 3.98--4.20 (4.06--4.13) \\
\enddata
\tablenotetext{a}{Days since explosion on 1987-02-23.}
\tablenotetext{b} {The total exposure time was computed by dividing the value of the EFFEXPTM header keyword from the final cube by the number of detectors (one for F070LP and two in all other cases). The science and leakcal observations had the same exposure times.} 
\tablenotetext{c}{Wavelength range where data are missing in part of the FOV. The range within brackets is where there are missing data at the center of the ejecta.}
\end{deluxetable*}

\subsection{NIRSpec Cycle 1 observations}

NIRSpec IFU observations of SN~1987A were obtained in Cycle 1 as part of the Guaranteed Time Observations (GTO) Program  1232 (PI: G. Wright). The observations were carried out on 2022 July 16 (12,927~days after the explosion) using the G140M/FP100LP, G235M/ F170LP, and G395M/F290LP gratings/filter combinations. All the details of the observations were described in \cite{Larsson2023}. These observations have lower spectral resolution compared to those in Cycle 2, but do not have any wavelength gaps. 

We reprocessed the observations using version 1.13.4 of the JWST Calibration Pipeline, with CRDS version 11.17.16 and CRDS context version “jwst\_1237.pmap". The Cycle 1 data had leakcal observations in G395M only, which were used to correct for light leakage through the MSA. A bright artifact caused by light leakage was identified in G140M and G235M, where it overlapped with part of the ER. It was removed as part of the outlier detection step of the pipeline processing. As for the Cycle 2 data, we extracted a background spectrum from small source-free regions located outside the ER and RS, which we subtracted from the source spectra only when studying the continuum.

\subsection{MRS Cycle 1 and 2 observations}

MIRI/MRS observations of SN~1987A were obtained as part of GTO programs 1232 (PI: G. Wright) and 2763 (PI: M. Meixner) in Cycles 1 and 2, respectively. The Cycle 1 observations were carried out on the same date as the NIRSpec Cycle 1 observations, while the Cycle 2 observations were obtained on 2023 August 04 (13,311 days after explosion), about six months before the NIRSpec Cycle 2 observations. The Cycle 1 observations were previously analyzed in  \cite{Jones2023}, while both data sets are presented in Kavanagh et al. (in prep.). We refer the reader to the latter paper for details about the data reduction, including reprocessing of the Cycle 1 data and background subtraction. 

These IFU observations provide spatially resolved spectroscopy over the 4.9--27.9~$\mu$m wavelength range, with spectral resolving power $R \sim$~4000--1500 \citep{Jones2023a}. The FOV increases with wavelength from $3\farcs{2}\times 3\farcs{7}$ to $6\farcs{6}\times 7\farcs{7}$, while the spatial resolution decreases from $\sim 0\farcs{25}$--$1^{\prime \prime}$ \citep{Law2023}. This implies that the emission from the innermost ejecta can be clearly separated from the ER only at the shorter wavelengths (see images of SN~1987A in the different MRS sub-bands in \citealt{Jones2023}, their Figure~1).


\section{Analysis and results}
\label{sec:analysis}

In the following analysis, we correct all velocities for the heliocentric systemic velocity of SN~1987A of 287~\kms\ \citep{Groningsson2008} and adopt a distance to the LMC of 49.6~kpc \citep{Pietrzynski2019}. At this distance, $1^{\prime \prime}$ is equivalent to $7.42\times 10^{17}$~cm. As the ejecta are expanding homologously (e.g., \citealt{Rosu2024}), distances can also be directly translated to ejecta velocities as $V_{\rm ej}=r/t_{\rm exp}$, where $r$ is the distance from the center and $t_{\rm exp}$ is the time since explosion. Fast ejecta that are just reaching the ER (semi-major axis $0\farcs{82}$, \citealt{Tegkelidis2024}) in the NIRSpec observations at 13,500~days have a velocity of 5200~\kms. The bulk of the ejecta are propagating at lower velocities and are still inside the ER, as is clear from Figure~\ref{fig:g140im}. For reference, one NIRSpec spaxel of 0\farcs{1} is equivalent to 635~\kms\ for the expanding ejecta at the time of these observations. 

Velocities also provide a way to disentangle different emission components in SN~1987A, which partly overlap  in images. Our analysis focuses on the innermost region of the ejecta, which is ionized by the compact object. The small emission region results in narrow emission lines (FWHM $\sim 150$~\kms), which were additionally observed to be blueshifted by $\sim -250$~\kms\ in the analysis of the Cycle 1 data (F24). Depending on the wavelength, these lines may be blended with emission from the following components: 
\begin{itemize}
\item[-] Emission from the surrounding ejecta. This is dominated by broad emission lines (FWHM $\sim 3500$~\kms), powered by the radioactive decay of $^{44}$Ti and X-rays from the ER \citep{Larsson2011,Larsson2013,Rosu2024}. There is also a weak continuum in the ejecta, discussed further in Section~\ref{sec:analysis:cont}.

\item[-] Emission from the RS. The RS extends from the inner edge of the ER to form a bubble-like structure above and below its plane \citep{Larsson2023}. Some of the high-latitude emission from the RS is projected at the center of SN~1987A. This includes line emission from fast ejecta that are excited by the RS, resulting in very broad H and He lines extending to $\pm 10,000$~\kms, as well as synchrotron continuum emission. 

\item[-] Emission from the ER. In this case, there is no direct spatial overlap, but the ER is extremely bright, which implies that emission in the tails of the PSF can contribute significantly even in the innermost ejecta. This effect is especially important at longer wavelengths, where the emission from the ER is brighter and the spatial resolution is lower. There are two main contributions to the emission from the ER: line emission from the shocked gas (FWHM $\sim 300$~\kms) and continuum emission from the dust \citep{Jones2023}. The latter primarily contributes above $\sim 4$~$\mu$m. 

\item[-] Emission from the northern outer ring, which passes through the center of the ejecta in projection (being physically on its far side, \citealt{Tziamtzis2011}), as well as any emission from diffuse CSM/ISM. These components have very low radial velocity and are spectrally unresolved in our observations. 
\end{itemize}

We use spatially resolved spectroscopy provided by the NIRSpec/IFU and MIRI/MRS to isolate the emission from the innermost ejecta from these other components. The line emission in NIRSpec and MRS are analyzed in Sections~\ref{sec:analysis:ns} and \ref{sec:analysis:mrs}, respectively, followed by an analysis of the continuum in Section~\ref{sec:analysis:cont}. A summary of the key results is provided in Section~\ref{sec:analyis:summary}. 

We refer to all spectral lines by their vacuum wavelengths and quote all uncertainties at 1$\sigma$. In the majority of cases, we present line profiles after subtracting the continuum level determined by fitting a linear function. We characterize the properties of emission lines by fitting Gaussian functions, but stress that these are phenomenological since the asymmetric ejecta will produce asymmetric line profiles.  

\subsection{Line emission in NIRSpec}
\label{sec:analysis:ns}

\subsubsection{The [\ion{Ar}{6}] 4.5292~$\mu$m line}
\label{sec:analysis:ar6}

\begin{figure}[t]
\centering
\includegraphics[width=\hsize]{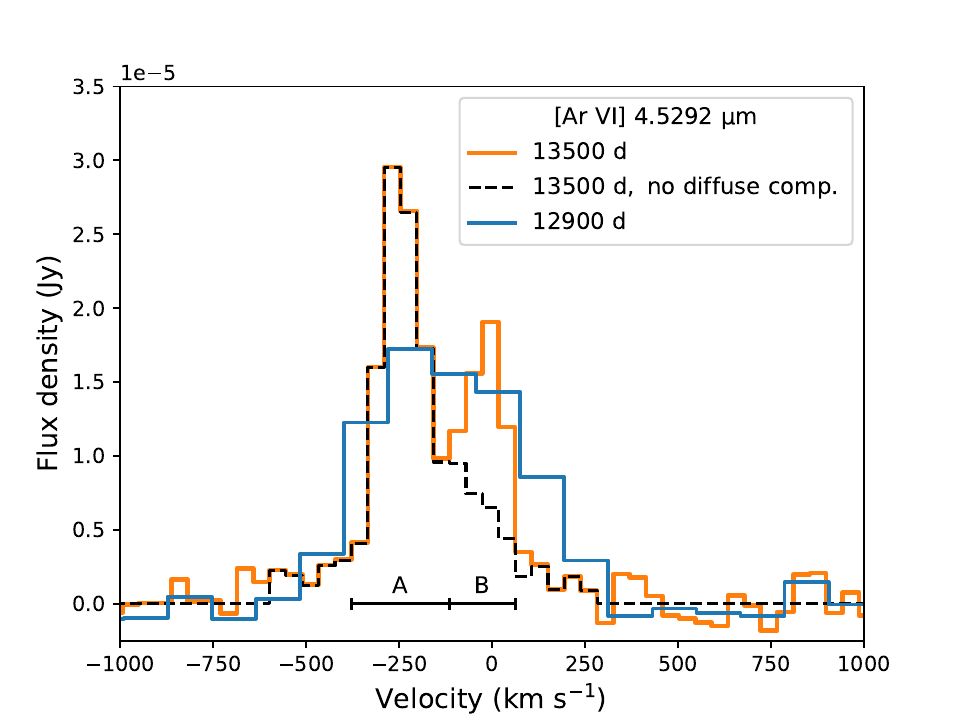}
\caption{Velocity profile of the [Ar~VI] line from the central region at 13,500\ and 12,900\ days. The extraction region is shown in Figure~\ref{fig:ar6ims}. The solid orange and dashed black lines show the velocity profiles at 13,500\ days before and after the removal of the diffuse 0-\kms\ line, respectively (see text for details). This component is blended with the emission from the central source in the spectrum at 12,900\ days due to the lower spectral resolution. The velocity intervals indicated by A and B were used for producing the images in Figure~\ref{fig:ar6ims}.}
\label{fig:ar6profiles}
\end{figure}

The new NIRSpec observations at 13,500 days confirm the presence of emission from [\ion{Ar}{6}]~4.5292\ $\mu$m close to the center of the system. Figure~\ref{fig:ar6profiles} shows the line profile compared to the one from 12,900~days (F24), illustrating the improvement in spectral resolution. The peak of the line at $\sim - 250$~\kms\ is consistent with the previous observation and other lines (including [\ion{Ar}{2}], F24), but there is also a narrow component close to 0~\kms, which is blended with the main peak in the lower-resolution data from 12,900~days.    

\begin{figure*}[t]
\centering
\includegraphics[width=\hsize]{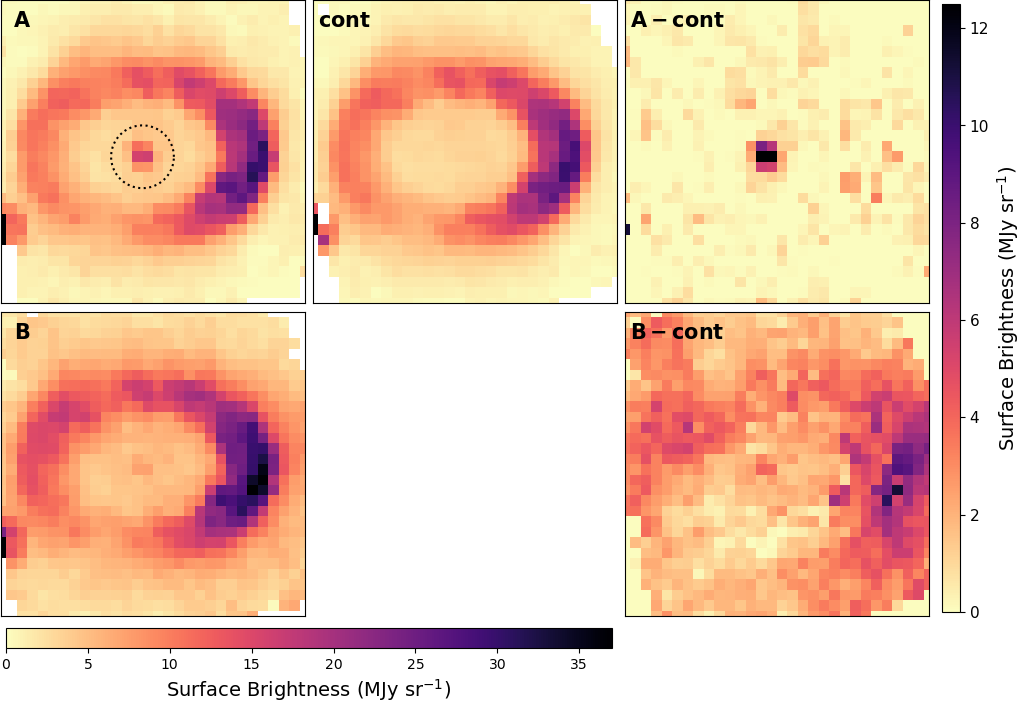}
\caption{Images of the [\ion{Ar}{6}] emission and adjacent continuum. Top left: Image integrated over the blue peak of the line (velocity interval A in Figure~\ref{fig:ar6profiles}). Top middle: Continuum image integrated over $\sim 300$\ \kms\ wide intervals on both sides of the full line profile. Top right: Result of subtracting the continuum image from image A. Bottom left: Image integrated over the narrow component around 0 \kms\ (velocity interval B in Figure~\ref{fig:ar6profiles}). Bottom right: Result of subtracting the continuum image from image B. The dotted black line in the top left panel shows the extraction region for the spectra in Figure~\ref{fig:ar6profiles}. The horizontal color bar at the bottom applies to the left and middle panels, while the color bar to the right applies to the right panels. The continuum-subtracted images in the right panels illustrate that the blueshifted component originates from the central source, while the spectrum at $\sim 0$ \kms\ has a significant contribution from extended diffuse emission.}
\label{fig:ar6ims}
\end{figure*}

To investigate the origin of the two peaks, we produce images integrated over the velocity intervals marked A and B in  Figure~\ref{fig:ar6profiles}. We also create a continuum image from the velocity intervals [-1060, -660]~\kms\ and [300, 700]~\kms. Figure~\ref{fig:ar6ims} shows these images, as well as the result of subtracting the continuum from intervals  A and B. For the blue peak (interval A), it is clear that only a compact central source remains after the continuum subtraction. For the peak at  $\sim 0$~\kms, there is instead significant diffuse emission in the whole field, with a particularly bright region in the west, and only a slightly increased surface brightness at the center. We will refer to this narrow line as the “0-\kms\ line"  from here on to avoid confusion with the narrow lines associated with the compact object.   

\begin{figure}[t]
\centering
\includegraphics[width=\hsize]{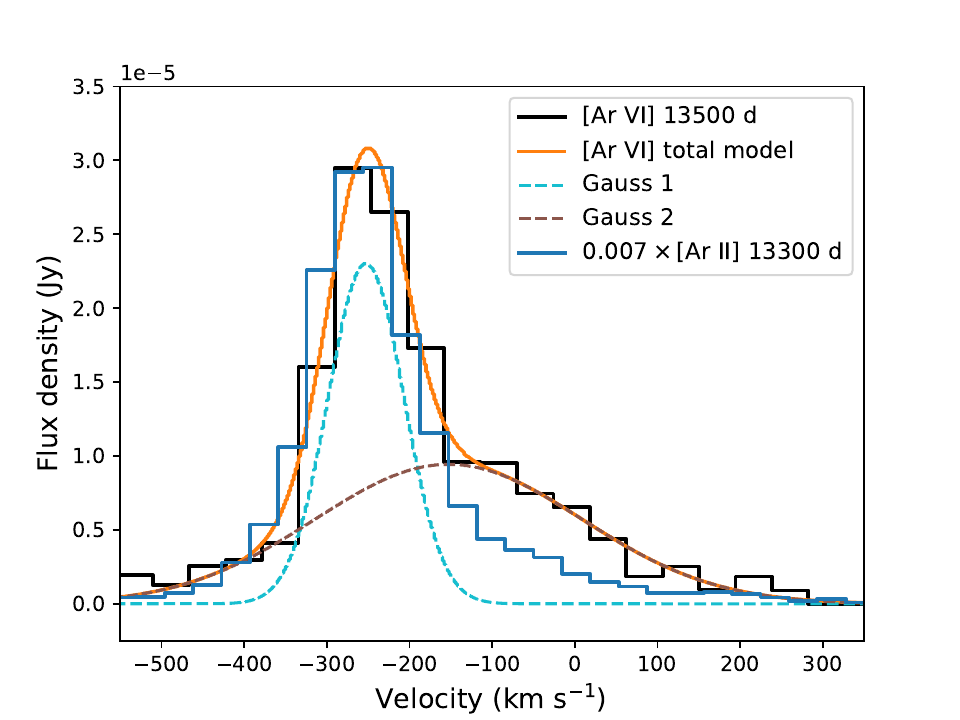}
\caption{Fit to the [\ion{Ar}{6}] profile from the central source at 13,500 days. The line was fitted with the sum of two Gaussians as indicated by the legend. The profile of the [\ion{Ar}{2}] line at 13,300 days is also shown for comparison.}
\label{fig:ar6specfit}
\end{figure}

The properties of the diffuse emission is investigated in more detail in Appendix~\ref{sec:appendix:diffuse}. For the analysis of the central [\ion{Ar}{6}] source, we simply wish to remove this component, as it is clearly not associated with the innermost ejecta and compact object. We therefore create a sub-cube around the [\ion{Ar}{6}] line, where we isolate the emission from the central source by subtracting the continuum and 0-\kms\ line. This was done by fitting the spectra in each spaxel, using a linear model for the continuum and then fitting a narrow Gaussian to account for the 0-\kms\ line as described in Appendix~\ref{sec:appendix:diffuse}. The final cube covers the velocity interval -600--300~\kms\ around the  [\ion{Ar}{6}] line. The line profile extracted from this cube is shown in Figure~\ref{fig:ar6profiles} together with that from the original cube, illustrating the removal of the 0-\kms\ line. The ``clean" line profile from the central source is asymmetric, showing an extended wing on the red side. We characterize the profile by fitting it with a model comprising two Gaussians. The best-fit model is shown in Figure~\ref{fig:ar6specfit} and the parameter values are reported in Table~\ref{tab:lines}. Figure~\ref{fig:ar6specfit} also shows the [\ion{Ar}{2}] line profile for comparison. The blue peaks of the two lines agree very well, and both lines show an extended red wing, though the latter is stronger for the [\ion{Ar}{6}] line. This is quantified by the fit to the [\ion{Ar}{2}] line (Section~\ref{sec:analysis:mrs} and Table~\ref{tab:lines}).

The [\ion{Ar}{6}] line profile in the observation from 12,900~days was also fitted with two Gaussians (F24, see their Table 1), but in this case the total line profile was contaminated by the 0-\kms\ line, which could not be identified and removed due to the lower spectral resolving power. The contamination primarily affected the centroid velocity of the broad component ($15\pm 33$~\kms\ at 12,900~days compared to $-153\pm 11$~\kms\ at 13,500~days) and lead to an overestimate of the line luminosity from the central  source. Specifically, the total luminosity in the two components was $\sim 40\%$ higher at at 12,900~days than at 13,500~days due to the diffuse background contained within the extraction region (similar to the one shown in the top left panel of Figure~\ref{fig:ar6ims}).  On the other hand, we find that the total luminosities are consistent within the uncertainties if we do not remove the 0-\kms\ line at 13,500~days, which implies that there is no evidence of time evolution of the central source on this time scale.  

\begin{figure}
\centering
\includegraphics[width=\hsize]{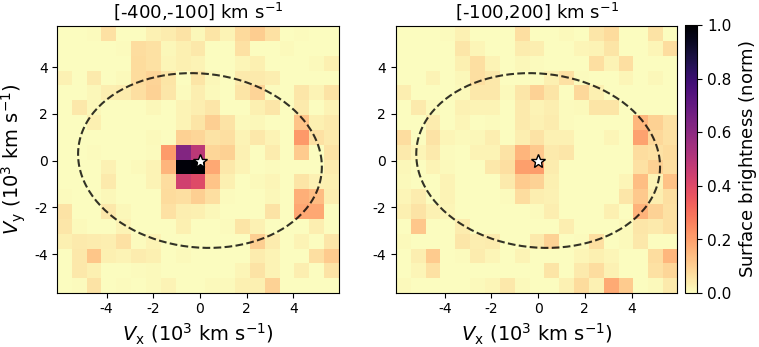}
\caption{Images of the [\ion{Ar}{6}] emission in the radial velocity intervals [-400, -100]~\kms\ (left) and [-100, 200]~\kms\ (right). The continuum and the diffuse 0-\kms\ line have been removed as described in the text. The spatial scale has been translated to a velocity scale for the freely expanding ejecta at the time of the observation. The white star symbol shows the center of the system (from \citealt{Alp2018}) and the black dashed line shows the position of the ER.}
\label{fig:ar6slices}
\end{figure}

In Figure~\ref{fig:ar6slices}, we show images produced from the sub-cube of the [\ion{Ar}{6}] emission from the central source. The images were integrated over the velocity intervals [-400, -100]~\kms\ and [-100, 200]~\kms, which are dominated by the blue peak and extended red wing, respectively. Fitting the central source in the two images with a 2D Gaussian results in positions that are slightly offset to the southeast of the center of the ER (taken from \citealt{Alp2018}), which we assume to coincide with the center of the explosion. The two images give consistent results, with offsets of ($30\pm 10$ mas south, $63\pm 10$ mas east)  for [-400, -100]~\kms\ and ($27\pm 19$ mas south, $71\pm 19$ mas east) for [-100, 200]~\kms. The corresponding position angles (PA), defined anticlockwise from the north, are $116\pm 8$\dg\ and $111\pm 14$\dg, respectively. The uncertainty in the position is dominated by the uncertainty in the absolute astrometry (Section \ref{sec:obs}) for the first image, while the uncertainties in the fit dominate for the second image, where the emission is much fainter.\footnote{The uncertainty in the absolute astrometry of 10~mas should be included when comparing different instruments and epochs, but not when comparing the positions of different lines/components in the G395H observation at 13,500~days. The statistical uncertainties in the fits to the two [\ion{Ar}{6}] images are 3.0 and 16.4~mas, respectively.}
These offsets are also consistent with the offset of ($31\pm 22$ mas south, $39\pm 22$ mas east, PA$=129\pm 25$\dg) obtained for the full profile from the observations at 12,900 days (F24), where the contribution from the 0-\kms\ line could not be removed. The connection between the position and the kick of the compact object is discussed in Section \ref{sec:disc-kick}.    

\begin{figure}
\centering
\includegraphics[width=\hsize]{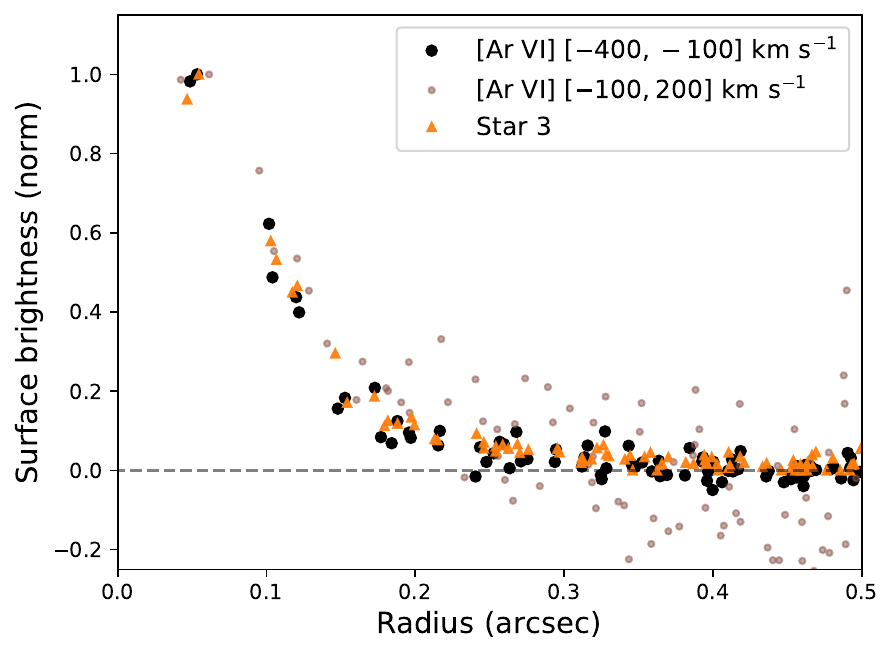}
\caption{Radial profiles produced from the images in Figure~\ref{fig:ar6slices} (black and gray points, respectively). The radial profile of ``Star 3'' is shown by the orange triangles. The latter was extracted from the observation at 12,900 days since the star is partly outside the FOV in the observation at 13,500 days.}
\label{fig:ar6radprofile}
\end{figure}

To investigate the spatial distribution of the [\ion{Ar}{6}] emission in more detail, we show the radial profiles of the images in the two velocity intervals in Figure~\ref{fig:ar6radprofile}. The radial profile of ``Star 3'' at the same wavelength is included for comparison. This star is partially outside the FOV in the most recent observation, so these values were taken from the observation at 12,900~days, which has the same spatial resolution. The comparison shows that the [\ion{Ar}{6}] emission is spatially unresolved, though the scatter in the profile is much larger for the faint emission in the [-100, 200]~\kms\ interval. The FWHM of an unresolved source is 0\farcs{21}, which implies an upper limit of $1.6 \times 10^{17}$ cm for the size of the emission region. The spatial analysis presented above was also carried out on an image produced by integrating over the full velocity interval of the line, which gave results within the 1$\sigma$ uncertainties of those for the blue interval.

\begin{deluxetable*}{llllllllll}[t]
\tablecaption{Summary of fits to emission lines from the innermost ejecta. \label{tab:lines}}
\tablecolumns{10}
\tablenum{2}
\tablewidth{0pt}
\tablehead{
\colhead{} &
\colhead{} &
\colhead{} &
\colhead{} &
\multicolumn3c{First component (highest blueshift)} &
\multicolumn3c{Second component\tablenotemark{a}}  \\ 
\colhead{Line} & 
\colhead{$\lambda$}  &
\colhead{R\tablenotemark{b}}  &
\colhead{Epoch \tablenotemark{c}} &
\colhead{$v_{\rm c}$} &
\colhead{FWHM} &
\colhead{L} &
\colhead{$v_{\rm c}$} &
\colhead{FWHM} &
\colhead{L} \\
\colhead{} & 
\colhead{($\mu$m)}  &
\colhead{(\kms)} &
\colhead{($10^3$ days)} &
\colhead{(\kms)} &
\colhead{(\kms)} &
\colhead{(${\rm 10^{30}\ erg\ s^{-1}}$)} &
\colhead{(\kms)} &
\colhead{(\kms)} &
\colhead{(${\rm 10^{30}\ erg\ s^{-1}}$)} 
}
\startdata
$[$\ion{Fe}{2}$]$\tablenotemark{d}  & 1.6440  & 93  & 13.5  &  $-253\pm 12$ & $< 152$  & $2.20 \pm 0.79$   & \nodata  & \nodata  & \nodata\\
$[$\ion{Ca}{5}$]$\tablenotemark{d}  & 4.1585  & 105  & 13.5  &  $-306\pm 10$ & $164\pm 31$  & $0.41 \pm 0.10$   & \nodata  & \nodata & \nodata\\
$[$\ion{Ar}{6}$]$   & 4.5292 & 96 & 13.5  &  $-252.3\pm 1.8$  & $107.9\pm 5.2$  & $1.724 \pm 0.097$  & $-153\pm 11$  & $380\pm19$  & $2.48\pm 0.14$  \\ 
$[$\ion{Ar}{2}$]$  & 6.9853 & 82  & 12.9+13.3 &  $-260.59\pm 0.42$  & $119.0 \pm 1.3$  & $203.2 \pm 3.4$  & $-188.2 \pm 4.9$  & $380 \pm 10$  & $129.4 \pm 4.1$  \\ 
$[$\ion{S}{4}$]$   & 10.5105 & 118 & 12.9+13.3 &  $-256 \pm 16$  & $196 \pm 74$  & $3.51 \pm 0.87$  & $-18.1 \pm 1.5$  & $100.2 \pm 3.6$  & $14.49 \pm 0.48$  \\ 
$[$\ion{Cl}{2}$]$  & 14.3678 & 112 & 12.9+13.3 & $-304 \pm 14$ & $285 \pm 54$ & $4.40 \pm 0.60$ & \nodata & \nodata & \nodata \\
$[$\ion{S}{3}$]$   & 18.7130 & 147 &  12.9+13.3 &  $-103 \pm 25$  & $330 \pm 33$  & $15.5 \pm 2.9$  & $-12.9 \pm 2.7$  & $124.4 \pm 6.2$  & $39.5 \pm 2.8$  \\ 
\hline 
    $[$\ion{Ca}{4}$]$\tablenotemark{e} & 3.2068  & 138  &  13.50 & $666 \pm 13$ & $138^{*}$ & $0.383 \pm 0.021$ & $789 \pm 18$ & $1086 \pm 44$  &  $6.42 \pm 0.35$ 
\enddata
\tablecomments{Parameter marked by $^{*}$ was kept fixed in the fits.}
\tablenotetext{a} {The second component is part of the emission from the central source for [\ion{Ar}{2}] and [\ion{Ar}{6}], while it is dominated by background for [\ion{S}{3}] and [\ion{S}{4}].}   
\tablenotetext{b} {The spectral resolving power at the relevant wavelength, expressed as the FWHM velocity of an unresolved line. This was obtained from \cite{Pontoppidan2024} for MRS and from the JWST user documentation \citep{jdox2016} for NIRSpec.}  
\tablenotetext{c} {Only the latest epoch is included for NIRSpec as the spectral components cannot be disentangled in the lower-resolution data from 12,900 days.} 
\tablenotetext{d}{Based on fits to spectra extracted from a small $0\farcs{2} \times 0\farcs{1}$ region. An aperture correction has been applied to the luminosity. There are systematic uncertainties due to blending with other lines. See text for details. }
\tablenotetext{e} {A different extraction region was used for this line, see Section~\ref{sec:analysis:ca} and Figure~\ref{fig:ca4slices}.}
\end{deluxetable*}


\subsubsection{Search for other narrow lines from the central region}

The photoionization and shock models presented in F24 predict that the Ar-line emitting innermost ejecta should also emit other lines in the NIR, including lines from Si, S, Ca and Fe. These lines carry diagnostic information about the properties of the ejecta and compact object, but are challenging to detect as they are expected to be faint and blended with strong broad emission lines in many cases. The higher spectral resolving power offered by the new NIRSpec observations allows us to assess the presence of these lines. 

\begin{figure*}
\centering
\includegraphics[width=\hsize]{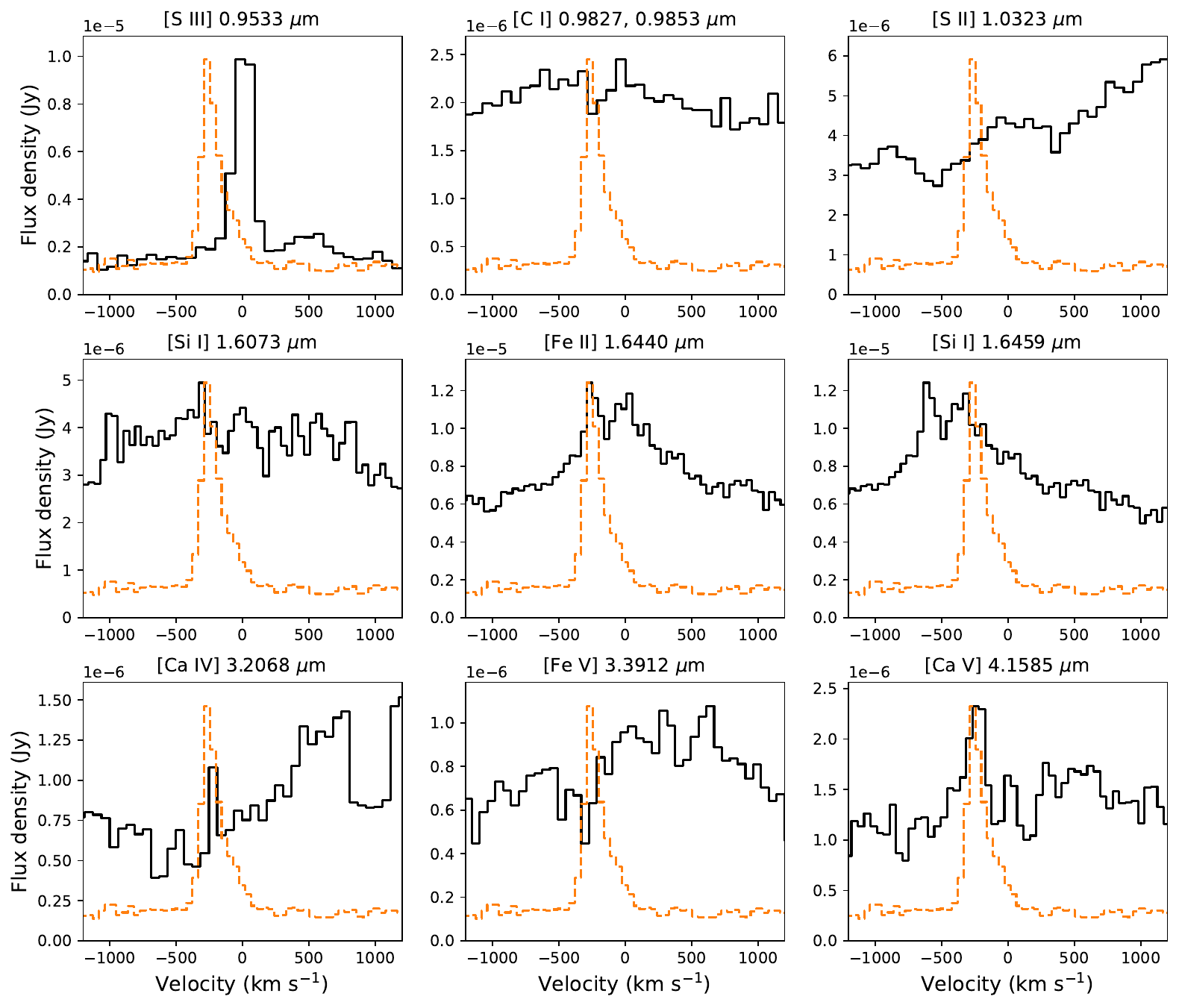}
\caption{Velocity profiles of lines that are predicted to be emitted from the innermost ejecta ionized by the compact object. The line identifications and wavelengths are given at the top of each panel. The spectra were extracted from a $0\farcs{2}\times 0\farcs{1}$ region (2 spaxels) where the [\ion{Ar}{6}] line is brightest. A scaled [\ion{Ar}{6}] profile from the same region is shown in all the panels for reference (dashed orange line). In the case of the [C I] doublet in the top middle panel, the dotted orange line shows the [\ion{Ar}{6}] profile shifted to the second component.}
\label{fig:linesarch}
\end{figure*}

Figure~\ref{fig:linesarch} shows spectra extracted from the central region at the wavelengths of the strongest lines predicted by the models. We used a small extraction region, comprising only the two spaxels where the [\ion{Ar}{6}] line is brightest, in order to maximize the signal from any narrow lines from the center compared to the ``background" from the surrounding ejecta. The profile of the [\ion{Ar}{6}] line extracted from the same region is shown for reference in all the panels in Figure~\ref{fig:linesarch}. This comparison indicates the presence of similar blueshifted narrow components in [\ion{Fe}{2}]~$1.6440$~$\mu$m and [\ion{Ca}{5}]~4.1585~$\mu$m.  The [\ion{Ca}{4}]~3.2068~$\mu$m line does not show such a component even though it is expected to be similar to [\ion{Ca}{5}] according to the photoionization models (Section~\ref{sec:disc:linemod}), but instead  exhibits a broader, redshifted excess (the spike at $\sim -230$~\kms, and sharp drop at $\sim$ 800--1100~\kms\ are artifacts). The properties of the [\ion{Fe}{2}] line and the two Ca lines are analyzed in Sections~\ref{sec:analysis:fe} and \ref{sec:analysis:ca} below, respectively. 

There are no narrow blueshifted components similar to [\ion{Ar}{6}] seen in any of the other line profiles in Figure~\ref{fig:linesarch}, including [\ion{S}{3}]~$0.9533$~$\mu$m, [\ion{C}{1}]~$0.9827, 0.9853$~$\mu$m, [\ion{S}{2}]~$1.0323$~$\mu$m, [\ion{Si}{1}]~$1.6073$~$\mu$m, [\ion{Si}{1}]~$1.6459$~$\mu$m, and [\ion{Fe}{5}]~$3.3912$~$\mu$m. In the case of the [\ion{S}{3}]~$0.9533$~$\mu$m line (upper left panel in Figure~\ref{fig:linesarch}), we note a  narrow line close to 0~\kms,  which is due to diffuse emission similar to that found in [\ion{Ar}{6}], as well as a bump at $\sim 600$~\kms, which we attribute to \ion{H}{1}~0.9549~$\mu$m. There is also a narrow feature at $\sim 0$~\kms\  in the [\ion{Fe}{2}]~$1.6440$~$\mu$m profile, or alternatively at $\sim -350$~\kms\ in the [\ion{Si}{1}]~$1.6459$~$\mu$m profile. From an investigation of images at the corresponding wavelengths, we find that this is primarily due to contamination by [\ion{Fe}{2}] emission from the northern outer ring and scattered light from the ER, while there is no evidence for a compact blueshifted source in any of the [\ion{Si}{1}] lines.  The non-detection of [\ion{Ar}{6}]-like emission components for all these lines is confirmed by running a point-source detection algorithm on images (Appendix~\ref{sec:appendix:ul}). We present upper limits on the luminosities of [\ion{Ar}{6}]-like components for all the lines in Appendix~\ref{sec:appendix:ul}. 

\subsubsection{The [\ion{Fe}{2}]1.6440~$\mu$m line}
\label{sec:analysis:fe}

\begin{figure*}
\centering
\includegraphics[width=\hsize]{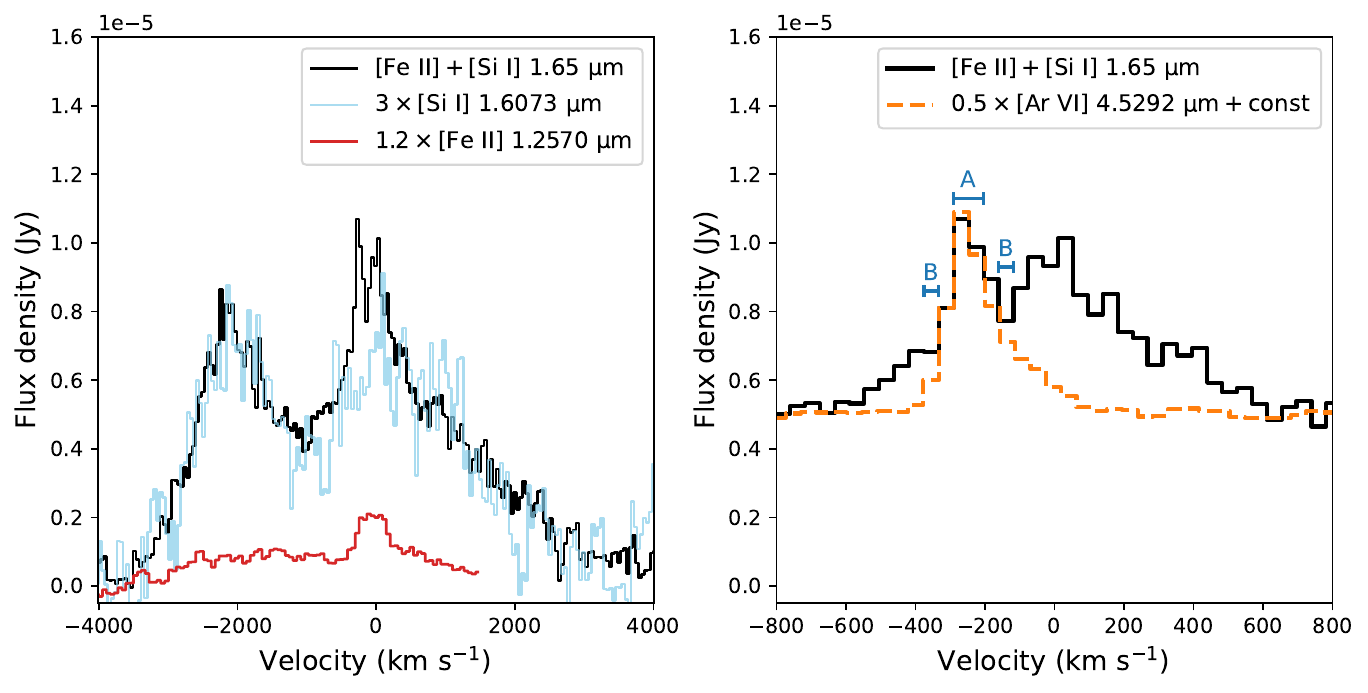}
\caption{Velocity profiles of the [\ion{Fe}{2}]+[\ion{Si}{1}]~1.65~$\mu$m blend, centered at 0~\kms\ for the [\ion{Fe}{2}]~1.6440~$\mu$m component. The spectrum was extracted from a $0\farcs{2}\times 0\farcs{1}$ region (2 spaxels) where the [\ion{Ar}{6}] line is brightest. Left: Comparison with the [\ion{Si}{1}]~1.6073~$\mu$m and [\ion{Fe}{2}]~1.2570~$\mu$m profiles extracted from the same region. The former is centered at 0~\kms\ for the [\ion{Si}{1}]~1.6459~$\mu$m component. The [\ion{Fe}{2}]~1.2570~$\mu$m line is only shown for velocities $< 1500$~\kms\  as it is contaminated by the blue wing of Pa$\beta$~1.2822~$\mu$m at higher redshifts.  Right: The [\ion{Fe}{2}]+[\ion{Si}{1}]~1.65~$\mu$m line profile zoomed in over a narrower velocity range to highlight the peak at $\sim -250$~\kms. The profile of the [\ion{Ar}{6}] line extracted from the same region is included for comparison. Velocity intervals marked by A and B were used for producing the images in Figure~\ref{fig:fe_ims}. 
}
\label{fig:fe_profiles}
\end{figure*}

To determine whether the narrow, blue peak in the [\ion{Fe}{2}]~1.6440~$\mu$m line seen in Figure~\ref{fig:linesarch} is consistent with [\ion{Ar}{6}], we need to consider the underlying broad profile of blended [\ion{Fe}{2}]~$1.6440$~$\mu$m and [\ion{Si}{1}]~$1.6459$~$\mu$m emission.  If the narrow feature is associated with [\ion{Si}{1}]~$1.6459$~$\mu$m, the inferred blueshift is $\sim -600$~\kms, which would make it inconsistent with the central [\ion{Ar}{6}] source. The left panel of Figure~\ref{fig:fe_profiles} shows the  [\ion{Fe}{2}]+[\ion{Si}{1}] profile from Figure~\ref{fig:linesarch} over a wider velocity interval, illustrating that the emission extends to $\sim \pm 4000$~\kms. To assess the relative contributions of Fe and Si to this profile, we use the [\ion{Fe}{2}]~1.2570~$\mu$m  and [\ion{Si}{1}]~1.6073~$\mu$m lines, which are the strongest and least blended  [\ion{Fe}{2}] and [\ion{Si}{1}] lines in the NIRSpec range. For [\ion{Fe}{2}]~1.2570~$\mu$m, we also improve the S/N by adding the spectra from G140H/F070LP and G140H/FI00LP.

As seen in Figure~\ref{fig:fe_profiles} (left), the [\ion{Si}{1}]~1.6073~$\mu$m line shows the best overall agreement with the full  [\ion{Fe}{2}]+[\ion{Si}{1}]~1.65~$\mu$m profile, being dominated by two broad, asymmetric features peaking around -2000 and 0~\kms, respectively. This [\ion{Si}{1}]  profile does not, however, exhibit narrow peaks matching those at  $\sim -250$ and 0~\kms\ in the  [\ion{Fe}{2}]+[\ion{Si}{1}]~blend.  The difference between the  peak at $\sim -250$~\kms\ in the  [\ion{Fe}{2}]+[\ion{Si}{1}]~blend and the [\ion{Si}{1}]~1.6073~$\mu$m profile $\times 3$ (where the factor 3 best matches the overall profile) is significant at 4$\sigma$, considering the noise level measured in the continuum close to both lines.  The peak at 0~\kms\ in the [\ion{Fe}{2}]+[\ion{Si}{1}]~blend is due to contamination of [\ion{Fe}{2}]~1.6440~$\mu$m emission from the ER and the northern outer ring.

The [\ion{Fe}{2}]~1.2570~$\mu$m profile from the central region is dominated by a flat-topped feature in the \mbox{$\sim \pm 300$}~\kms\ range, without any narrow, blueshifted [\ion{Ar}{6}]-like component (Figure~\ref{fig:fe_profiles}).  The [\ion{Fe}{2}]~1.2570~$\mu$m line was multiplied by a factor 1.2 in this figure, which is the observed ratio of [\ion{Fe}{2}]~1.6440~$\mu$m / [\ion{Fe}{2}]~1.2570~$\mu$m in the ER (not corrected for extinction). This line ratio is expected to be the same in the ejecta, with the major caveat that it could be affected by  wavelength-dependent scattering effects due to dust in the ejecta, which could remove a possible narrow component in the [\ion{Fe}{2}]~1.2570~$\mu$m profile  (Section~\ref{sec:disc:dust}). 

\begin{figure*}
\centering
\includegraphics[width=\hsize]{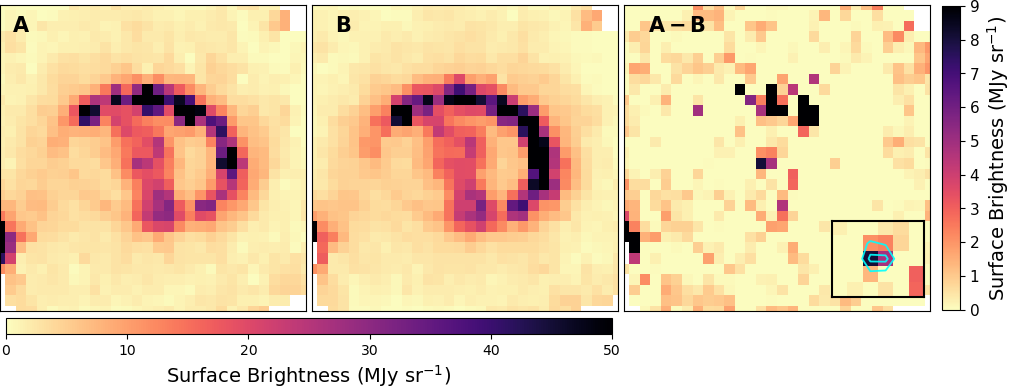}
\caption{Images of the [\ion{Fe}{2}]+[\ion{Si}{1}]~1.65~$\mu$m line blend. Left: Image integrated over the blue peak of the line (velocity interval A in Figure~\ref{fig:fe_profiles}). Middle: Average image of the emission on both sides of the blue peak (velocity intervals marked by B in Figure~\ref{fig:fe_profiles}). Right: Result of subtracting image B from image A. There is a bright source at the center due to the narrow blueshifted component of [\ion{Fe}{2}]~1.6440~$\mu$m. The inset shows the central region of the image with contours from the [\ion{Ar}{6}] line superposed (from Figure~\ref{fig:ar6slices}, left). The horizontal color bar at the bottom applies to the left and middle panels, while the color bar to the right applies to the right panel. }
\label{fig:fe_ims}
\end{figure*}

We also produce images to investigate the spatial properties of the  narrow [\ion{Fe}{2}]~1.6440~$\mu$m component. Figure~\ref{fig:fe_ims} shows an image integrated over the peak of the line, an average image of the emission on both sides of the peak, as well as the difference between the two. The velocity intervals used for the images are marked in the line profile in the right panel of  Figure~\ref{fig:fe_profiles}. The difference image in Figure~\ref{fig:fe_ims} shows a point-like source at the center, as well as bright emission in the northern part of the ER. The latter is due to shocked gas, which is known to be blueshifted in the north (e.g., \citealt{Jones2023}), while the former is consistent with the position of the central [\ion{Ar}{6}] source. This is seen from the comparison with the [\ion{Ar}{6}] contours (inset in Figure~\ref{fig:fe_ims}) as well as a fit with a 2D Gaussian, which gives a best-fit position offset from the center by ($19\pm 6$ mas south, $72\pm 3$ mas east, PA$=107\pm 5$\dg), consistent with the fit to the [\ion{Ar}{6}]  position within the 2$\sigma$ uncertainties.\footnote{Here and in Section~\ref{sec:analysis:ca} we only report the statistical uncertainties in the positions as we are interested in the relative positions of different lines from the same observation.} The central source is also detected as a point source at $5\sigma$ using the \texttt{DAOStarFinder} tool from \texttt{photutils} \citep{Bradley2023}. 

To determine the total luminosity in the narrow [\ion{Fe}{2}] component, we need to consider a larger region than the $0\farcs{2}\times 0\farcs{1}$ region used to extract the spectra in Figure~\ref{fig:fe_profiles}. However, the narrow peak is not detected in the spectrum from the larger $0\farcs{3}$ radius circular region used for the [\ion{Ar}{6}] analysis (Section~\ref{sec:analysis:ar6}) due to the strong emission from the broad [\ion{Si}{1}]~1.6459~$\mu$m line. A comparison of the [\ion{Ar}{6}] line profiles extracted from the two different regions reveals very similar shapes, but a factor $\sim 3$ difference in luminosity. We therefore fit the narrow [\ion{Fe}{2}] peak using the spectrum from the small region (after subtracting the ``continuum" from intervals B in Figure~\ref{fig:fe_profiles}), but apply an aperture correction to the luminosity based on the [\ion{Ar}{6}]  line. The best-fit parameters of the Gaussian model are reported in Figure~\ref{fig:fe_profiles}, showing that the blueshifted line centroid is consistent with the narrow component of the [\ion{Ar}{6}] line, while we only obtain an upper limit on the FWHM. 

In summary, we find that the narrow component of the [\ion{Fe}{2}]~1.6440~$\mu$m line has spectral and spatial properties consistent with the central [\ion{Ar}{6}] source associated with the compact object. The feature has a formal significance of $>4\sigma$ in spectra and images, but  we stress that this depends on  assumptions about the background from the overlapping broad emission lines. Finally, the lack of a similar peak in  [\ion{Fe}{2}]~1.2570~$\mu$m adds further uncertainties regarding the origin of the [\ion{Fe}{2}]~1.6440~$\mu$m feature.  

\subsubsection{The [\ion{Ca}{4}]~3.2068~$\mu$m and [\ion{Ca}{5}]~4.1585~$\mu$m lines}
\label{sec:analysis:ca}

\begin{figure}
\centering
\includegraphics[width=\hsize]{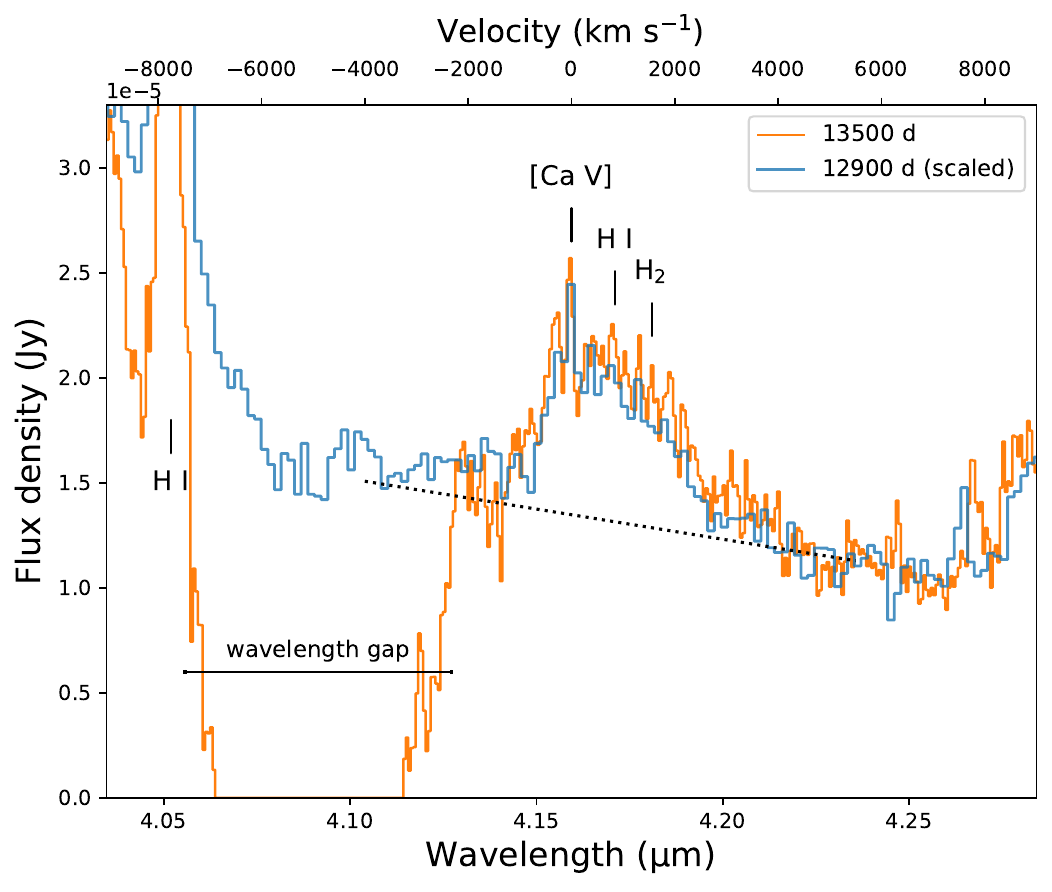}
\caption{Spectrum of the central ejecta in the wavelength region around the [\ion{Ca}{5}] line at 13,500 and 12,900~days. The extraction region is shown in Figure~\ref{fig:ar6ims}. The drop between $\sim$~4.06--4.12~$\mu$m at 13,500 days is due to the wavelength gap of the G395H grating. The strongest lines expected in this wavelength region are identified by the labels. The dotted line shows the continuum level used to isolate the blended emission feature around [\ion{Ca}{5}] in Figure~\ref{fig:ca5profiles}.}
\label{fig:ca5specreg}
\end{figure}

We start our investigation of the Ca lines with the [\ion{Ca}{5}] line, as Figure~\ref{fig:linesarch} reveals a blueshifted peak that coincides with the [\ion{Ar}{6}] line. The analysis of the [\ion{Ca}{5}] line is complicated by blending with other lines, as well as the fact that it is located close to the wavelength gap of the G395H grating. This is illustrated in Figure~\ref{fig:ca5specreg}, which shows the spectrum from the central region ($0\farcs{3}$ radius circle, see Figure~\ref{fig:ar6ims}) in a 0.25~$\mu$m wide wavelength interval centered on the line. The comparison with the G395M spectrum from 12,900~days, which does not have any wavelength gap, shows that the full line profile from [\ion{Ca}{5}] is captured in the G395H spectrum, but that the continuum on the blue side of the line is not included. We therefore use the G395M spectra to determine the slope of the continuum around the broad emission feature that includes [\ion{Ca}{5}], as well as likely \ion{H}{1}~4.1708~$\mu$m and H$_2$~4.1811~$\mu$m lines (see Figure~\ref{fig:ca5specreg}). The continuum level was normalized by a factor 0.83, which accounts for the variation in the background and continuum between the observations. We note that the linear model for the continuum is a simplification, as we expect the emission at these wavelengths to have contributions from both the extended red tail of the Br$\alpha$~4.0523~$\mu$m line from ejecta interacting with the RS and synchrotron emission from the RS (see \citealt{Larsson2023,Jones2023}). The RS is physically far away from the innermost ejecta, so this emission is due to the projection from high latitudes.    

\begin{figure}
\centering
\includegraphics[width=\hsize]{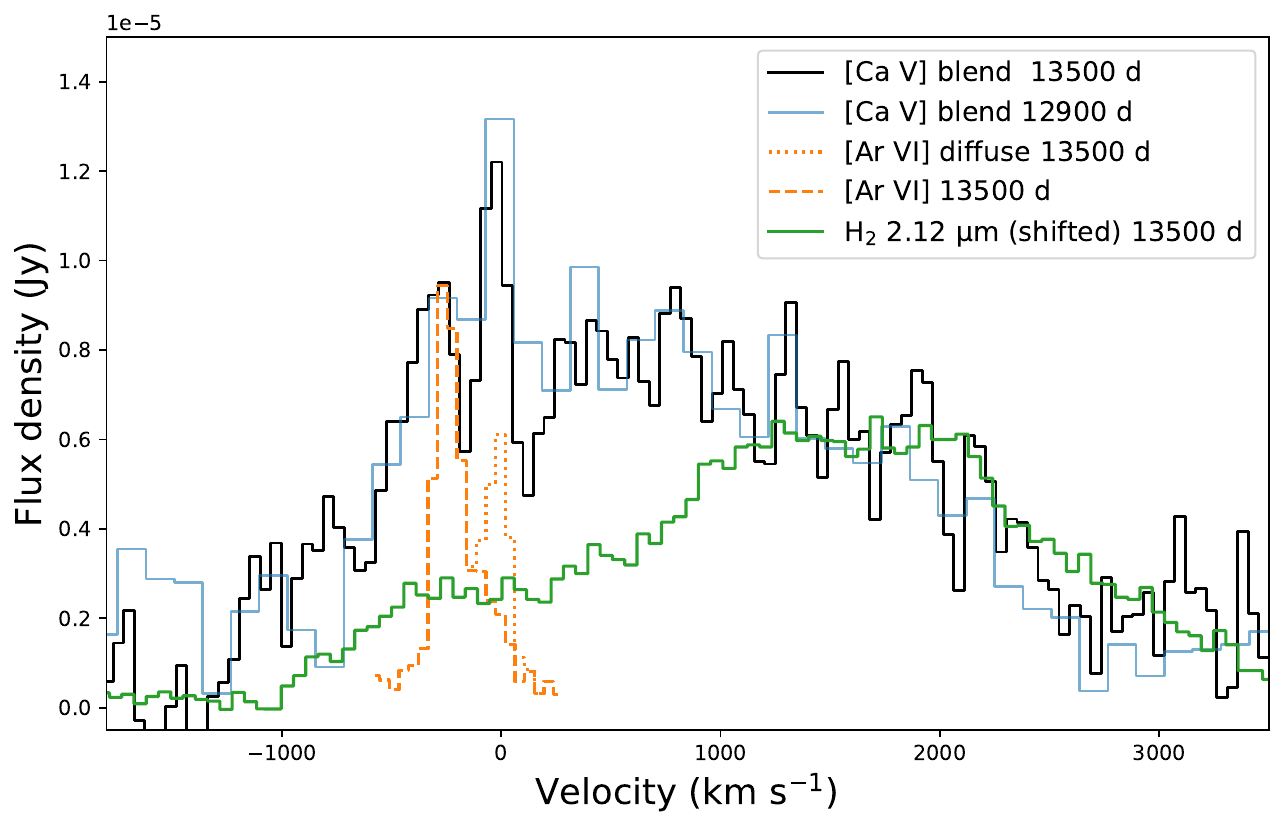}
\caption{Velocity profile of the broad, blended emission feature including the [\ion{Ca}{5}] line at 13,500 and 12,900~days (see also Figure~\ref{fig:ca5specreg}). The spectrum was extracted from the central ejecta, as shown in Figure~\ref{fig:ar6ims}. The [\ion{Ar}{6}] profiles from 13,500~days with and without the diffuse 0-\kms\ line are included for comparison (from Figure~\ref{fig:ar6profiles}). To illustrate a plausible level of contribution from the H$_2$~4.1811\ $\mu$m line, we also show the line profile of the H$_2$~2.1218$~\mu$m line, shifted to 4.1811~$\mu$m and scaled by a factor of 0.24 to match the flux level around the peak of the line.}
\label{fig:ca5profiles}
\end{figure}

Figure~\ref{fig:ca5profiles} shows the resulting continuum-subtracted emission feature at both epochs. The [\ion{Ar}{6}] profile from 13,500~days is included for reference, illustrating that the peaks at $\sim -300$~\kms\ (compact source) and $\sim 0$~\kms\ (diffuse component) appear also in the [\ion{Ca}{5}] profile, albeit with different relative intensities. The lower-resolution G395M spectrum from 12,900~days does not clearly separate these components, though there is a peak at $\sim 0$~\kms. 

The lines most likely to be blended with the [\ion{Ca}{5}] line are expected to have their velocity peaks at 837~\kms\  (\ion{H}{1} 13--6 4.1708~$\mu$m) and 1558~\kms\ (H$_2$ 0--0 S(11) 4.1811~$\mu$m). We estimate a plausible contribution from the latter by extracting the profile of the H$_2$ (1,0) S(1) 2.1218~$\mu$m line from the same region and shifting it to the velocity corresponding to 4.1811~$\mu$m (see Figure~\ref{fig:ca5profiles}). This comparison indicates that H$_2$ can account for all the emission at velocities $\gtrsim 1000$~\kms\ and about half of the flux in the broad feature at lower velocities. In the case of \ion{H}{1}~4.1708~$\mu$m, there is no unblended line that can be used as a template in a similar way as for H$_2$. However, we can use the Br$\alpha$ line to estimate its maximal contribution. We use the observation from 12,900~days to avoid the wavelength gap and measure the flux ratio of Br$\alpha$ and \ion{H}{1}~4.1708~$\mu$m in the ER. The ratio is found to be 50, in agreement with the theoretical ratio for Case B recombination (I(\ion{H}{1} 5--4)/I(\ion{H}{1} 13--6) = 52.6 for electron number density $N_{\rm e} =10^4\ \rm{cm^{-3}}$ and temperature $T_{\rm e} = 10^4$~K; \citealt{Storey1995}). This ratio is weakly dependent on $N_{\rm e}$ and $T_{\rm e}$, and we expect it to be very similar in the ejecta. Based on the peak of the Br$\alpha$ line in the central ejecta at $\sim -1700$~\kms, the expected peak flux density from \ion{H}{1} in the [\ion{Ca}{5}] profile is then $5.4 \times 10^{-7}$~Jy at $\sim -860$~\kms, which corresponds to $\sim 1/6$ of the observed flux density at that velocity (Figure~\ref{fig:ca5profiles}).
 
This raises the question of the origin of the excess in the broad feature between $\pm 1000$~\kms\ in Figure~\ref{fig:ca5profiles}. There are several possible explanations, all of which could contribute to some extent. First, the ratio of the H$_2$ 2.1218 and 4.1811~$\mu$m lines is dependent on density and temperature (as inferred from the models for photodissociation regions by \citealt{Draine1996}). This would result in different total line profiles when integrating over regions of ejecta with different physical conditions. More of the blueshifted emission than indicated in Figure~\ref{fig:ca5profiles} could thus be due to H$_2$. Another possibility is that the uncertain continuum subtraction has contributed to creating an artificial excess at these velocities. Finally, it is possible that some of the excess is due to [\ion{Ca}{5}], which would imply that the extent of the emission regions and/or the impact of dust are slightly different for [\ion{Ca}{5}] and [\ion{Ar}{6}]. 

\begin{figure}
\centering
\includegraphics[width=\hsize]{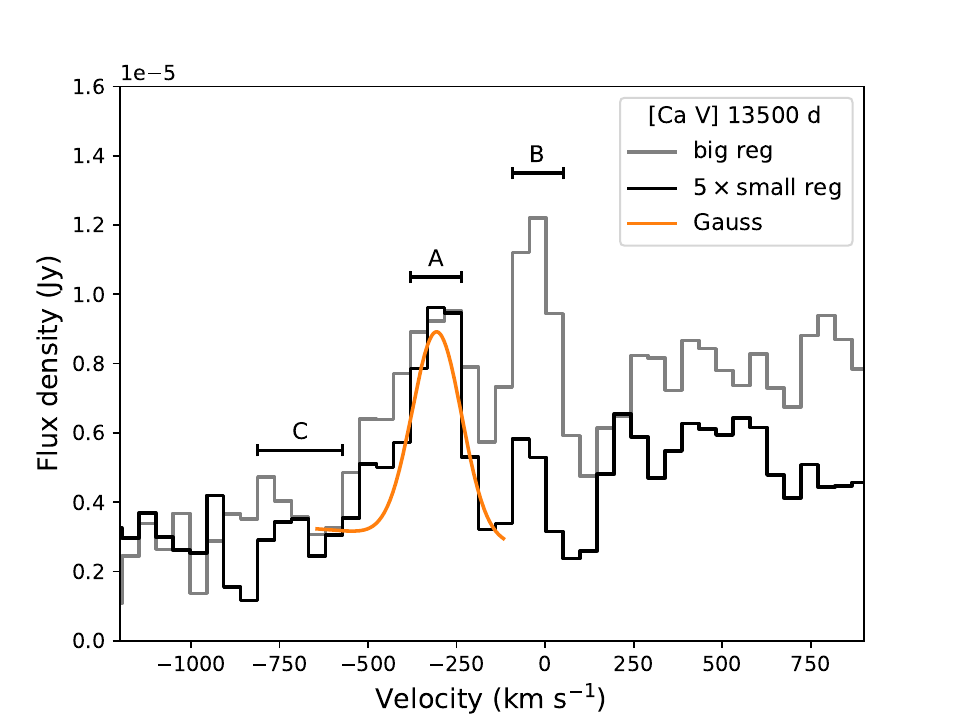}
\caption{Close-up view of the narrow peaks from [\ion{Ca}{5}] in the velocity profile in Figure~\ref{fig:ca5profiles} (gray) together with a spectrum covering the same velocity interval, but extracted from a small $0\farcs{2}\times 0\farcs{1}$  region where the [\ion{Ar}{6}] emission peaks (black). The region around the main peak in the latter spectrum was fitted by a Gaussian (orange) and a straight line. The velocity intervals indicated by A, B, and C were used for producing the images in Figure~\ref{fig:ca5im}.}
\label{fig:ca5fit}
\end{figure}

\begin{figure*}
\centering
\includegraphics[width=\hsize]{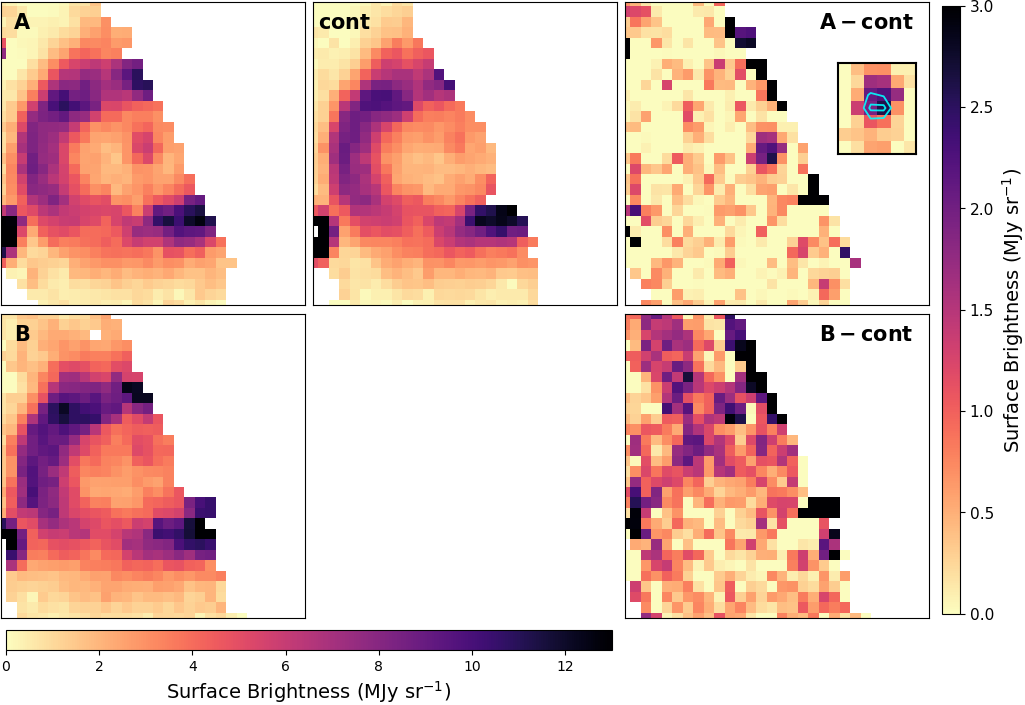}
\caption{Images of the [\ion{Ca}{5}] emission and adjacent ``continuum". The missing data in the upper right of all images is due to the wavelength gap of the G395H grating. Top left: Image integrated over the blue peak of the line (velocity interval A in Figure~\ref{fig:ca5fit}). Top middle: Image integrated over velocity interval C in Figure~\ref{fig:ca5fit}. This interval is dominated by continuum, but also contains weak line emission (see text for details). Top right: Result of subtracting the continuum image from image A. The inset shows the central region of the image with contours from the [\ion{Ar}{6}] line superposed (from Figure~\ref{fig:ar6slices}, left). Bottom left: Image integrated over the 0-\kms\ line (velocity interval B in Figure~\ref{fig:ca5fit}). Bottom right: Result of subtracting the continuum image from image B. The horizontal color bar at the bottom applies to the left and middle panels, while the color bar to the right applies to the right panels. The continuum-subtracted images in the right panels show similarities with the [\ion{Ar}{6}] lines images in Figure~\ref{fig:ar6ims}, showing that the blueshifted component originates from a compact central source, while the spectrum at $\sim 0$ \kms\ is dominated by extended diffuse emission.
}
\label{fig:ca5im}
\end{figure*}

Due to the uncertainties regarding blended lines, we investigate the spatial properties of the [\ion{Ca}{5}] emission by producing images integrated over only the three brightest spectral bins in each of the two peaks, thereby maximizing the signal from [\ion{Ca}{5}]. These velocity intervals are marked A and B in Figure~\ref{fig:ca5fit}, and the resulting images are shown in the left column of Figure~\ref{fig:ca5im}. We also produce an image integrated over a nearby wavelength region (marked by C in Figure~\ref{fig:ca5fit}) and subtract it from the images of the line peaks, the results of which are also shown in  Figure~\ref{fig:ca5im}. Region C is not a true continuum region, but it is useful as an illustration of the emission close to the peaks, and it works well for subtracting the strong emission from the ER, as evident from Figure~\ref{fig:ca5im}.    

The images show that the blue peak originates from a compact source near the center of the system, while the 0-\kms\ line is dominated by diffuse emission.  Although there is missing data in part of the images due to the G395H wavelength gap, it is clear that the spatial structure of the diffuse emission in the north-east at  $\sim 0$~\kms\  is similar to that observed for [\ion{Ar}{6}] and [\ion{Mg}{4}] (Figure~\ref{fig:ar6ims} and Appendix~\ref{sec:appendix:diffuse}).  Fitting the compact source in the continuum-subtracted image of the blue peaks gives a position ($30 \pm 14$ mas north, $48 \pm 14$ mas east, PA$=58\pm 14$\dg) from the center, which is further north than [\ion{Ar}{6}] at 3$\sigma$. We note, however, that this result is dependent on systematic uncertainties in the continuum subtraction, as there is also some emission in the ejecta region of the continuum image (see Figure~\ref{fig:ca5im}). The position in the image without continuum subtraction is further north at ($89 \pm 5$ mas north, $29 \pm 5$ mas east, PA$=18\pm 3$\dg). 

We fit the blue peak of the [\ion{Ca}{5}] profile with a single Gaussian to characterize its main properties, using a linear function to approximate the underlying blended emission.  As for the [\ion{Fe}{2}] line, we perform this fit using a spectrum extracted from the small $0\farcs{2}\times 0\farcs{1}$  region in order to maximize the signal from the central source. The fit is shown in Figure~\ref{fig:ca5fit} and the best-fit parameters are reported in  Table~\ref{tab:lines}, where we have applied the same aperture correction for the luminosity as for [\ion{Fe}{2}]. We note that there is an excess on the blue side of the  [\ion{Ca}{5}]  line, which results in a slightly larger FWHM and lower centroid velocity than obtained for the narrow component of the [\ion{Ar}{6}] line. This excess is  more pronounced in the spectrum extracted from the larger region (Figure~\ref{fig:ca5fit}), which indicates that it may be due to contamination by other lines.

The [\ion{Ca}{4}] line is more straight-forward to analyze as it is not affected by significant blending with other lines or the wavelength gap. However, it is clear that this line does not show any centrally located narrow blueshifted [\ion{Ar}{6}]-like emission component (Figure~\ref{fig:linesarch}). It instead exhibits a redshifted emission region just north of the center, as previously noted in the observations from 12,900~days in F24. This result is puzzling considering that Ar and Ca are expected to be co-located in the ejecta, and because the [\ion{Ca}{4}] and [\ion{Ca}{5}] emission is expected to originate from the same region according to the models in F24. We therefore investigate the [\ion{Ca}{4}] line in more detail.  

\begin{figure*}
\centering
\includegraphics[width=\hsize]{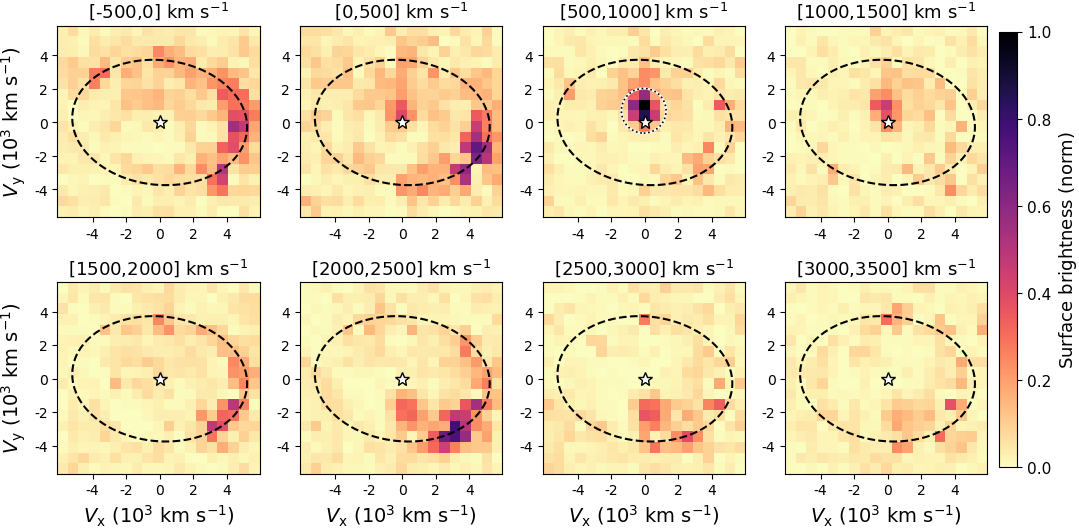}
\caption{Images of the [\ion{Ca}{4}] emission in radial velocity intervals of 500~\kms, spanning the range -500--3500~\kms. There is no significant emission at higher or lower velocities. The continuum was subtracted based on fits to the spectra in each spaxel. The spatial scale has been translated to a velocity scale for the freely expanding ejecta at the time of the observation. The white star symbol shows the center of the system (from \citealt{Alp2018}), and the black dashed line shows the position of the ER. The black/white dotted line in the third panel in the upper row shows the region used to extract the spectra in Figure.~\ref{fig:ca4profile}.}
\label{fig:ca4slices}
\end{figure*}

Figure~\ref{fig:ca4slices} shows images of the [\ion{Ca}{4}] emission integrated over 500~\kms\ wide intervals in radial velocity. The continuum spectrum in each spaxel was fitted by a straight line and subtracted before producing these images. Notably, the results show no significant [\ion{Ca}{4}] emission at blueshifts $<-500$~\kms. The interval [-500, 500]~\kms\ is dominated by emission from shocked gas in the ER, with similar properties as other lines from the ER, being blueshifted in the north and redshifted in the south (e.g., \citealt{Jones2023}). This velocity interval also shows some diffuse emission, but we do not detect any clear narrow 0-\kms\ line similar to the the ones identified for the [\ion{Ca}{5}], [\ion{Mg}{4}], and [\ion{Ar}{6}] lines discussed above and in Appendix~\ref{sec:appendix:diffuse} (see also Figure~\ref{fig:ca4profile}). 

In the northern part of the ejecta, we see emission in the velocity bins extending between 0--1500~\kms\ in Figure~\ref{fig:ca4slices}. The emission is dominated by a small region just north of the center, but there is also a fainter region extending all the way to the northern part of the ER. In the south, there are two emission regions seen at radial velocities $\sim$~2000--3500~\kms, one in the southern ejecta and one superposed on the southwest part of the ER. The latter exhibits similar properties as observed for many other emission lines (\citealt{Larsson2023}, Kavanagh et al., in prep.) and can be attributed to the dense inner ejecta interacting with the RS and ER. The emission in the southern ejecta is likely photoionized by X-rays from this interaction region. 

The emission in the northern ejecta is more likely to have a connection with energy input from the compact object. The line profile from the brightest region is shown in Figure~\ref{fig:ca4profile}, together with the profile from 12,900~days, which is similar. We characterize the line profile by fitting it with two Gaussians and report the results in Table~\ref{tab:lines}. The peak of the line is at $666\pm13$~\kms, significantly offset from the [\ion{Ar}{6}] peak at $-252.3 \pm 1.8$~\kms. This implies that the emission regions of the two lines are physically distinct, despite the fact that there is some overlap in projection in the images (cf. Figures~\ref{fig:ar6slices} and \ref{fig:ca4slices}). The peak of the [\ion{Ca}{4}] emission in the sky plane is ($136 \pm 7$ mas north, $22 \pm 7$ mas east, PA$=9\pm 3$\dg) from the center, based on fits with a 2D Gaussian in an image integrated around the peak of the line. This position is offset from the [\ion{Ar}{6}] line by more than 5$\sigma$.

\begin{figure}
\centering
\includegraphics[width=\hsize]{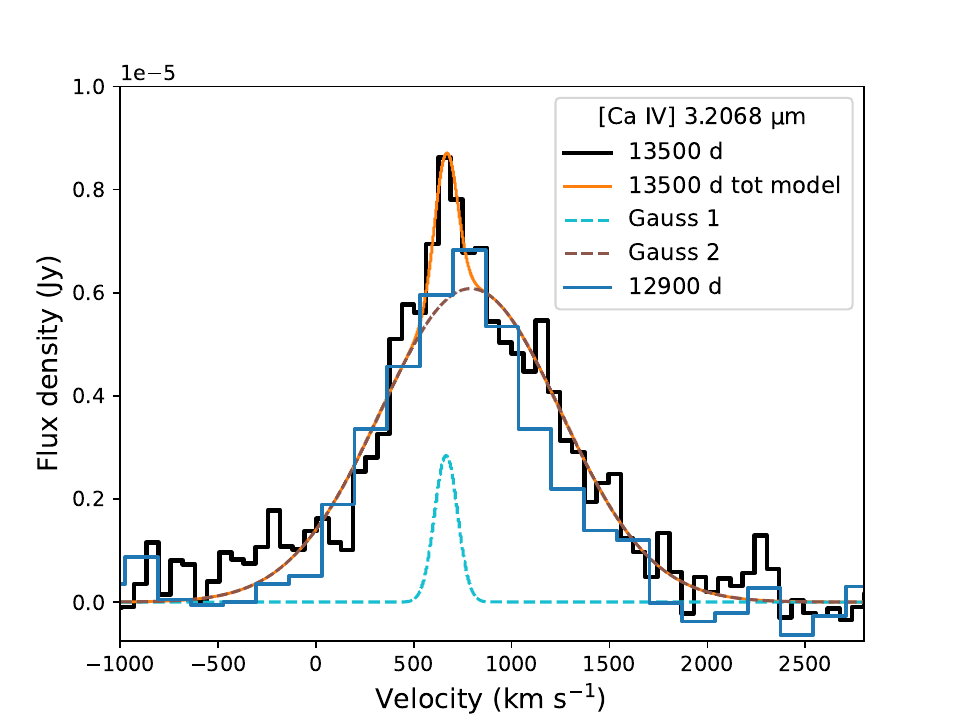}
\caption{Velocity profile of the [\ion{Ca}{4}] line from near the center at 13,500 and 12,900~days. The extraction region is shown in Figure~\ref{fig:ca4slices}. The profile at 13,500~days was fitted with the sum of two Gaussians as indicated by the legend.}
\label{fig:ca4profile}
\end{figure}

\subsection{Line emission in MIRI/MRS}
\label{sec:analysis:mrs}

\begin{figure*}[t]
\centering
\includegraphics[width=\hsize]{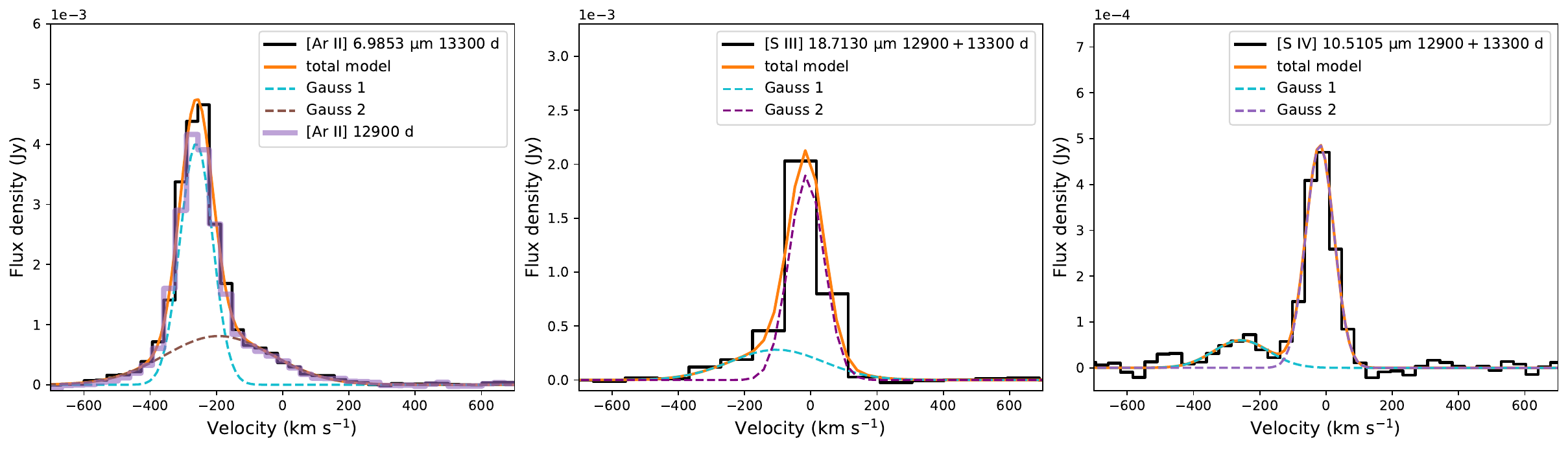}
\caption{Velocity profiles of narrow lines from the central ejecta detected with the MRS. Panels from left to right show [\ion{Ar}{2}]~$6.9853~\mu$m, [\ion{S}{3}]~$18.7130~\mu$m, and [\ion{S}{4}]~$10.5105~\mu$m, respectively. The [\ion{Ar}{2}] profiles are shown separately for the observations at  13,300 and 12,900 days, while the profiles combined from both observations are shown for the [\ion{S}{3}] and  [\ion{S}{4}] lines, which have lower S/N. All panels also show the best-fit models comprising two Gaussian components (for clarity, only the fit to the data at 13,300 days is shown for [\ion{Ar}{2}]). Both Gaussian components originate from the central source for [\ion{Ar}{2}], while only the blueshifted component (“Gauss 1")  can be linked to the central source for [\ion{S}{3}] and  [\ion{S}{4}]. The 2nd components for the latter lines are instead dominated by scattered light from the ER and/or contributions from the northern outer ring and diffuse emission.}
\label{fig:mrsprofiles}
\end{figure*}

The temporal evolution of all emission lines in SN~1987A observed with the MRS at 12,900 and 13,300 days is investigated in Kavanagh et al. (in prep.) This shows that there is no significant time evolution of the narrow, blueshifted components from the central ejecta for the three lines previously identified in F24, i.e. [\ion{Ar}{2}]~6.9853~$\mu$m, [\ion{S}{4}]~10.5105~$\mu$m, and [\ion{S}{3}]~18.7130~$\mu$m. This is illustrated for the [\ion{Ar}{2}] line in the left panel of Figure~\ref{fig:mrsprofiles}. Given the lack of time evolution, we added the spectra from the two observing epochs to improve the S/N. No background spectrum was subtracted. We fitted the continuum-subtracted line profiles with two Gaussians, where the two components capture the shape of the line profile of the central source in the case of [\ion{Ar}{2}], while the second component (at $\sim 0$~\kms) is dominated by diffuse background and contamination by the ER and northern outer ring for [\ion{S}{3}] and [\ion{S}{4}]. The best-fit parameters for the co-added spectra are reported in Table~\ref{tab:lines}, while the line profiles and models are plotted in Figure~\ref{fig:mrsprofiles} ( where we show the co-added spectra and corresponding models for the S-lines, but the spectra from the two epochs and the model for the latest epoch for [\ion{Ar}{2}] to allow for a comparison). 

The results in  Table~\ref{tab:lines} show some differences compared to the fits presented in F24 (see their Table~1). Most importantly, the luminosities of the blue components ( “Gauss 1" in Figure~\ref{fig:mrsprofiles}) are somewhat higher (by factors $1.42 \pm 0.04$, $1.6 \pm 0.5$, and $3.4 \pm 0.9$ for [\ion{Ar}{2}], [\ion{S}{4}] and [\ion{S}{3}], respectively) and the centroid of the [\ion{S}{3}] line has a lower blueshift  ($-103 \pm 25$~\kms\ compared to $-288 \pm 13$~\kms\ in F24).  These differences are due to the combined effects of the data reprocessing, improved S/N in the total spectrum from the two epochs, and the fact that no background spectrum was subtracted. The latter mainly affect the [\ion{S}{3}] and [\ion{S}{4}] lines,  which have a strong diffuse background (Kavanagh et al., in prep) seen as the narrow peaks at 0-\kms\  (“Gauss 2" in Figure~\ref{fig:mrsprofiles}). Including this emission in the fits rather than subtracting it from a different region make the fits more stable. We also note that the [\ion{S}{3}] line has the most uncertain luminosity (Table~\ref{tab:lines}) and that its large increase compared to F24 is partly driven by a larger FWHM in the new fits ($330 \pm 33$~\kms\ compared to $158 \pm 39$~\kms\ in F24). 

\begin{figure}[t]
\centering
\includegraphics[width=\hsize]{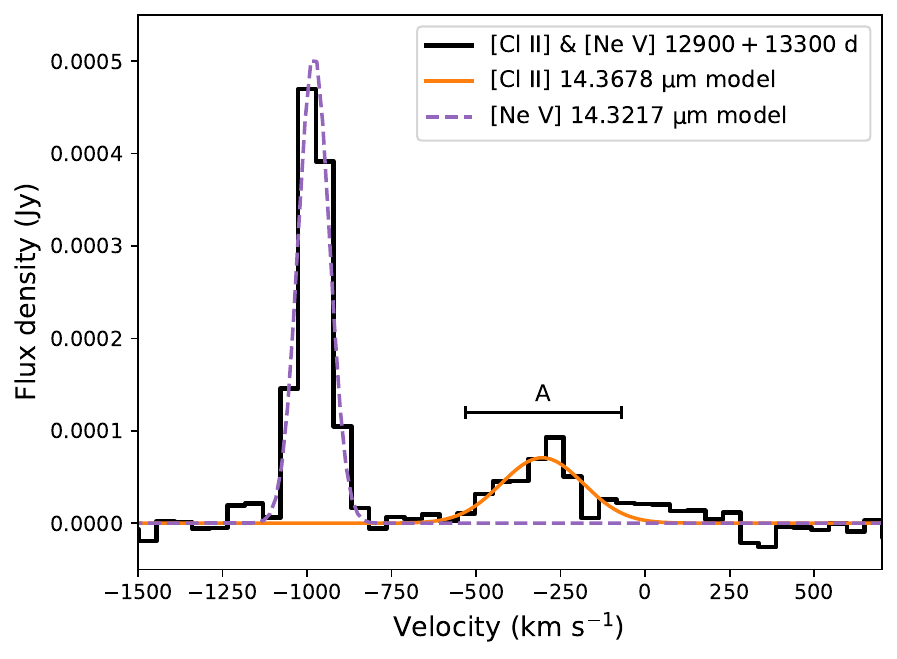}
\caption{Velocity profile of the [\ion{Cl}{2}]~$14.3678~\mu$m line from the central ejecta in the combined MRS spectrum from 13,300 and 12,900 days. The nearby [\ion{Ne}{5}]~$14.3217~\mu$m line from the diffuse background is also included in the plotted velocity interval.  The velocity interval indicated by A was used for the images in Figure~\ref{fig:cl2location}.}
\label{fig:cl2profile}
\end{figure}

In addition to the lines previously identified in  F24, we identify a narrow line from the central ejecta at $14.354\ \mu$m, which we attribute to [\ion{Cl}{2}]\ 14.3678\ $\mu$m. This implies a  blueshift of $\sim -300$~\kms, consistent with the other lines from the center. The only other line in the same wavelength region is [\ion{Ne}{5}]~14.3217\ $\mu$m, which is unresolved and peaks close to the systemic velocity of SN~1987A, making it consistent with the diffuse background. The line profiles of the  [\ion{Ne}{5}] and [\ion{Cl}{2}] are shown in Figure~\ref{fig:cl2profile}, and the best-fit parameters for the [\ion{Cl}{2}]  line are reported in Table~\ref{tab:lines}. 

\begin{figure*}[t]
\centering
\includegraphics[width=\hsize]{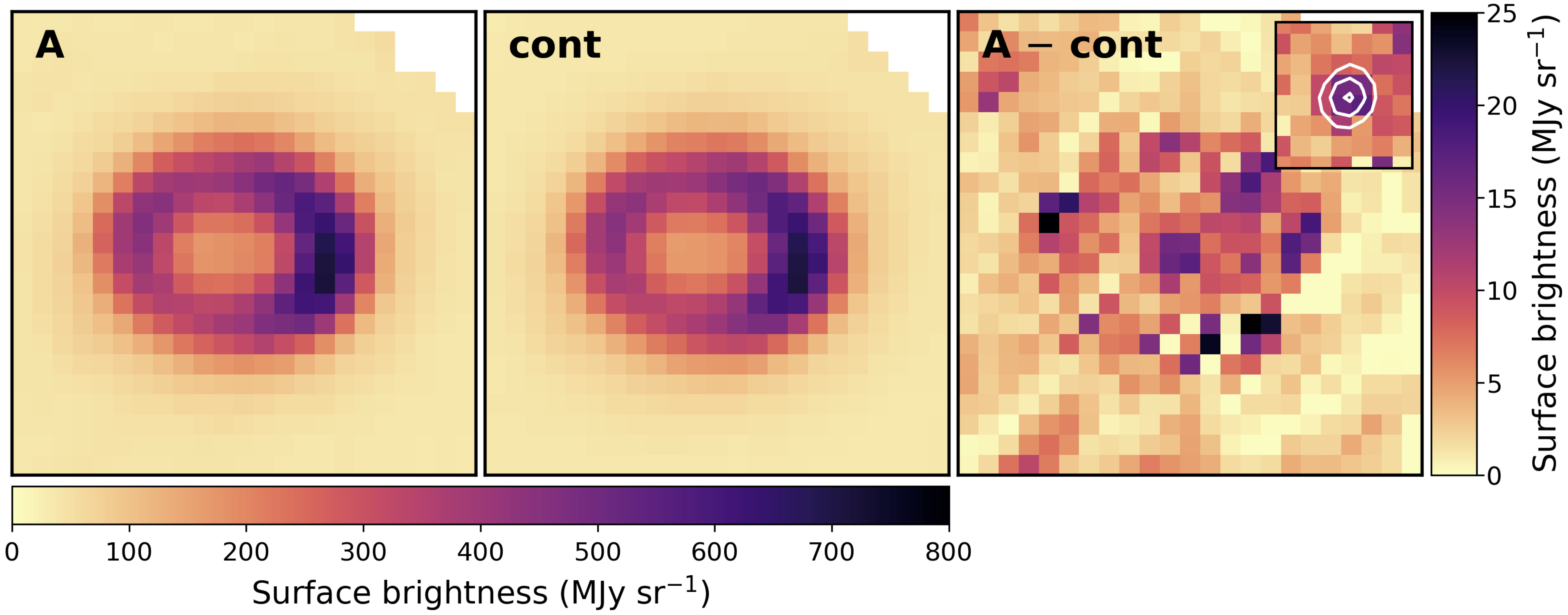}
\caption{Images of the [\ion{Cl}{2}] emission. Left: Image of the [\ion{Cl}{2}] line region in the [-527, -58]~km~s$^{-1}$ velocity interval (interval A in Figure~\ref{fig:cl2profile}). The faint [\ion{Cl}{2}] line is overwhelmed by the bright continuum emission. Middle: Continuum image adjacent to the [\ion{Cl}{2}] line. Right: The continuum subtracted [\ion{Cl}{2}] line map. As noted in the text, the apparent [\ion{Cl}{2}] from the ER is a residual of the continuum subtraction. Right-inset: Location of the [\ion{Ar}{2}] emission in relation to the [\ion{Cl}{2}] emission in the central part of the image. The [\ion{Ar}{2}] emission is indicated by the white contours. The contour levels were chosen arbitrarily to highlight the position of the [\ion{Ar}{2}].
}
\label{fig:cl2location}
\end{figure*}

The spatial properties of the [\ion{Cl}{2}] line are shown in  Figure~\ref{fig:cl2location}, illustrating that a faint point source consistent with the location of [\ion{Ar}{2}] appears at the center after subtracting the strong continuum. Residuals from the subtraction of the bright continuum are present in the image at the location of the ER, which results from either low-contrast residual fringing, fringe-like resampling noise from the cube building process \citep{Law2023}, or both. [\ion{Cl}{1}]~11.3334\ $\mu$m was not detected in spectra extracted from the ER in either \citet{Jones2023} or Kavanagh et al., (in prep).  We note that [\ion{Cl}{1}]~11.3334\ $\mu$m was seen in the spectra of the ejecta $\sim 1$ year after explosion, although blended with H$_{9-7}$ \ 11.3087 $\mu$m and \ion{Ni}{1} $11.3073 \mu$m \citep{Roche1993}. [\ion{Cl}{2}] $\lambda 8627$ has also been detected in the optical in Cas A \citep{Fesen2001}, while  [\ion{Cl}{2}] $\lambda \lambda  8578.7, 9123.6$ and  [\ion{Cl}{3}] $\lambda 5517.7$ have been observed in other SN remnants \citep{Fesen1996}. 

\begin{figure}[t]
\centering
\includegraphics[width=\hsize]{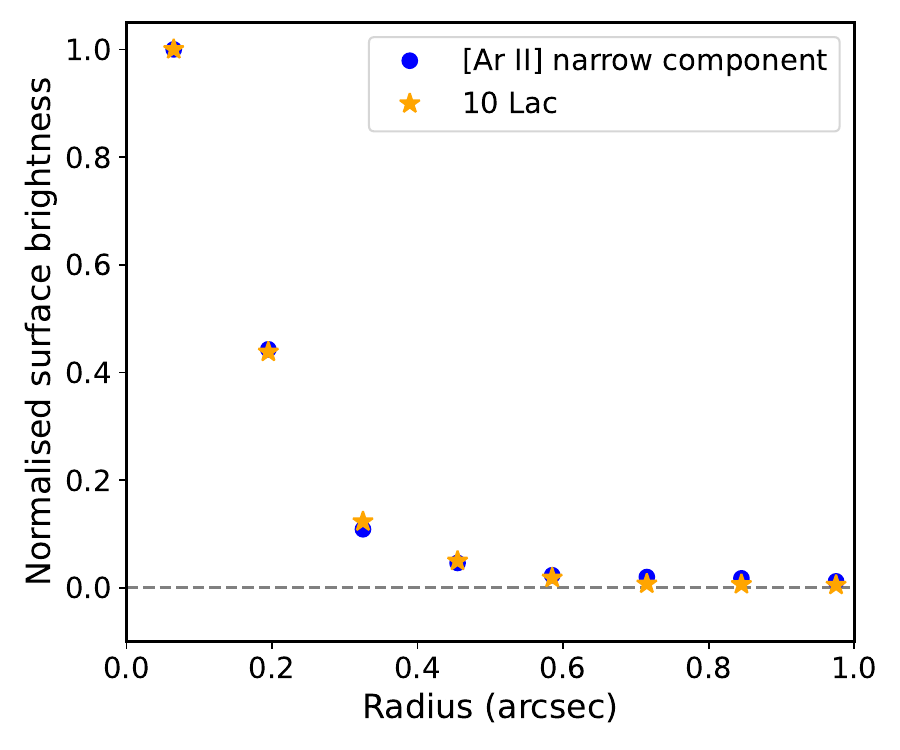}
\caption{Radial profiles of the narrow component of the [\ion{Ar}{2}] line measured at 13,300~days (blue circles). The profile of 10 Lac (orange stars) is also shown to indicate the profile of an unresolved point source in MRS band 1C.}
\label{fig:ar2radialprofile}
\end{figure}

F24 reported that the [\ion{Ar}{2}] line is spatially unresolved in the MRS observation at 12,900~days. To assess whether this is also the case at 13,300~days, we followed the same procedure outlined in F24, i.e., to compare the spatial profile of the [\ion{Ar}{2}] line to that of the unresolved point source 10 Lac (RA=22:39:15.67, Dec=+39:03:00.97). We reprocessed these data (PID 1524, PI: D. Law) following the same method described in Kavanagh et al. (in prep.). The resulting radial profiles are shown in Figure~\ref{fig:ar2radialprofile}. The spatial profile of the [\ion{Ar}{2}] line is almost identical to that of the unresolved point source, indicating that the [\ion{Ar}{2}] line is still spatially unresolved at 13,300~days. Limits on [\ion{Ar}{2}]-like components for other emission lines in the MRS wavelength range are presented in Appendix~\ref{sec:appendix:ul}.

\subsection{Continuum emission}
\label{sec:analysis:cont}

In the scenario where a PWN is at the center of SN~1987A, there should also be a synchrotron continuum in the IR (as for the well-studied Crab PWN, e.g.,\ \citealt{Lyutikov2019}). We use the NIRSpec/G395H grating to constrain any such emission. The continuum is more prominent in this wavelength interval ($\sim $2.9--5.2~$\mu$m) compared to the shorter wavelengths observed with G140H and G235H, where line emission dominates \citep{Larsson2023}.  For the longer wavelengths covered by the MRS,  the identification of any PWN synchrotron component  is hampered by the rising background from the ER and lower spatial resolution. 

\begin{figure*}
\centering
\includegraphics[width=\hsize]{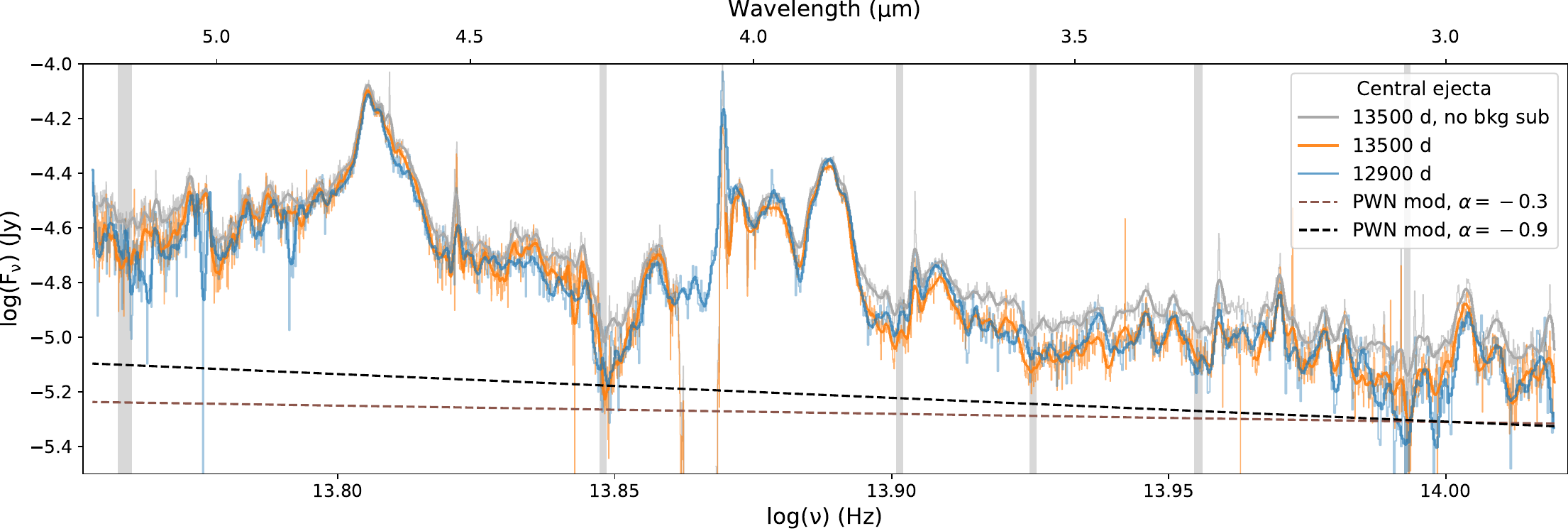}
\caption{Spectra of the central ejecta in G395H at 13,500 and 12,900 days. The thin solid lines show the original spectra (used for analysis) while the thick lines show the same spectra smoothed with a Savitzky-Golay function \citep{Savitzky1964}. The extraction region is shown in Figure~\ref{fig:ar6ims}. The spectrum from 13,500~days is shown with and without background subtraction, illustrating the noise added by subtracting the low S/N background spectrum. The observation from 12,900 days is shown only after the background subtraction. Wavelength intervals used for the continuum fits are marked by the shaded gray regions. The two dashed lines show the maximal contribution from a PWN synchrotron component for two different values of the spectral index, $\alpha$.}
\label{fig:contspec}
\end{figure*}

Figure~\ref{fig:contspec} shows the full G395H spectrum extracted from the central region for the observations at 13,500 and 12,900 days (see extraction region in Figure~\ref{fig:ar6ims}). Background spectra have been subtracted  for this comparison, as the background is non-negligible in the faint central ejecta region and slightly different in the two observations. The small areas of clean background available within the FOV imply that this subtraction adds considerable noise to the spectra, as also illustrated in Figure~\ref{fig:contspec}.  It is clear, however, that the background-subtracted continuum is consistent in the two epochs.  In Figure~\ref{fig:contspec},  we have also marked six wavelength intervals that are free of strong lines in the whole FOV, determined from visual inspection and comparison with the spectral model in \cite{Larsson2023}. We use these intervals to further characterize the continuum, but caution that there is likely some contribution from weak blended lines also in these intervals. 

The synchrotron emission from a PWN is expected to be described by a power law,  $F_{\nu} \propto {\nu}^{\alpha}$, where $\alpha$ is the spectral index. It is clear from inspection of Figure~\ref{fig:contspec} that this simple model  does not capture the complex continuum shape in G395H. The continuum in the central region is instead expected to include several different components: high-latitude emission from the RS (ALMA observations in the sub-mm range gives $\alpha=-0.70\pm 0.06$ for the RS, \citealt{Cigan2019}), H and He continuum (free-free, bound-free, and two-photon emission),  scattered emission from the hot gas and dust in the ER (i.e.\ emission in the tails of the PSFs that extend to the ejecta region), as well as any possible hot dust in the ejecta and PWN synchrotron emission. The first three of these were included in the continuum model for the full ejecta region in \cite{Larsson2023}. 

To place constraints on PWN synchrotron emission, we use the spectral indices measured for the Crab nebula and PWN0540, both of which are well-observed in the IR. For the Crab, the IR spectral index in the inner region measured with Spitzer is $\alpha \sim -0.3$, consistent with the values measured at radio frequencies \citep{Lyutikov2019}. For  PWN0540, NIR to MIR wavelengths observed with VLT/XSHOOTER, Spitzer, and AKARI give  $\alpha \sim -0.9$  \citep{Lundqvist2020,Tenhu2024}. We plot the highest possible contributions from these power laws in Figure~\ref{fig:contspec}, obtained by requiring that the models do not overshoot the continuum anywhere. The luminosity integrated over the full G395H wavelength interval is $7.4 \times 10^{32}$ and  $8.5 \times 10^{32}\ \rm{erg\ s^{-1}}$ for the Crab-like and PWN0540-like spectra, respectively. These limits are very conservative since the PWN is likely considerably smaller than the $0\farcs{3}$ circular extraction region.  If we instead consider the smaller $0\farcs{2} \times 0\farcs{1}$ region where the [\ion{Ar}{6}] emission peaks, the limits are $4.1 \times 10^{31}$ and  $3.6 \times 10^{31}\ \rm{erg\ s^{-1}}$, respectively.  These limits are approximate due to the uncertain value of $\alpha$, which may be different in a very young PWN compared to the Crab and PWN0540, which are both about 1000~years old.  

\begin{figure}
\centering
\includegraphics[width=\hsize]{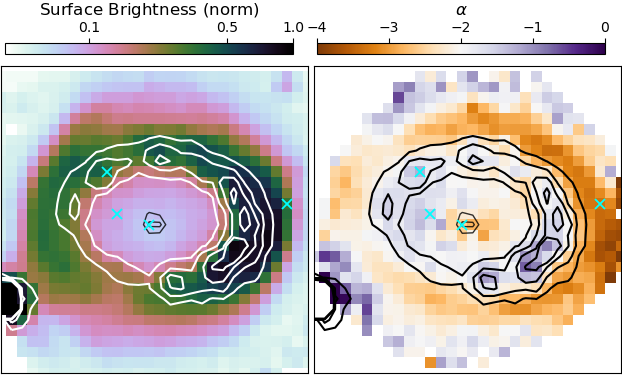}
\caption{Properties of the continuum emission in G395H. Left: Average image of the continuum based on the six continuum intervals in Figure~\ref{fig:contspec}. Right: Spectral index map, obtained by fitting the spectra in each spaxel with a power-law model. The contours at the center of both panels are for [\ion{Ar}{6}] (left panel of Figure~\ref{fig:ar6slices}), while the contours along the ER are for the
 \ion{H}{1}~Pa$\beta$~1.2822~$\mu$m line. The latter traces the shocked gas in the hotspots of the ER. Spectra from the four spaxels marked by cyan crosses are shown in Figure~\ref{fig:contin_pixfit}.}
\label{fig:contmap}
\end{figure}

Figure~\ref{fig:contmap} (left) shows an image of the continuum, obtained by taking an average of images integrated over the six continuum intervals in Figure~\ref{fig:contspec}. As expected from the spectral analysis above, there is no enhancement of the continuum at the center. Instead, there is a region with lower surface brightness that is slightly elongated in the east-west direction. To further analyze the spatial variations of the continuum, we fit the power-law model to the spectra in each spaxel, using the same continuum intervals as above. We stress that this model, while not an accurate description of the shape of the continuum, is useful for highlighting general trends of the spatial variations. We do not subtract a background in these fits, as this results in many spectra with low S/N. The impact of this on the spectral index is typically $\lesssim 0.1$.   

\begin{figure*}
\centering
\includegraphics[width=\hsize]{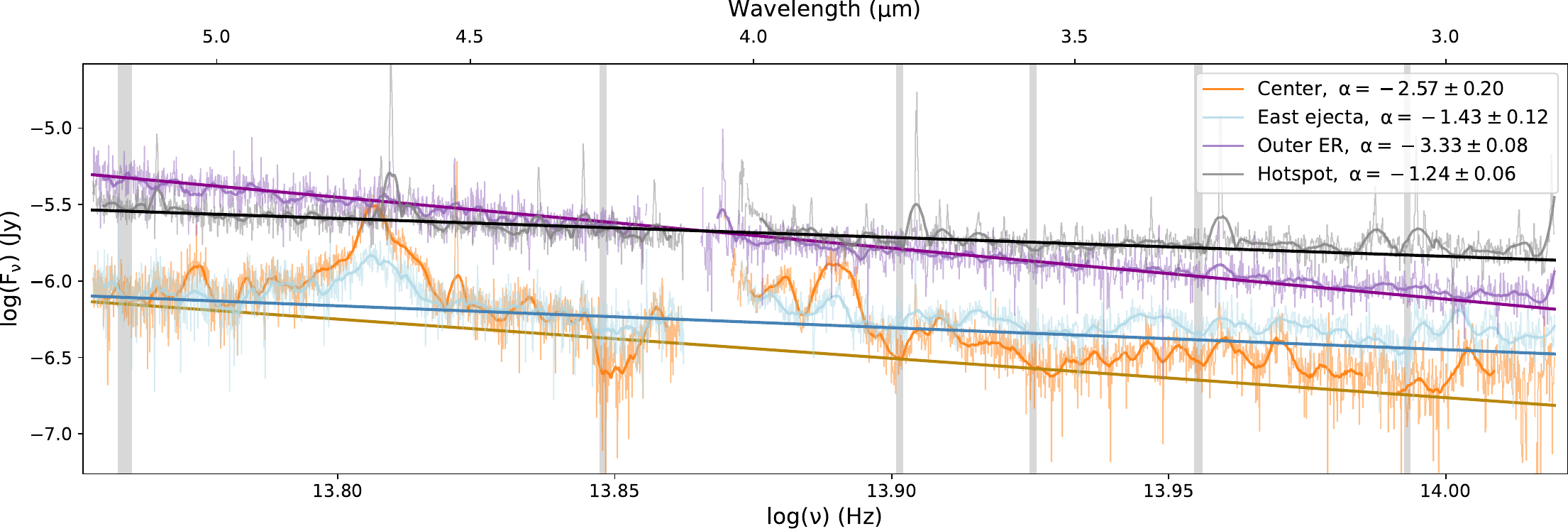}
\caption{Examples of spectra from individual spaxels in different regions of SN~1987A. The positions of the spaxels are marked by cyan crosses in Figure~\ref{fig:contmap}. The thin solid lines show the original spectra (used for analysis) while the thick lines show the same spectra smoothed with a Savitzky-Golay function. The straight lines show the models fitted to the continuum intervals of the four spectra. The continuum intervals are marked by the shaded gray regions.}
\label{fig:contin_pixfit}
\end{figure*}

The resulting spectral index map is shown in Figure~\ref{fig:contmap} (right), with contours for the central [\ion{Ar}{6}] source and the Pa$\beta$ line from the hotspots in the ER superposed for comparison. Examples of fits from four different spaxels are shown in Figure~\ref{fig:contin_pixfit} to illustrate how the spectrum changes between the different regions. We see clear spectral index variations in the ejecta, with steeper indices in a region elongated in the east-west direction. The shape of this region is similar to that of the low-surface brightness region seen in images from NIRCam/F356W \citep{Matsuura2024} and MIRI/F560W \citep{Bouchet2024}. The lower values of $\alpha$ in this region are primarily due to a lower flux level below $\sim 3.7~\mu$m and a drop around $4.25~\mu$m, as illustrated in Figure~\ref{fig:contin_pixfit}. The steeper spectral index at the center is opposite to what would be expected in the case of a significant PWN contribution. It is likely caused by a combination of spatial variations in projection of high-latitude emission, scattered light from the ER, and dust emission/absorption in the ejecta.

Finally, though not the focus of this study, we note that there are clear spectral index variations also in the region of the ER (Figures~\ref{fig:contmap} and \ref{fig:contin_pixfit}). An index of  $\alpha \sim -1.5$ is observed in the regions of shocked gas traced by the hotspots, while the spectrum clearly steepens outside, where the emission is dominated by the RS (which is especially bright in the east, \citealt{Larsson2019b,Matsuura2024}) and hot dust (which is brightest in the west, \citealt{Jones2023}). These trends are also in agreement with spatial variations in the spectral index measured from NIRCam images in the 3--4~$\mu$m range \citep{Matsuura2024}. 

\subsection{Summary of emission from the central region}
\label{sec:analyis:summary}

The key results from the analysis presented above can be summarized as follows:

\begin{itemize}
\vspace{-0.2cm}
\item [-] The innermost ejecta show narrow emission lines from [\ion{Fe}{2}]\ $1.6440\ \mu$m,  [\ion{Ca}{5}]\ $4.1585\ \mu$m,  [\ion{Ar}{6}]\ $4.5292\ \mu$m, [\ion{Ar}{2}]\ $6.9853\ \mu$m,  [\ion{S}{4}]\ $10.5105\ \mu$m, [\ion{Cl}{2}]\ $14.3678\ \mu$m, and [\ion{S}{3}]\ $18.7130\ \mu$m.  Out of these, [\ion{Fe}{2}], [\ion{Ca}{5}], and [\ion{Cl}{2}] were identified from our new observations and not previously discussed in F24. We also determined upper limits for several other lines predicted by photoionization models.

\vspace{-0.2cm}
\item  [-]  The lines are characterized by a blueshifted peak in the range $\sim -250$  to $-300$~\kms\ and FWHM $\sim$ 100--200~\kms, as summarized in Table~\ref{tab:lines}.  The [\ion{Ar}{6}] and  [\ion{Ar}{2}] lines also exhibit weaker red wings extending to $\sim 200$~\kms, which are not seen for the other lines due to lower S/N and/or blending with other emission components. 

\vspace{-0.2cm}
\item  [-]  The spatial properties of the emission region are best constrained from the [\ion{Ar}{6}] line, which is unresolved (FWHM=0\farcs{21}) and located south-east of the geometric center of the ER ($30\pm 10$ mas south, $63\pm 10$ mas east, PA$=116\pm 8$\dg).  

\vspace{-0.2cm}
\item  [-]  Emission from [\ion{Ca}{4}]~$3.2068\ \mu$m is also detected near the center, but this line is redshifted by $\sim 700$~\kms\ and significantly displaced to the north compared to the other lines. 

\vspace{-0.2cm}
\item  [-]  There is no significant variability in the flux, line profiles, or spatial properties of these lines on the timescales of 400--600~days probed by the observations. 

\vspace{-0.2cm}
\item  [-]  The continuum in the innermost ejecta does not show any sign of a PWN in terms of enhanced flux or spectral index variations.
\vspace{-0.2cm}

\end{itemize}

\section{Discussion}
\label{sec:disc}

\subsection{Impact of dust in the ejecta}
\label{sec:disc:dust}
In F24, it was argued that dust absorption by silicates was responsible for the weak lines above $\sim 8$ $\mu$m from the central source. We also proposed that scattering by the expanding dust could be important at shorter wavelengths. This would redshift the wavelengths, analogous to the effects of electron scattering in an expanding ejecta, neglecting the broadening by the thermal velocities of the electrons. Below, we first provide a general description of the dust scattering, before discussing the impact on the line profiles in Section~\ref{sec:disc:line_profiles} and comparing with the ALMA dust maps in Section~\ref{sec:disc:alma}. 

\begin{figure*}
\centering
\includegraphics[width=8.7cm]{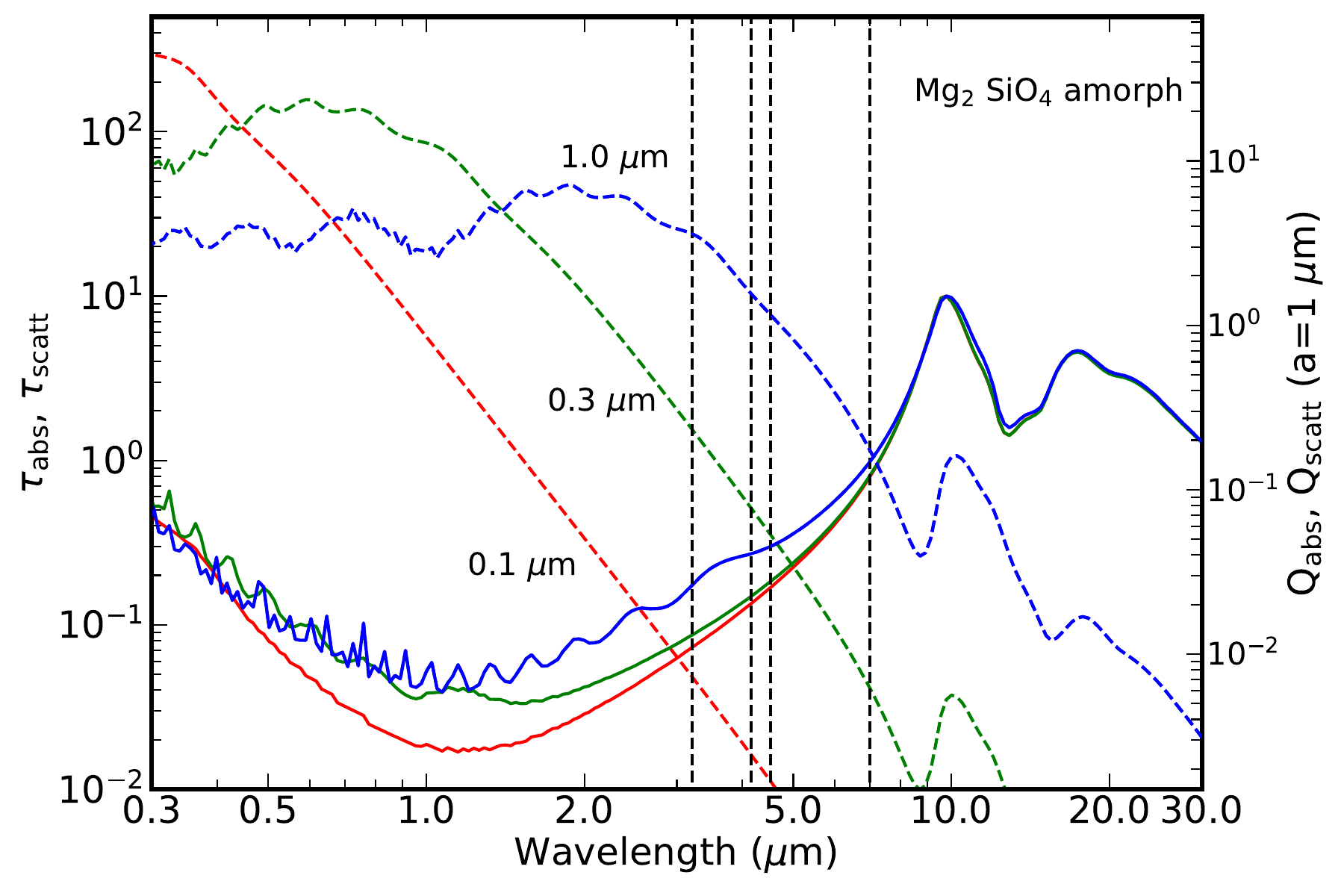}
\includegraphics[width=8.7cm]{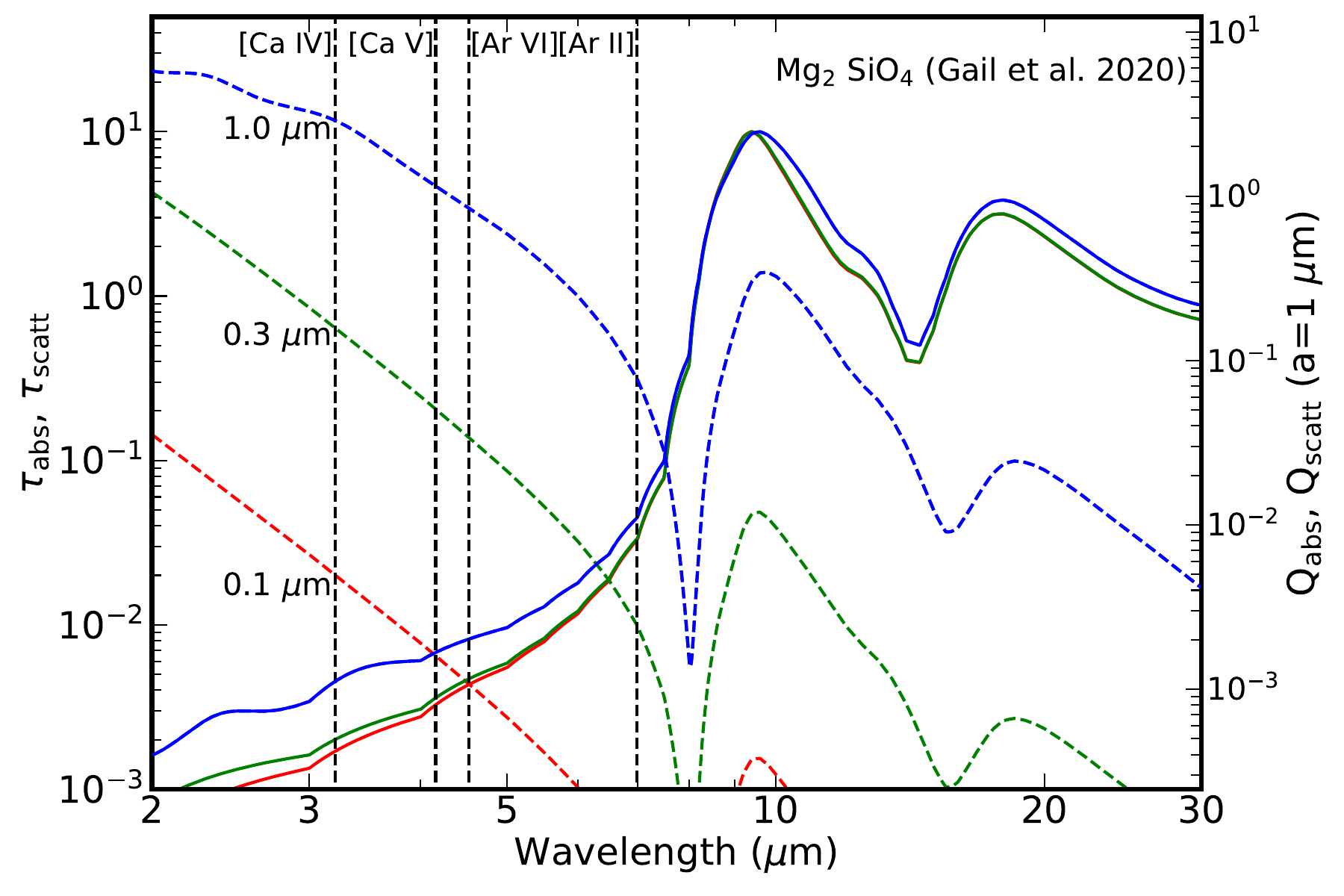}
\includegraphics[width=8.7cm]{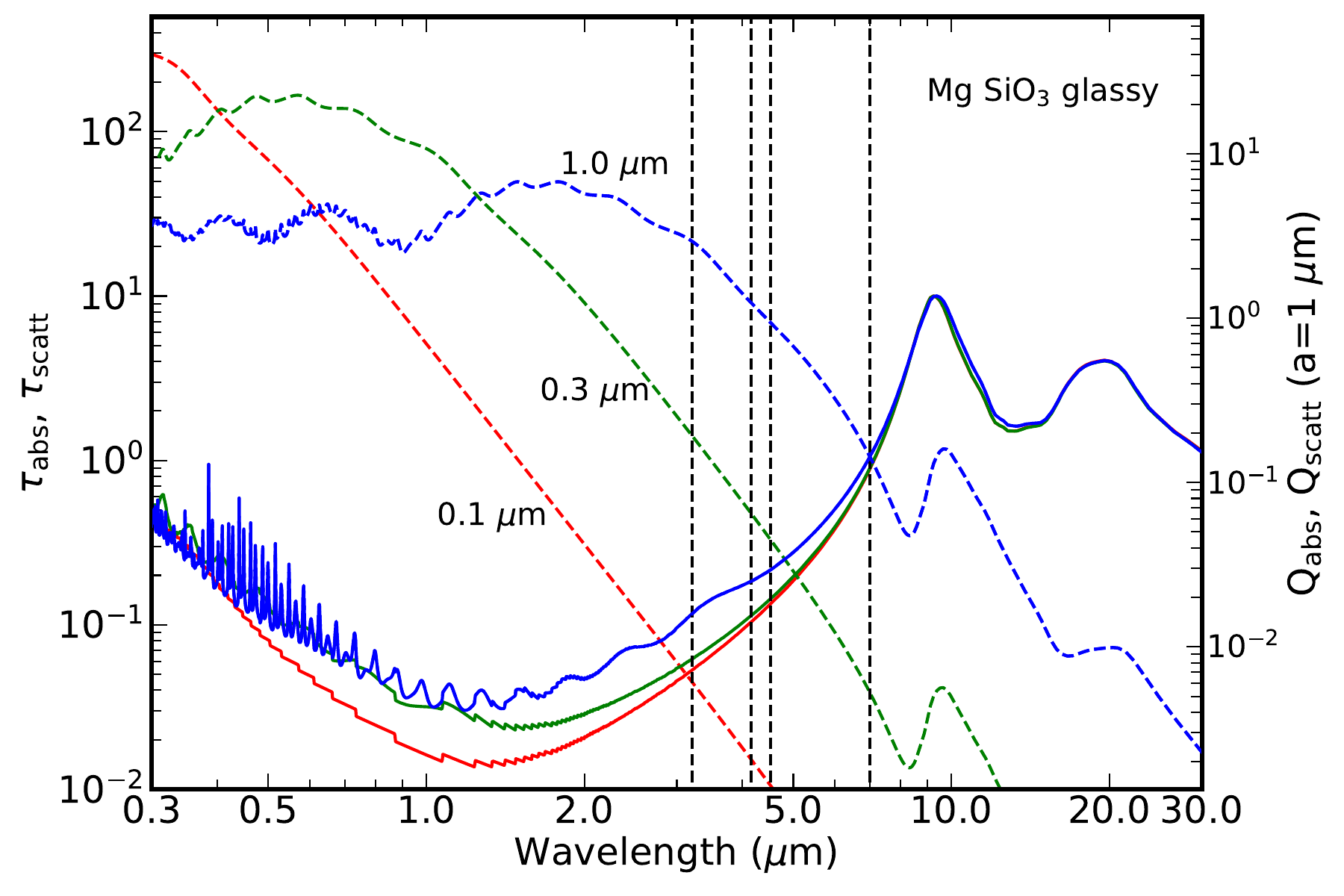}
\includegraphics[width=8.7cm]{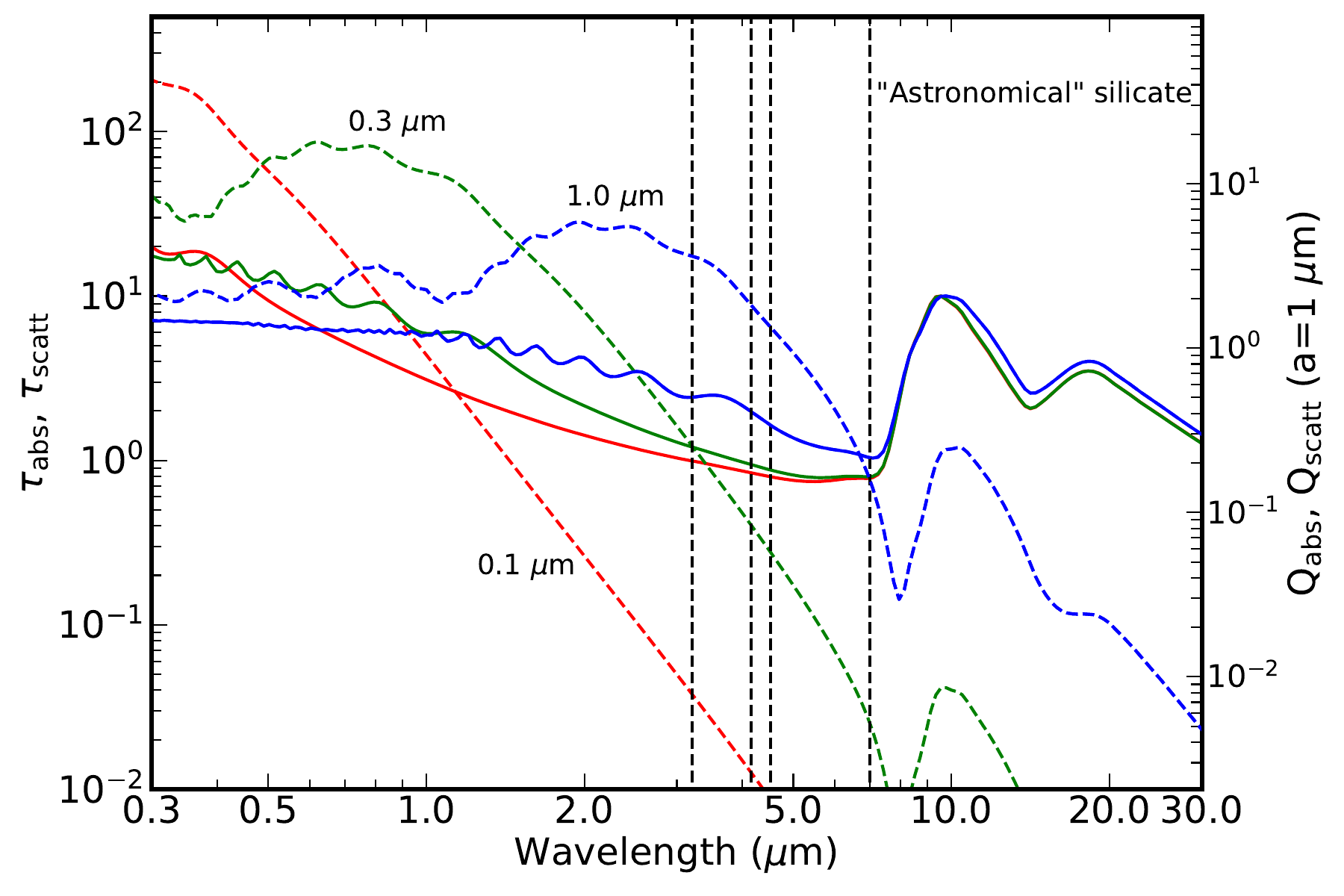}
\caption{Absorption optical depth, $\tau_{\rm abs}$, (solid lines) and scattering optical depth, $\tau_{\rm scatt}$, (dashed lines) for spherical grains with radii $a=0.1, 0.3$ and $1.0 \ \mu$m. The top panels show two different samples of forsterite (Mg$_2$SiO$_4$) from  \cite{Jager2003} and \cite{Gail2020}, respectively, the bottom left panel shows glassy enstatite (MgSiO$_3$; \citealt{Dorschner1995}) and the bottom right panel shows “astronomical silicates" \citep{Draine1984}. The absorption optical depth has been normalized to $\tau_{\rm abs} = 10$ at $10 \ \mu$m. The right-hand axis gives the scale for the absorption and scattering efficiencies, $Q_{\rm abs, scatt}$ for  $a=1.0 \ \mu$m (only). The vertical, dashed lines show the wavelengths for some of the most important, observed narrow lines from the center. Note the different x- and y-scales for the forsterite in the top right panel.}
\label{fig:dust_abs_scatt}
\end{figure*}

To estimate the scattering effects, the scattering efficiency, $Q_{\rm scatt}$, is needed for different compositions and and grain sizes. For these calculations, we used the Mie scattering theory, assuming spherical dust grains with different radii, $a$ \citep[e.g.,][]{Bohren1983}. While the shape of the grains is also important \citep[e.g.,][]{Min2003}, the qualitative aspects should mainly be determined by the composition.  Because the inner regions close to the neutron star is devoid of carbon (e.g., Fig. S7 in F24), we concentrate on silicates, although carbonaceous grains may be formed in the outer envelope \citep{Sarangi2015}.
As representative examples of silicates, we show in Figure~\ref{fig:dust_abs_scatt} two different determinations of the absorption and scattering for amorphous forsterite (Mg$_2$SiO$_4$) and glassy enstatite (MgSiO$_3$), with optical constants from \cite{Jager2003}, \cite{Gail2020} and \cite{Dorschner1995}, respectively. \cite{Gail2020} only gives data for wavelengths longer than 2 $\mu$m, which explains the truncation of the curves at shorter wavelengths. For reference, we also show the often-used “astronomical silicates" from \cite{Draine1984}. As these authors discuss, the optical constants for “astronomical silicates" have been derived from observations of different ISM and CSM objects and are unlikely to be representative of the newly formed dust in SN ejecta. In particular, the composition includes “impurities" such as iron and corundum, which especially alter the absorption at short wavelengths, as seen in the fourth panel of Figure~\ref{fig:dust_abs_scatt}. In contrast, the dust in the central region of the SN ejecta may be the result of dust condensation from unmixed material with abundances characteristic of the different nuclear burning zones as shown by \cite{Sarangi2015}, specifically zones with high abundances of O, Mg and Si.

The optical depths are given by $\tau_{\rm abs, scatt} = \pi a^2 N_{\rm dust} Q_{\rm abs, scatt}$, where $N_{\rm dust}$ is the column density of dust. We, however, know neither $N_{\rm dust}$, which probably varies along different lines of sights, the size of the grains, or the exact composition of the dust in the ejecta of SN 1987A. In Figure~\ref{fig:dust_abs_scatt}, we show the resulting $\tau_{\rm abs, scatt}$  for these silicates, where we have arbitrarily normalized the optical depths to $\tau_{\rm abs} = 10$ at 10 $\mu$m. This is also close to the estimated optical depths for the models in Section \ref{sec:disc:linemod}.

Comparing these four dust compositions, we note that the $\tau_{\rm abs}$  below $\sim 8\ \mu$m differ by at least an order of magnitude, while $\tau_{\rm scatt}$  shows less variation and a similar behavior with wavelength. For $\lambda \gg 2 \pi a$  (the Rayleigh limit),  $Q_{\rm scatt} \propto \lambda^{-4}$, and hence $\tau_{\rm scatt}  \propto \lambda^{-4}$, while these become nearly constant in the opposite limit. When we compare $\tau_{\rm scatt}$ for different grain radii, we also note the high sensitivity to the grain size.

Comparing the two forsterite plots, we see that there are large differences in optical depth (especially  $\tau_{\rm abs}$) even for similar compositions and structures of the grains. For the same $\tau_{\rm abs}$(10 $\mu{\rm m}$),  $\tau_{\rm abs}\approx 1$ at the [\ion{Ar}{2}] line in the left two panels of Figure~\ref{fig:dust_abs_scatt}, while $\tau_{\rm abs}\approx 0.04$ for the case in the upper right panel. Also the scattering optical depth differs substantially in the different cases for the same grain size. 
Furthermore, because $\tau_{\rm scatt}$ increases by 1--2 orders of magnitude from $\sim 7 \ \mu$m to $1-2 \ \mu$m, scattering is expected to affect the flux and line widths, as well as the spatial distribution, of all lines at short wavelengths, unless there are lines of sight with much lower $N_{\rm dust}$ or the grains are small.

\subsubsection{Line profiles}
\label{sec:disc:line_profiles}
The fact that scattering is likely to be important in the NIRSpec range and that the gas and dust are expanding homologously means that the line profiles will be affected (F24, \citealt{Chugai2024}). Because the scattering by dust is coherent in the frame of the grains, this is similar to the case of electron scattering if we neglect the thermal motion of the electrons. This results in a redshift of the photons as shown by  \cite{Fransson_Chevalier1989}, who demonstrated that even a single scattering can increase the wavelength of the photon by up to twice the velocity of the scattering medium, $\Delta \lambda / \lambda \approx 2 V_{\rm ejecta}/c$. Line scattering by dust has also previously been discussed for expanding spherical circumstellar envelopes by \cite{Romanik1981} (for absorption line profiles) and \cite{Lefevre1992} (for emission profiles), using Monte Carlo simulations. 

The most interesting indication of this effect in SN~1987A comes from the profile of the [\ion{Ca}{4}] line (Figure~\ref{fig:ca4profile}). This line, as well as the [\ion{Ca}{5}] line, were both expected from the photoionization models in F24. However, while the [\ion{Ar}{2}],  [\ion{Ar}{6}], and  [\ion{Ca}{5}] lines  have blueshifted peaks at $\sim -250$~\kms, the [\ion{Ca}{4}] line has a broad redshifted peak at $\sim 700$~\kms. In addition, the peak position is north of the center (Figure \ref{fig:ca4slices}), clearly different from the [\ion{Ar}{6}] position.

There may be two explanations for this redshift. It could simply be that the emitting gas is expanding away from us from the “back" of the ejecta. However, this raises the problem of why we do not see other lines, like [\ion{Ar}{2}] and [\ion{Ar}{6}] from this position, which are expected from the same zone as the [\ion{Ca}{4}]. The other explanation is based on dust scattering. This was invoked by \cite{Chugai2024} to explain the red wing of the [\ion{Ar}{2}] line (seen in Figure~\ref{fig:ar6specfit}), but the [\ion{Ca}{4}] line profile and position may be an even more clear case. The fact that the [\ion{Ca}{4}] emission is spatially offset from the [\ion{Ar}{6}] emission can then be explained as a result of the increasing optical depth to dust scattering as the wavelength of the lines decreases. If there is a clump of dust in the direction of the observed  [\ion{Ca}{4}] emission, it will reflect the [\ion{Ca}{4}] emission from a source at a different location, e.g., the [\ion{Ar}{6}] location, while the clump may be transparent to the [\ion{Ar}{6}] emission.

We performed Monte Carlo simulations to investigate the effect of the scattering on the line profiles. Because we are mainly interested in the effects on the line emission from the center of the ejecta, we assumed a point-like source. We also assumed coherent scattering in the comoving frame of the expanding dust. The phase function of the scattering is often described by the Henyey-Greenstein function \citep{Henyey1941} 
\begin{equation}
    P(\theta)= \frac{1}{4 \pi} \frac{1-g^2}{(1+g^2-2 g  \cos{\theta})^{3/2}}
\end{equation}
where $g=< \cos{\theta} >$ is the asymmetry factor between forward and back scattering. However, other functions have been proposed \citep[e.g.,][]{Baes2022}. The forward - back scattering depends on the composition, shape and size of the grains. In general, forward scattering dominates in the optical, $g = 0.5-0.7$, but decreases above $\sim 2 \ \mu$m in the near- and mid-IR, $g < 0.1$ \citep[e.g.,][]{Shen_Draine2009}.

\begin{figure*}
\centering
\includegraphics[width=8.5cm]
{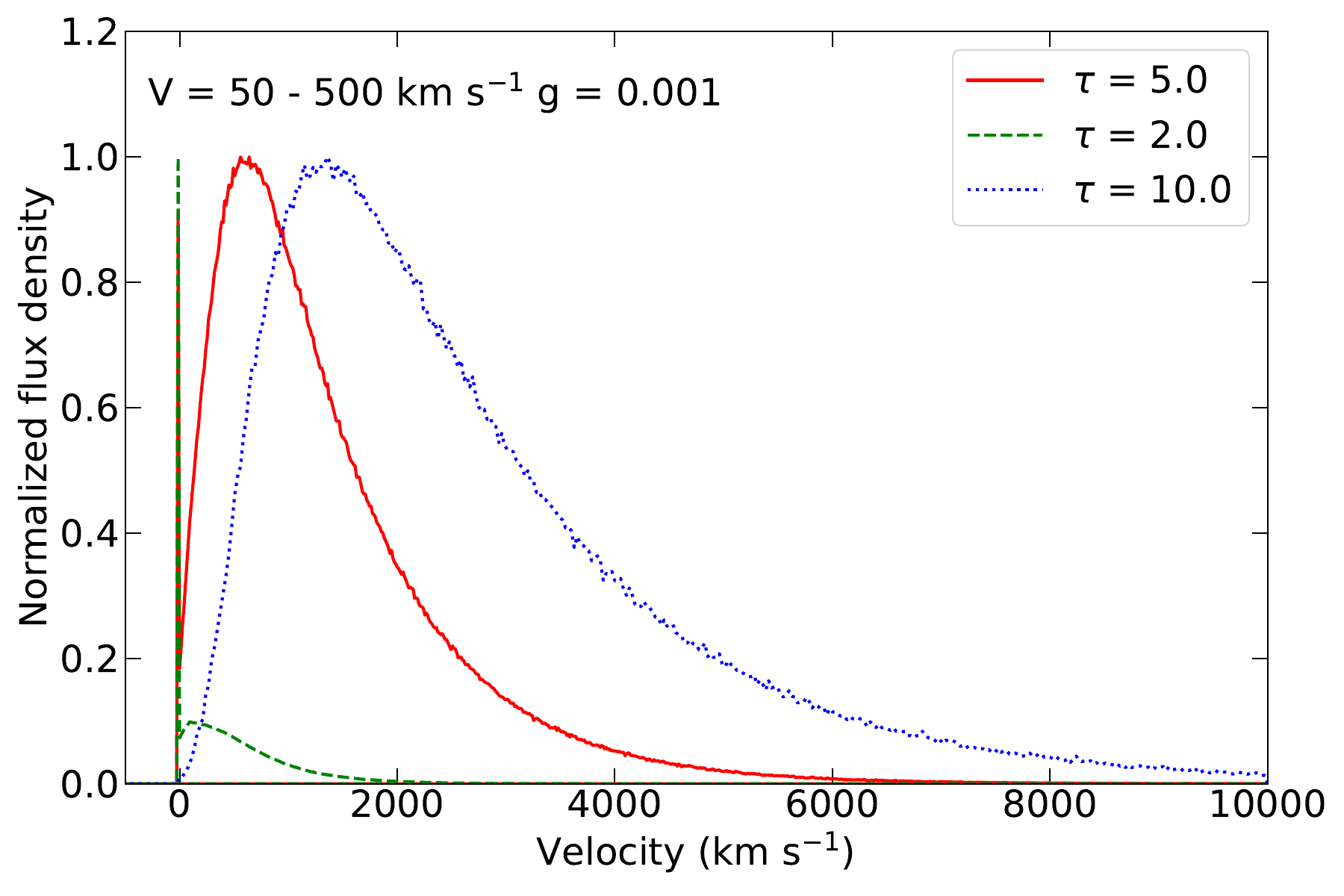}
\includegraphics[width=8.5cm]{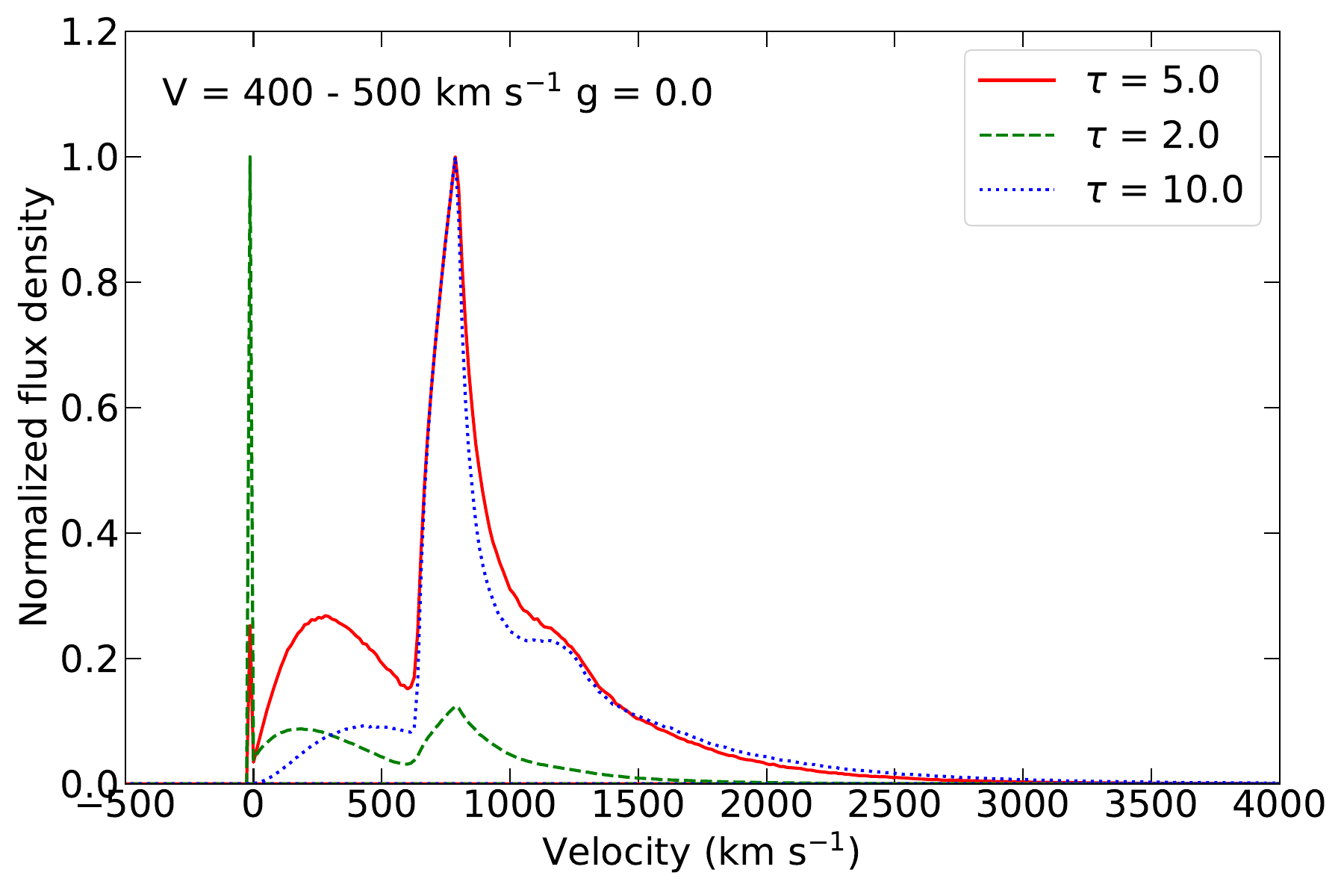}
\includegraphics[width=8.5cm]
{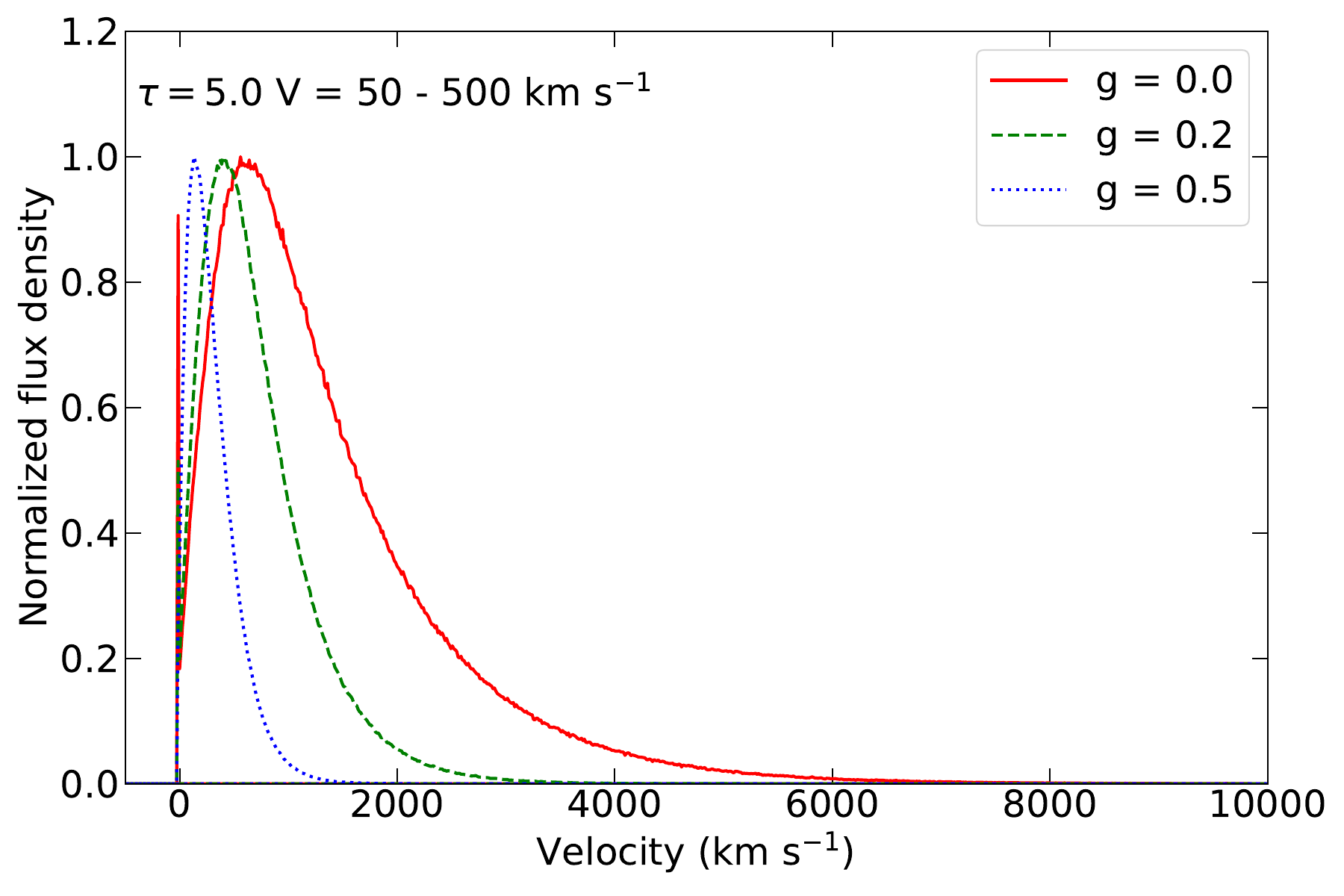}
\includegraphics[width=8.5cm]
{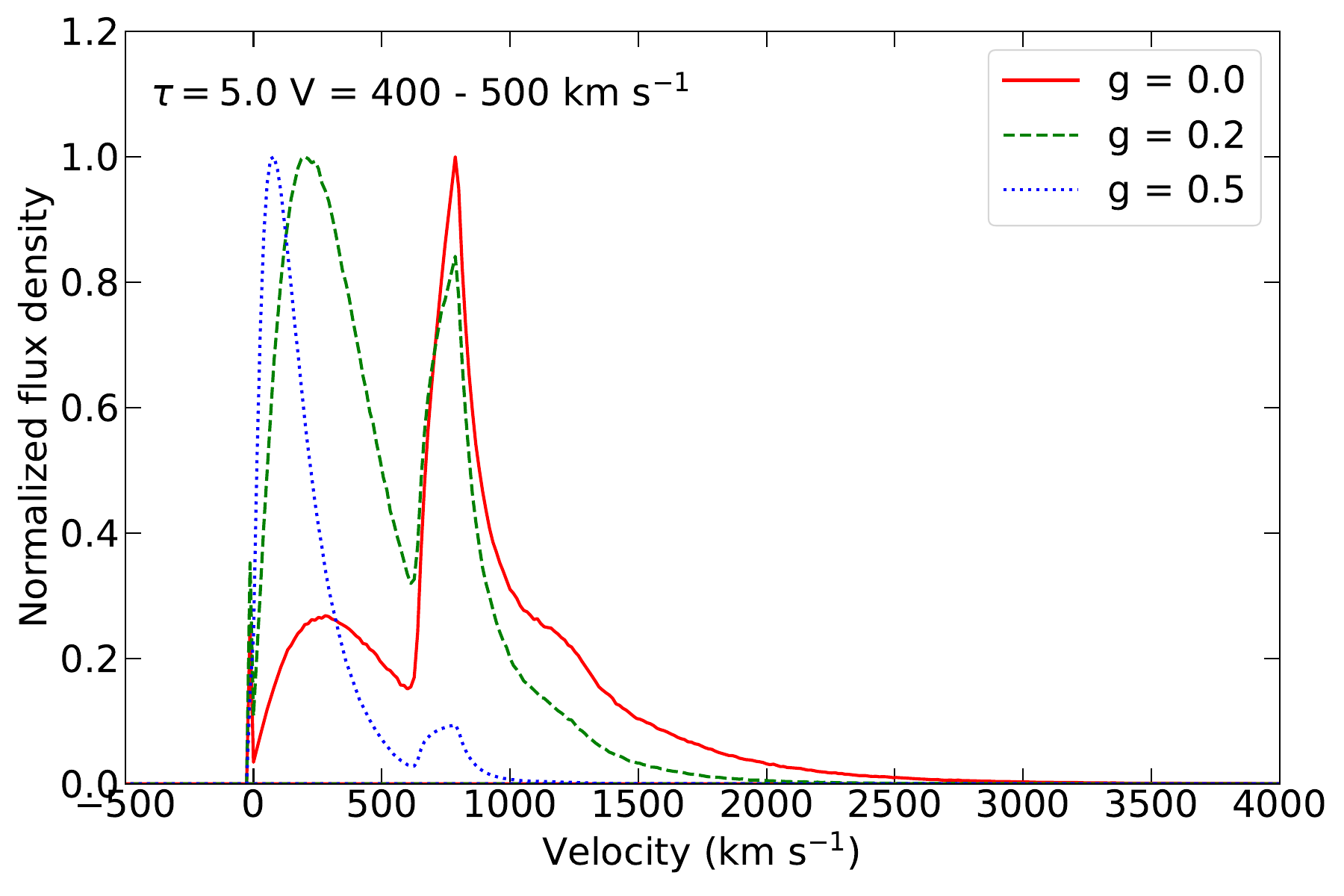}
\caption{Line profiles affected by dust scattering. The panels to the left and right are for a thick and thin shell of dust, respectively. The top panels show the scenario with isotropic scattering, while the bottom panels show the effect of varying the anisotropy parameter $g$. All cases assume a point-like source at the center, emitting a narrow line at 0~\kms. The unscattered component of this can be seen as the narrow spike at this velocity.}
\label{fig:dust_line_prof}
\end{figure*}

As examples of the dust scattering effects, we show in Figure \ref{fig:dust_line_prof} Monte Carlo simulations of two different spatial distributions of the dust. The code is a simplified version of the code used in \cite{Fransson2014} for electron scattering in an expanding medium, but neglecting the thermal motions of the dust particles. In both cases, we assumed a spherical geometry, but in the first case the dust is homogeneously distributed between radii corresponding to velocities of  50--500~\kms, while in the second case we assumed a homogeneous shell between 400--500~\kms. The velocity is assumed to be homologous, $V(r) \propto r$, where $r$ is the distance from the center. In all cases, we assume a central point like source of emission. 

In the top, left panel of Figure \ref{fig:dust_line_prof} we show the line profiles from the thick shell with inner boundary close to the emitting source for $\tau_{\rm dust} = 2, 5, 10$ and for isotropic scattering of the dust. This results in smooth redshifted line profiles with peaks at 130, 600, and 1400 \kms, respectively, illustrating the effect of multiple scatterings. The top, right panel shows the same for the detached, thin shell. In this case, the line profiles are more complicated. In addition to the peak from the un-scattered photons at zero velocity, there are two peaks, where the one at higher velocities becomes increasingly dominant with increasing optical depth. The anisotropy parameter $g$ is uncertain and is also expected to increase with decreasing wavelength. In the bottom panels of Figure \ref{fig:dust_line_prof} we show the effect of varying this parameter for $g=0, 0.2$ and $0.5$ and for $\tau_{\rm dust} = 5$. As the forward scattering increases (i.e., increasing $g$), the photons escape more easily and undergo fewer scatterings, resulting in line profiles that are less redshifted compared to the $g=0$ case.

A comparison of these models with observations can only be made qualitatively because of the limitations of the model and uncertainties in the line profiles. The most important assumptions are spherical symmetry and equal dust optical depth in all directions. A clumpy and/or anisotropic distribution can of course result in even more complicated line profiles, although they will in general be redshifted compared to the unscattered emission.

The [\ion{Ca}{4}] line observed north of the center peaks at $\sim 700$~\kms\ and has a FWHM $\sim 1000$~\kms\ (Table \ref{tab:lines}). Compared to the line profiles in Figure~\ref{fig:dust_line_prof}, this is most similar to the case of a thick shell with $\tau \sim 5$, though there are clear differences, including a remaining weak, unscattered narrow peak in the models. This highlights that asymmetries most likely play an important role, as also expected from the observed asymmetric morphology of the ejecta. The general trend of a stronger scattered red wing for lines at shorter wavelengths is also compatible with the more prominent red wing seen in  [\ion{Ar}{6}] compared to  [\ion{Ar}{2}] (Figure~\ref{fig:ar6specfit}). In the case of [\ion{Ca}{5}], we are not able to assess the presence of a red wing due to blending with other broad lines. The same is true for the other lines predicted by the models at shorter wavelengths. However, the absence of narrow, blueshifted components in these lines can be used to obtain a rough estimate of the typical grain size of the dust, while the possible [\ion{Ar}{6}]-like narrow peak identified for [\ion{Fe}{2}]~1.6440~$\mu$m is more difficult to explain (see Section~\ref{sec:disc:linemod}).

\subsubsection{Comparison with ALMA images}
\label{sec:disc:alma}

\begin{figure*}
\centering
\includegraphics[width=\hsize]{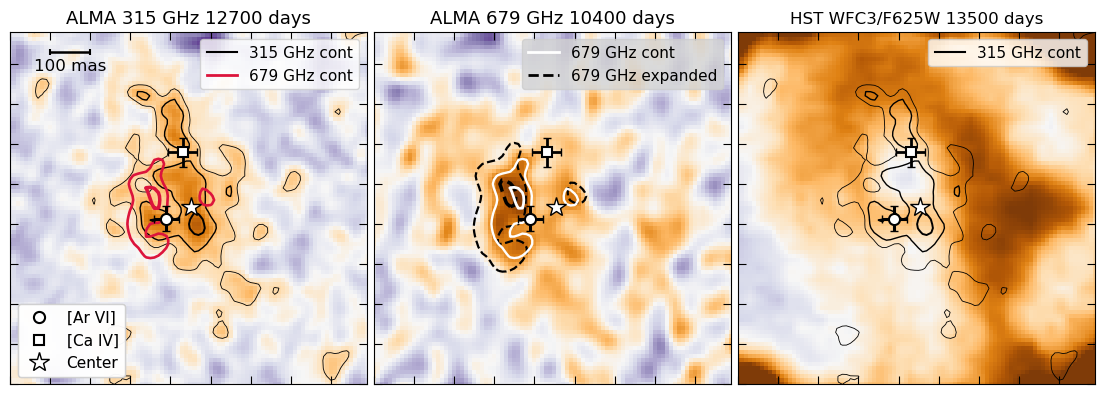}
\caption{Comparison of emission from the central ejecta at different wavelengths. North is up and East is to the left. Color images from left to right show the ALMA 315~GHz dust emission at 12,700~days \citep{Matsuura2024}, the ALMA 679~GHz dust emission at 10,400~days \citep{Cigan2019}, and the HST WFC3/F625W image at 13,500~days. The latter is dominated by H$\alpha$ emission and was obtained only a week after the NIRSpec observations.  In all panels, the center of the ER is marked by a star symbol, the position of the blue peak of the [\ion{Ar}{6}] line by a filled circle, and the centroid of the [\ion{Ca}{4}] emission by a filled squared, as indicted by the legend at the bottom of the left panel. The black error bars on the latter two positions are 3$\sigma$ confidence intervals, including the absolute astrometric uncertainty. Solid black contours are for the 315~GHz emission (left and right panels), while red and white contours are for the 679~GHz emission (left and middle panels, respectively). The dashed black contours in the middle panel show the effect of the expansion of ejecta between the time of the ALMA observations at 10,400~days and the NIRSpec and  HST observations at 13,500 days, assuming homologous expansion from the center of the ER. The tick marks in all panels are separated by 100~mas. }
\label{fig:almacomp}
\end{figure*}

Figure~\ref{fig:almacomp} shows ALMA images of the ejecta dust at 315 and 679~GHz,  together with the positions of the [\ion{Ar}{6}] source and the peak of the [\ion{Ca}{4}] emission from the NIRSpec observations (Sections~\ref{sec:analysis:ar6} and \ref{sec:analysis:ca}, respectively). An HST WFC3/F625W image of the ejecta, which is dominated by H$\alpha$ emission, is also included for comparison. The HST image was obtained only a week after the NIRSpec observations at 13,500~days (PID 16996), while the ALMA observations were obtained earlier, at 12,700 days for the 315~GHz image \citep{Matsuura2024}, and 10,400~days for the 679~GHz image \citep{Cigan2019}. 
All images in  Figure~\ref{fig:almacomp} were resampled to the 10~mas pixel scale of the 315~GHz image using \texttt{astropy/reproject} \citep{reproject2020} with bilinear interpolation. We estimate the uncertainty in the absolute astrometry of the HST image to be $\sim 1$~mas (1$\sigma$, based on fits to stars in Gaia DR3), while \cite{Cigan2019} report an astrometric uncertainty of 15~mas for the ALMA 679 image. No astrometric uncertainty has been reported for the 315~GHz image, but previous ALMA images of SN~1987A have had uncertainties in the range 10--15~mas \citep{Cigan2019}. 

The left panel of Figure~\ref{fig:almacomp} shows that the [\ion{Ar}{6}] and [\ion{Ca}{4}]  positions are within the region of bright dust emission at 315~GHz, supporting the scenario discussed above that these lines are affected by dust absorption and scattering. The 315~GHz dust emission is strongest just below the center of the ER, in a region of bright H$\alpha$ emission just below the “hole"  (right panel) and clearly extends further north than south from the center. This asymmetry is likely partly due to flux loss in the ALMA image, as a previous 315~GHz image at lower resolution shows dust emission extending further to the south \citep{Cigan2019}. It is thus possible that the faint redshifted [\ion{Ca}{4}]  emission in the south (Figure~\ref{fig:ca4slices}) is also caused by dust scattering. 

The comparison with the 679~GHz image is particularly interesting since it shows a bright region near the center, which has been interpreted as being caused by dust heated by the compact object \citep{Cigan2019}. The middle panel of Figure~\ref{fig:almacomp} shows that the peak of this dust blob is offset to the north-east of the [\ion{Ar}{6}] source, as previously discussed in F24. We find that the offset is within the 3$\sigma$ confidence intervals (but outside the 2$\sigma$ intervals), considering the reported astrometric uncertainty for the 679~GHz image and the astrometric plus statistical uncertainties of the  [\ion{Ar}{6}]  position. However, a major complication in this comparison is that the observations were obtained $\sim 3,100$~days apart, during which time the ejecta would have expanded by nearly 30\%. The dashed black contours in the middle panel of Figure~\ref{fig:almacomp} illustrate this effect, assuming homologous expansion from the center, showing how this would move the blob further away from the [\ion{Ar}{6}] source. Considering these uncertainties, we cannot draw any firm conclusions about the connection between the dust blob and line-emitting region. If  a spatial offset is confirmed with higher significance in future observations, it would either mean that the dust blob is unrelated to the compact object, or that the affected region is spatially extended. 

\subsection{Modeling of the line emission} 
\label{sec:disc:linemod}
\subsubsection{Summary of code}
The modeling of the different line luminosities in the NIR and mid-IR ranges is similar to that in F24, with the most important differences discussed below.  
 
The abundances used for the calculation are the same as those of the O -- Si -- S -- Ar -- Ca zone  in Table S2 in F24, with the addition of Cl, where we have used the average abundance in the same zone for the 19 \msun model in \cite{WoosleyHeger2007}, which is $7.6 \times 10^{-4}$ by number. These elements are produced by O burning and are expected to reside in the innermost ejecta (F24). Some of the elements in this zone may be locked up in dust, but we do not attempt to account for this given the considerable uncertainties regarding the properties of the dust. Depletion into dust is most likely to affect Si, and to a lesser extent also O and Ca. Most silicates also need Mg, which is not abundant in the O-Si-S region (F24), which suggests that these dust species may form at the interface to the O-Ne-Mg zone or inside this.

Updates of the code used for these calculations also include more detailed recombination data, especially for dielectric recombination for several ions. Details of this, as well as discussions of other abundance zones, will be given in Fransson et al. (in prep.). Collision strengths and radiative transition rates for \ion{Cl}{2}, \ion{Cl}{3} and \ion{Cl}{4} were taken from the Chianti Atomic Database \citep{DelZanna2021},  while data for the [\ion{Cl}{1}] $11.3334 \ \mu$m fine-structure line was taken from \cite{Hollenbach1989}. The collision strength of this transition is highly uncertain, while the other Cl data should be more accurate. In particular, the important [\ion{Cl}{2}] $14.3678 \ \mu$m collision strength is estimated to have an accuracy of $\sim 10 \%$ \citep{Wilson2002}.

For the ionizing flux, we have used the same spectra as in F24, either a power law synchrotron spectrum with $\alpha=1.1$ from a PWN or that from a young cooling NS (CNS), with a surface temperature of $3 \times 10^6$~K (see F24 for details). Except for the abundances, the most important parameter is the ionization parameter, $\xi=L_{\rm ion}/n_{\rm ion} r^2$, where $L_{\rm ion}$ is the ionizing luminosity above 13.6~eV, $n_{\rm ion}$ is the number density of ions, and $r$ is the distance to the ionizing source. The ion density especially affects the collisional de-excitation of the different lines. We have used the same volume filling factor, $0.1$, as in F24, resulting in $n_{\rm ion} = 2.6 \times 10^4 \  {\rm cm}^{-3}$. The other parameters in $\xi$ are discussed below.

\subsubsection{Modeling results}

The best observational constraints on the line luminosities were obtained for [\ion{Ar}{2}] and [\ion{Ar}{6}], while the other lines have lower S/N and also higher systematic uncertainties due to blending with other ejecta lines (especially for [\ion{Fe}{2}] and [\ion{Ca}{5}]) and prominent lines in the diffuse background (especially for [\ion{S}{3}] and [\ion{S}{3}]). Additionally, the Ar-line profiles were best fit by two Gaussians, where we use only the blue peak for model comparison, while the others were fit with single Gaussians, which adds further uncertainties. Considering these systematic uncertainties, we only aim to reproduce the observed luminosities of the weaker lines within a factor $\sim 2$.

The ionization parameter is mainly determined by the [\ion{Ar}{6}] / [\ion{Ar}{2}] ratio. In the PWN case, we find $\xi \approx 0.18$ and in the CNS case $\xi \approx 0.26$. These are marginally lower than those found in F24, which is a result of the slightly lower [\ion{Ar}{6}] / [\ion{Ar}{2}] ratio in the new observations. More specifically, we take $L_{\rm ion}= 3.7 \times 10^{34}$\ erg\ s$^{-1}$ and $r=2.79 \times 10^{15}$~cm for the PWN case, corresponding to 25~\kms\ for the freely expanding ejecta.  $L_{\rm ion}/r^2$ is mainly fixed by the ionization parameter, and that the bolometric luminosity should be $\lesssim 5.3 \times 10^{35}$ erg s$^{-1}$ \citep{Alp2018}. In the CNS case, the surface temperature and luminosity are constrained to $1-3 \times 10^6$~K and  $3 \times 10^{34} - 3\times 10^{35}$~erg~s$^{-1}$, respectively \citep{Beznogov2021}. We take $L_{\rm ion} = 3 \times 10^{35} $ erg s$^{-1}$ and $r = 6.7 \times 10^{15}$ cm (corresponding to 60 \kms). 

\begin{figure*}
\centering
\includegraphics[width=8.9cm]
{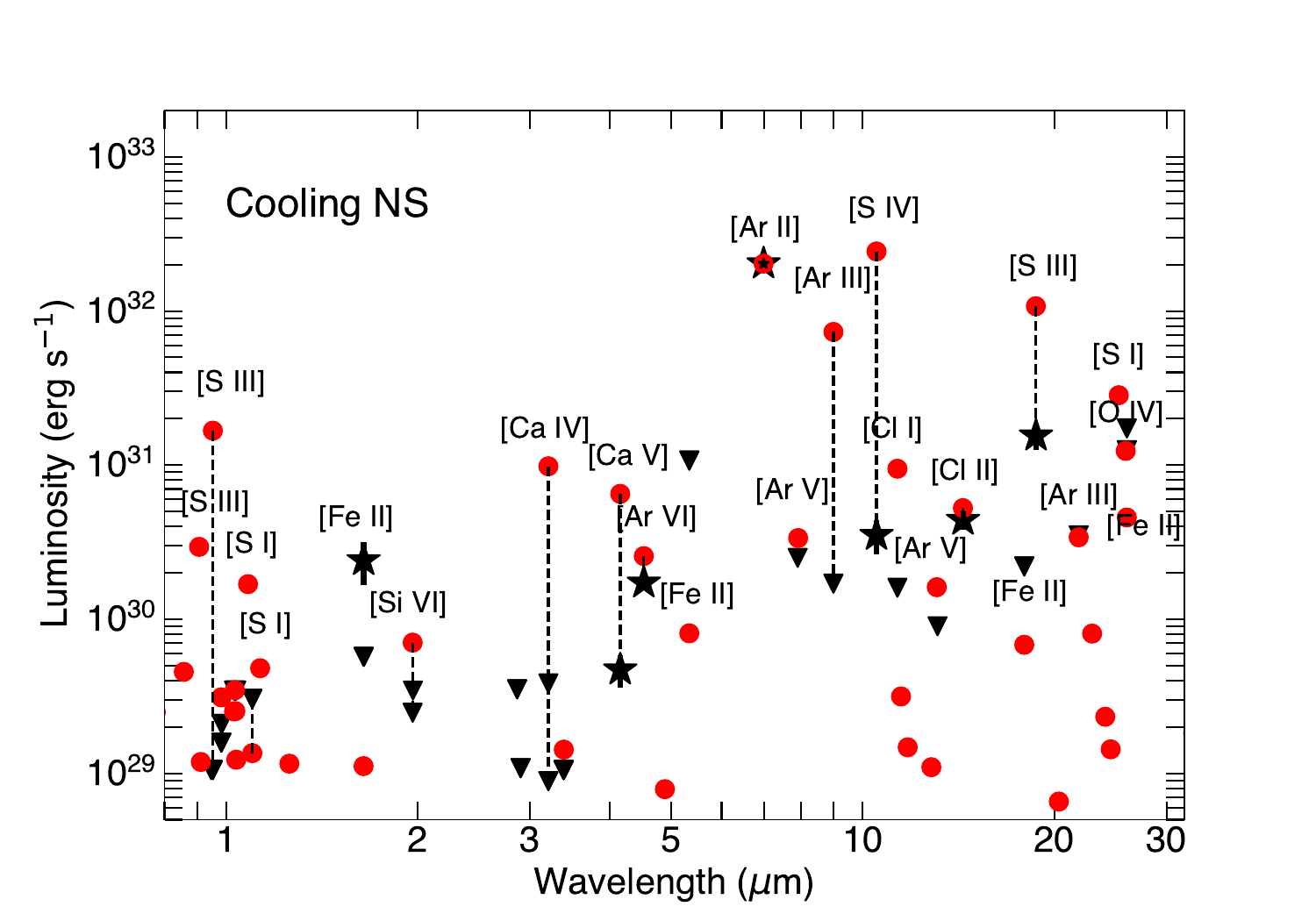}
\includegraphics[width=8.9cm]
{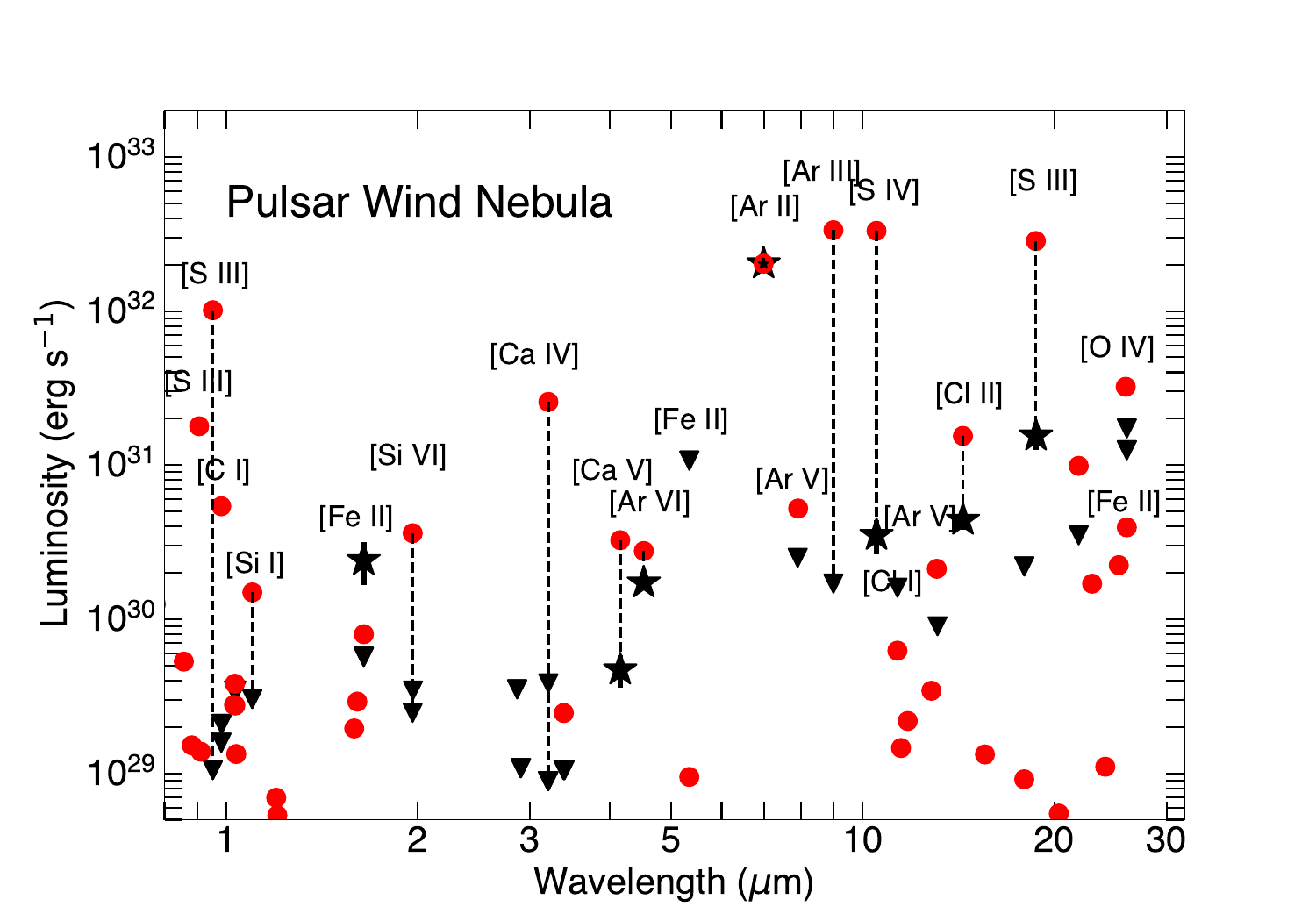}
\includegraphics[width=8.9cm]
{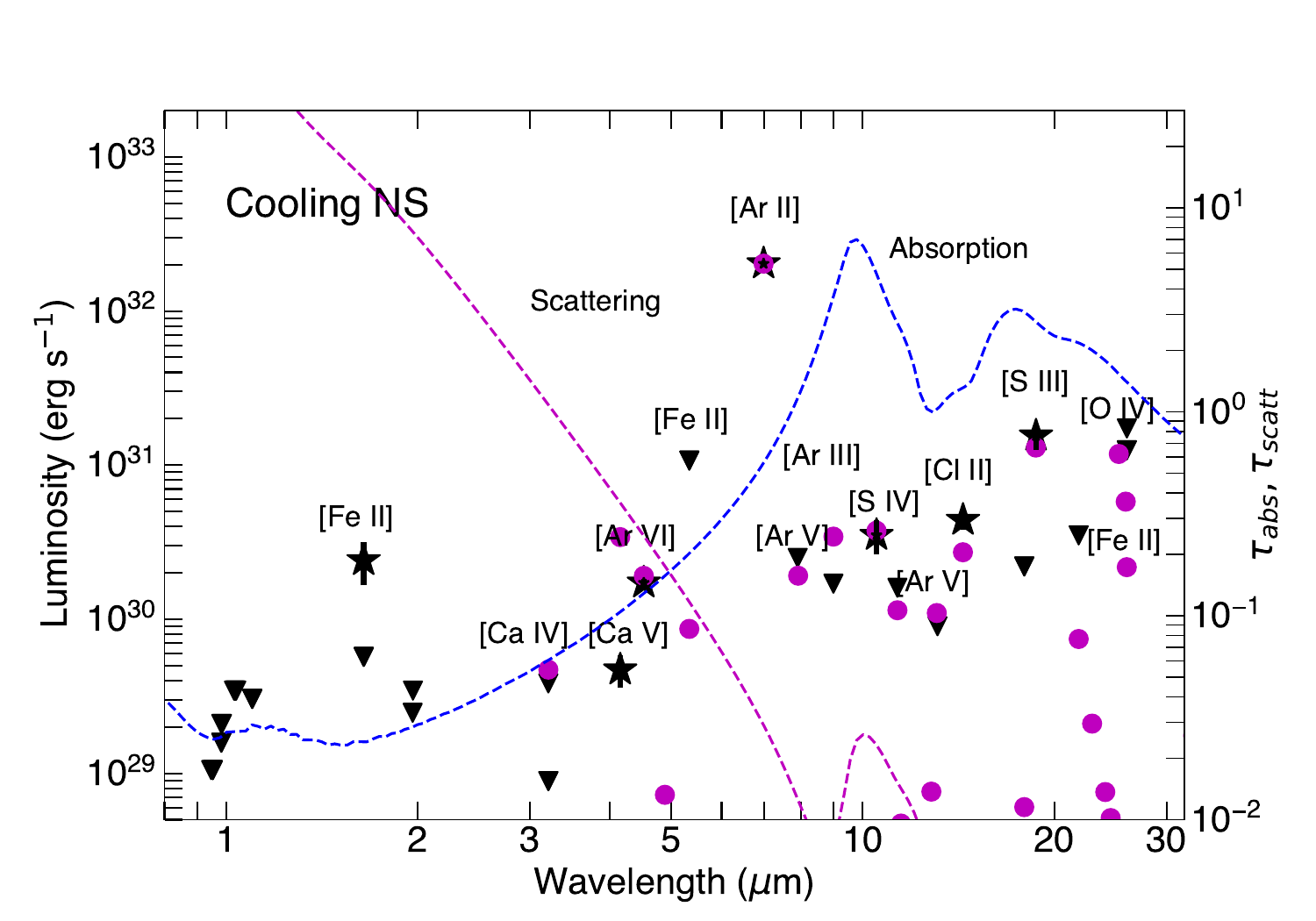}
\includegraphics[width=8.9cm]
{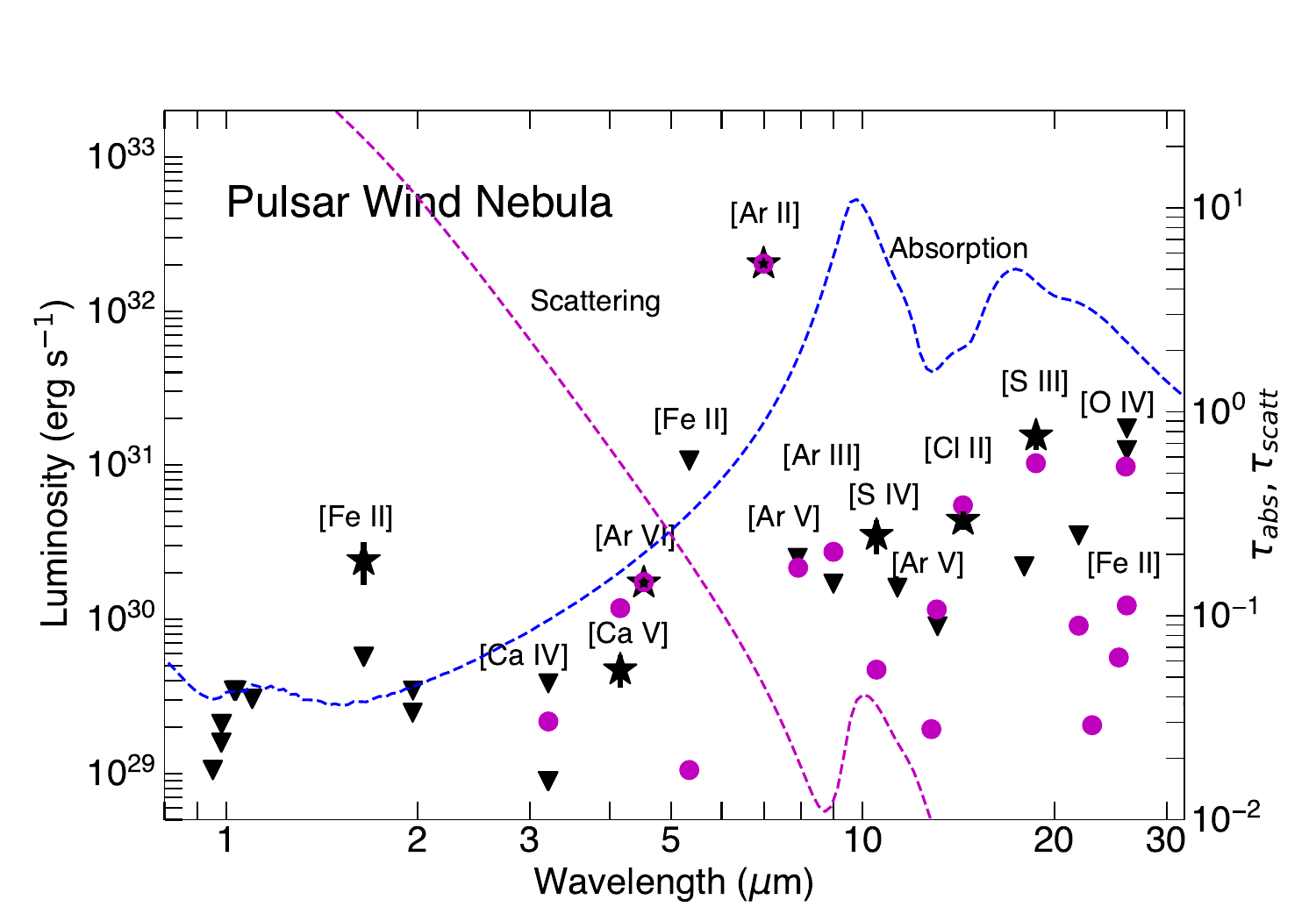}
\caption{Photoionization model results compared to observed line luminosities. Top left panel: Model of the oxygen-burning layers expected in an SN
ionized by a young CNS. The red dots show the predicted luminosities of different lines, while the stars show the observed luminosities from  Table~\ref{tab:lines}. Triangles show the corresponding upper limits from Tables~\ref{tab:limits} and \ref{tab:mrslimits}. The dashed vertical lines in the upper plots connect the observed luminosities or limits with the predicted line luminosities for a selection of lines to illustrate the discrepancy between these. No correction for extinction has been applied. Top right panel: The same, but with ionization by a PWN. Bottom panels: The same models as the top panels, but with luminosities (purple dots) now corrected for absorption by silicate dust, specifically Mg$_2$ SiO$_4$ (forsterite), shown in the top left panel of Figure \ref{fig:dust_abs_scatt}. The dashed blue and purple lines show the corresponding $\tau_{\rm abs}$ and $\tau_{\rm scatt}$, respectively, with the scale given on the right-hand axis. The optical depth is $\tau_{\rm abs}=7$ at 10~{\textmu}m in the CNS case and $\tau_{\rm abs}=11$ in the PWN case. A grain radius $a = 0.3 \ \mu$m is assumed in both cases. Note the suppression of the luminosities for $\lambda \gtrsim 8 \ \mu$m and for $\lambda \lesssim 4 \ \mu$m, where $\tau_{\rm abs}$ and $\tau_{\rm scatt}$ are large, respectively.  }
\label{fig:pred_lines_cns_pwnt}
\end{figure*}

In the upper panels of Figure \ref{fig:pred_lines_cns_pwnt}, we show the results for the two different ionization models without any correction for dust absorption or scattering. The normalization of the predicted line luminosities has been determined from the observed luminosity of the [\ion{Ar}{2}] line. In the case of  [\ion{Ca}{4}], we have for consistency only plotted the upper limit from Table \ref{tab:limits} for the narrow component with the same blueshift as the Ar lines. 

When we compare the predicted and observed luminosities, we see that with the exception of the [\ion{Ar}{6}]  line, which together with the [\ion{Ar}{2}] line determines the ionization parameter, the models over-predict the luminosities of lines at both longer and shorter wavelengths. This is in agreement with the results in F24. However, if we include absorption from dust as in F24, where we use the same dust absorption as in the first panel of Figure~\ref{fig:dust_abs_scatt} for amorphous forsterite (Mg$_2$ SiO$_4$) with an optical depth of $\tau_{\rm abs}=11$ at 10 $\mu$m in the PWN case and $\tau_{\rm abs}=7$ in the CNS case,  the improvement at wavelengths $\gtrsim 8~\mu$m is dramatically improved for both models, which is in agreement with F24. 

At shorter wavelengths, absorption by silicates is much lower, as can be seen from the dashed blue curves in the bottom panels. Therefore, the absorption  does not affect the  over-estimate of the luminosities of the lines shortwards of $\sim 2 \ \mu$m, apparent in the top panels. In F24 we suggested that this  may be explained by scattering of the same dust as it is providing the absorption. As discussed in Section~\ref{sec:disc:line_profiles}, the scattering is more complicated to treat than pure absorption, as it affects both the position of the emission and the wavelength. Considering the unknown dust distribution and properties, we simplify our treatment of the scattering by including it as an absorption, which results in an extinction of the flux from the central source. This implies that we disregard the scattered, redshifted emission seen in Figure~\ref{fig:dust_line_prof}. This is also justified by the fact that we only include the narrow, blueshifted emission component in the comparison with our models, which comes from a spatially unresolved region.

While the absorption is fairly independent of the size of the grains, the scattering cross section, $\sigma$, is roughly proportional to the geometrical cross section of the grain (i.e., $\sigma \propto a^2 $) in the Rayleigh limit ($\lambda \gg a$), as seen in Figure \ref{fig:dust_abs_scatt}.  To get a suppression of lines shortward of the [\ion{Ca}{4}] and [\ion{Ca}{5}] lines, the grain radii have to be relatively large, $a \gtrsim 0.3 \ \mu$m. In Figure \ref{fig:pred_lines_cns_pwnt} we take $a = 0.3 \ \mu$m, which is close to the peak in the size distribution predicted from dust nucleation models in the innermost Si-rich zones for SN 1987A \citep{Sarangi2015}. This should be seen as a “typical" grain size, while in reality, a distribution of sizes is expected \citep[Fig.\ 7 in ][]{Sarangi2015}. 

With this absorption and scattering included, we see a major improvement in the model predictions compared to the observations. While this agrees with F24 at long wavelengths, scattering was not explicitly included in that paper. The fact that extinction increases steeply with decreasing wavelengths below $\sim 5 \ \mu$m, roughly as $\tau \propto \lambda^{-4}$, nearly completely suppresses the narrow lines below $\sim 2 \ \mu$m. 

In general, our results are fairly similar to those in F24, although the exact luminosities of the lines have changed with the new, more accurate observations, as well as some changes in the atomic data. One encouraging result is that the newly detected [\ion{Cl}{1}] $11.3334 \ \mu$m line is reproduced in both models, with a predicted luminosity within $\sim 50 \%$ of the observed values.

The main discrepancy is now  the [\ion{Fe}{2}] $1.6440 \ \mu$m line, which is severely underpredicted. This is surprising because this line should be weak even without scattering (Figure \ref{fig:pred_lines_cns_pwnt}), unless the Fe abundance in the model is several orders of magnitude too low. With a large Fe abundance, we would, however, also expect to see other Fe lines, such as [\ion{Fe}{2}]~1.2570\ $\mu$m, which is not detected. The latter could potentially be suppressed due to scattering by a population of smaller dust grains (cf.\ Figure~\ref{fig:dust_abs_scatt}), but this requires fine tuning of the distribution of grain sizes to also match the ratios of the other lines. Another possibility is that the [\ion{Fe}{2}] line does not originate from the same region as the other lines. Given that the spatial locations and line profiles of all the lines are consistent, this would imply that there are clumps of ejecta  with different abundance patterns on a size scale smaller than the resolution. We note that in Cas A there are knots highly enriched with iron, coming from silicon burning regions \citep{Hughes2000}. As discussed in Section~\ref{sec:analysis:fe}, the significance of the narrow [\ion{Fe}{2}] peak depends on the details of the strong background from the radioactively powered surrounding ejecta, particularly the broad line from  [\ion{Si}{1}] $1.6459 \ \mu$m. It therefore needs to be confirmed with future observations. 

In summary, we find that, with the exception of the possible  [\ion{Fe}{2}] $1.6440 \ \mu$m line, both the CNS and PWN models give good agreement with the observed luminosities, as well as upper limits for other predicted lines.  This, however, requires strong extinction effects from silicates, where absorption dominates in the mid-IR and scattering in the NIR. This therefore gives indirect evidence for silicate dust from the inner ejecta in SN 1987A.  

In F24, other possible emission from the compact object were discussed, including the dust excess in \cite{Cigan2019} and the possible hard X-ray excess in \cite{Greco2021,Greco2022}.  The former was discussed further in Section \ref{sec:disc:alma} and is compatible with the need for dust in our models, though the connection with the IR line emission is inconclusive. The X-ray excess reported in \cite{Greco2022} is based on a fit to the observed NuSTAR spectra with a thermal plus power law model. The latter, which represents the PWN, has an energy power law index $\alpha=1.8$ and a luminosity $3.0 \times 10^{34}$~erg s$^{-1}$ between 10--20~keV. This power law index is, however, considerably steeper than found for a sample of Galactic young PWNs, which are in the range $\sim$ 0--1.2 \citep{Gotthelf2003,Li2008}.

As also discussed in F24, if a power law with $\alpha = 1.8$ and a 10--20~keV luminosity $3.0 \times 10^{34} $ erg s$^{-1}$ is extended to 13.6 eV, it would correspond to a luminosity of $1.4 \times 10^{37} $ erg s$^{-1}$. The upper limit to the total bolometric luminosity of the central  object in SN 1987A is 138 $L_\odot$ or $5.2 \times 10^{35} $ erg s$^{-1}$ \footnote{Note that in F24, this number was a factor 10 too large, resulting in the factor 25  discrepancy below.} \citep{Alp2018}. Therefore, the extrapolated luminosity for the constant $\alpha=1.8$ power-law spectrum is a factor $\sim 25$ too large. This ignores the non-ionizing luminosity below 13.6 eV. A break in the power law below 10 keV could relax this discrepancy. However, to both produce a  10--20~keV luminosity of $3.0 \times 10^{34} $ erg s$^{-1}$ and a total ionizing 13.6 eV to 20 keV luminosity $\lesssim 5.2 \times 10^{35} $ erg s$^{-1}$ requires the break energy to be at $\sim 11$ keV. Including the PWN spectrum in the radio to UV below 13.6~eV would require the break to be at even higher energies, in the observed NuSTAR range.

Overall, it is clear that that the PWN-dominated scenario is much less constrained than the one dominated by the CNS. Given that the observed line ratios in the mid-IR  are mainly determined by $\xi$, and less by the spectral shape of the ionizing source, it is important to have independent constraints on $L_{\rm ion}$, $n_{\rm ion}$ and $r$. Unfortunately, there are no density sensitive pairs of lines among the observed lines from the central source, and only indirect estimates can be used to constrain $n_{\rm ion}$ (F24).  For the CNS, cooling models limit the maximal temperature and hence $L_{\rm ion}$ \citep{Beznogov2021}. The relatively low $L_{\rm ion}$ implies that the NS needs to be close to the line-emitting ejecta for a plausible density. This is supported by the observed line widths; assuming that the narrow component of the line profile corresponds to the size of the emission region, the [\ion{Ar}{6}] FWHM corrected for the spectral resolution is 56~\kms, equivalent to  $r \sim 6.5 \times 10^{15}$~cm, which is close to the value used for the model calculations. 

For the PWN scenario, $L_{\rm ion}$ is less constrained, limited only by the constraint on the total bolometric luminosity from \cite{Alp2018}, which leaves more freedom for $n_{\rm ion}$ and $r$. F24 estimated the expansion velocity of the PWN shell to be in the range $\sim$150--300~\kms, which is clearly larger than the size of the emission region inferred from the line widths. This implies that the size of the PWN is either smaller than assumed, which suggests that the PWN is less energetic, or that we are only seeing emission from a fraction of the ejecta photoionized by the PWN. The PWN shock is, however, highly unstable, which will give rise to a clumpy and filamentary structure \citep{Chevalier1992,Blondin2017}, making the above size estimate uncertain. Dust extinction, as inferred from the photoionization models, may also suppress the flux from a large fraction of the PWN, explaining the small emission region.  In this scenario, the properties of the dust would need to be tuned to explain the lack of wavelength dependence of the peak velocity of the observed lines (Table~\ref{tab:lines}).

An alternative scenario, discussed in F24, is emission from a shock caused by a pulsar jet (similar to the SN remnant G54.1+0.3, \citealt{Temim2010}). The shock models in F24 have, however, problems explaining the luminosity of some of the observed lines, although a shock scenario cannot be completely ruled out. Finally, we note that the CNS must contribute to the ionization also if there is a PWN, but its contribution may be sub-dominant if the CNS has a lower surface temperature and/or is located further away from the emission region.

\subsection{Position and relation to the neutron star kick velocity}
\label{sec:disc-kick}

NSs created in core-collapse SNe are expected to receive a “kick" due to asymmetries in the explosion (e.g., \citealt{Janka2024}). We can estimate the kick velocity of the NS in SN~1987A by assuming that it is located close to the line-emitting region. As discussed above, this assumption is more likely to hold if the ionizing radiation is dominated by thermal emission from the NS surface rather than a PWN. The [\ion{Ar}{6}] line provides the best constraints on the kick, with the Doppler shift and offset from the center translating to a  3D kick velocity of $510 \pm 55$~\kms\ ($252.3\pm 1.8$~\kms\ towards the observer and $443  \pm 64$~\kms\ to the south-east in the sky plane). This is consistent with the kick velocity of $416 \pm 206$~\kms\  estimated from the NIRSpec observations at 12,900~days (F24), but has a smaller uncertainty due to the improved spectral resolution in the new observations. 

In this calculation, we have assumed that the center of the explosion coincides with the center of the ER as reported in \cite{Alp2018}. Tegkelidis et al. (submitted) have recently investigated the position of the center in more detail using a larger set of HST observations registered to Gaia DR3. They found a systematic difference of $\sim 25$~mas between the peak of the early, marginally resolved ejecta and the center of the ER, while the statistical uncertainties on the positions are small in comparison. The final favored position, which is taken as the average of the different methods, is very close to the position from \cite{Alp2018} used here and does not change the main conclusions regarding the magnitude and direction of the kick. However, the systematic uncertainty, which is equivalent to $\sim 160$~\kms\ in the sky plane at 13,500~days, should be kept in mind for the interpretation.

The inferred magnitude of the kick in SN~1987A is close to the typical 3D velocity of $\sim 400$~\kms\ obtained from population studies of radio pulsars \citep{Hobbs2005, Faucher2006}. Numerical simulations of core-collapse SNe have shown that there are two contributions to the total kick velocities: the asymmetric ejection of matter  \citep{Scheck2006,Wongwathanarat2013,Janka2017b,Burrows2024} and the asymmetric neutrino emission  \citep{Fryer2006,Nagakura2019}. The former typically produces kicks in the range $\sim 300-1000$~\kms, while the latter gives lower kick velocities ($\lesssim  200$~\kms) and is subdominant except for very low-mass progenitors \citep{Burrows2024,Janka2024}. The kick inferred for the NS in SN~1987A is thus expected to be dominated by the asymmetric ejection of matter.  

In this scenario, momentum conservation implies that the NS velocity vector, $\bf v_{\rm NS}$, is given by
\begin{equation}
m_{\rm NS}{\bf v_{\rm NS}}= - \int \rho({\bf r}) {\bf v({\bf r})} d^3x
\end{equation}
where $m_{\rm NS}$ is the mass of the NS, $\rho$ the density and ${\bf v(r)}$ the velocity at radius r and the integral is over the whole ejecta.  In the case of SN~1987A, there are 3D emissivity maps of several different emission lines \citep{Larsson2016,Abellan2017,Larsson2019a,Larsson2023}, but to translate these into a total density distribution would require a complete spectral model for the ejecta, accounting for all relevant energy sources as well as the effects of dust. We therefore make the simplifying assumption that the 3D density distribution is directly proportional to the 3D emissivity, and limit ourselves to calculating the direction of motion for the NS. 

We base our calculations on the 3D map of the [\ion{Fe}{1}]~1.444~$\mu$m line obtained from the NIRSpec observations at 12,900 days \citep{Larsson2023}. This shows a similar overall 3D morphology as other lines, in particular H$\alpha$ and [\ion{Fe}{2}]+[\ion{Si}{1}]~1.65~$\mu$m \citep{Larsson2016}, but offers better S/N and has minimal contamination by other emission lines from the ejecta and ER. We find that the resulting kick direction is $\rm{\bf \bar{v}_{NS}} = (-0.34,0.94,-0.01)$, which corresponds to motion to the east, north and towards the observer. The latter component is very small, so the predicted motion is almost in the plane of the sky, at PA$=20^{\circ}$. The statistical uncertainty on the predicted direction is negligible ($\lesssim 0.005$ in all directions) as the S/N in the 3D map is high. 

For comparison, the position and blueshift of the  [\ion{Ar}{6}]  source translates to a kick direction $\rm{\bf \bar{v}_{obs}} = (-0.78 \pm 0.12,  -0.37 \pm 0.12, -0.49 \pm 0.004)$, i.e. to the east, south and towards the observer, with a PA of $116^{\circ}$ in the sky plane. This does not agree with the prediction from the 3D map, with the difference along the line of sight being the most significant due to the well-constrained Doppler shift.  The discrepancies are likely due to the major simplifying assumptions noted above, but could also indicate that the assumed center of explosion is wrong, that the [\ion{Ar}{6}]  emission region is further away from the compact object, and/or that the neutrino mechanism contributed significantly to the total kick. 

An independent estimate of the NS kick velocity in SN~1987A has been presented by \cite{Jerkstrand2020}, who used the redshifted centroids of the $^{56}$CO decay lines to place a lower limit on the kick of $500$~\kms, which is compatible with the value inferred from the [\ion{Ar}{6}] line. Predictions for the NS kick have also been obtained from  3D simulations of explosions that resemble SN~1987A, where the models were oriented to match previous 3D-maps of the Fe/Si-emitting ejecta. Two different explosion models, one presented in \cite{Janka2017} (with more details on the kick provided in \citealt{Page2020}), and another one presented in \cite{Ono2020} and \cite{Orlando2020}, find similar results with a NS kick velocity of $\sim 300$~\kms\ directed to the north and towards the observer. It is interesting to note that both the observed [\ion{Fe}{1}] map and these models predict a significant NS velocity component to the north, contrary to the direction inferred from the [\ion{Ar}{6}] line, which may indicate that there is significant momentum carried by ejecta not captured in the observed 3D maps, or that the NS is not very close to the line-emitting region.


\section{Summary and conclusions}
\label{sec:conclusions}

We have presented an analysis of Cycle 2 JWST NIRSpec IFU observations of SN~1987A  obtained 13,500~days post-explosion, complemented by MRS observations obtained at 13,300~days. Our analysis is focused on the innermost ejecta, motivated by the discovery of narrow emission lines from Ar and S from this region in the Cycle 1 observations at 12,900~days, which provided strong evidence for ionization by a compact object (F24). The new NIRSpec observations were obtained with the high-resolution gratings, offering a factor $\sim 3$ higher spectral resolving power than the Cycle 1 observations in the $\sim$0.9--5.2~$\mu$m wavelength range.  The main results from the analysis can be summarized as follows:

\begin{itemize}

\item[-] The new NIRSpec data enable a better characterization of the [\ion{Ar}{6}]~4.5292\ $\mu$m line profile. Importantly, we identified and removed a diffuse background component that emits an unresolved line at the systemic velocity of SN~1987A. The line profile of the compact central source is dominated by a narrow peak (FWHM $\sim 110$~\kms) with blueshift $-252 \pm2 $~\kms, in good agreement with the [\ion{Ar}{2}]~$6.9853\ \mu$m line in the MRS. Both Ar lines also show a weak red wing extending to $\sim +200$~\kms. No significant time variability of the Ar lines was detected between the two observing epochs ($\Delta t \sim$400--600~days). 

\item[-]  We identified the presence of similar, blueshifted, narrow emission lines from the central ejecta for [\ion{Fe}{2}]\ $1.6440\ \mu$m,  [\ion{Ca}{5}]\ $4.1585\ \mu$m, [\ion{S}{4}]\ $10.5105\ \mu$m, [\ion{Cl}{2}]\ $14.3678\ \mu$m, and [\ion{S}{3}]\ $18.7130\ \mu$m.  The [\ion{S}{4}] and [\ion{S}{3}] lines were previously discussed in F24, while the others have been identified based on the new observations. The  properties of all these lines are more uncertain than the Ar lines due to lower S/N and/or blending with other emission components. We also placed upper limits on many undetected lines predicted by the photoionization models, including several Si lines at short wavelengths. 

\item[-]  Emission from [\ion{Ca}{4}]~$3.2068\ \mu$m is  detected near the center, though significantly displaced to the north compared to the lines discussed above. The line is also broader (FWHM $\sim 1100$~\kms), with its peak redshifted by $\sim 700$~\kms.

\item[-] The spatial properties of the emission region are best constrained from the [\ion{Ar}{6}] line, which is unresolved and located south-east of the geometric center of the ER ($30\pm 10$ mas south, $63\pm 10$ mas east, PA$=116\pm 8$\dg). This offset, together with the blueshift of the line, translates to a 3D kick velocity of $510 \pm 55$~\kms\ for the NS, assuming that it is located close to the line-emitting ejecta. The [\ion{Ar}{6}] source is located to the south of the peak in the ALMA 679~GHz dust map, which has been interpreted as dust heated by a compact object, though the positions are consistent within the 3$\sigma $ astrometric uncertainties. The significant expansion of ejecta in the $\sim 10$~years between the ALMA and JWST observations adds further uncertainties to the comparison and prevents us from drawing conclusions about the possible connection between the dust blob and IR lines. 

\item[-] The continuum in the innermost ejecta does not show any signs of a PWN in terms of enhanced flux or spectral index variations. The upper limit on a Crab-like PWN spectrum with $\alpha = -0.3$ in the G395H grating (2.9--5.2~keV) is  $7.4 \times 10^{32}\ \rm erg\ s ^{-1}$, considering a circular region with 0\farcs{3} radius. A possible PWN is most likely considerably smaller, so this is a conservative limit.  
\end{itemize}

We find that these observed results are strongly affected by dust in the ejecta. Absorption by silicates  at $\lambda \gtrsim 8~\mu$m is required to explain the line ratios, as previously discussed in F24, and from the new results, it is clear that the spectrum is also affected by dust scattering at shorter wavelengths. We show that dust scattering in the expanding ejecta leads to a broadening and redshift of the line profiles. This can explain the red wings of the [\ion{Ar}{2}] and [\ion{Ar}{6}] lines, as well as the redshifted [\ion{Ca}{4}] profile, considering the sharp increase in scattering optical depth between the wavelengths of these lines. Scattering by the same dust also explains the absence of many strong narrow lines predicted by the models at shorter wavelengths, as the broadened scattered profiles will be blended with the other broad lines in the spectrum. We find that a typical grain size of $\sim 0.3~\mu$m provides a good match to the observations. 

We compared the observed line luminosities with photoionization models similar to those in F24, where we accounted for the effects of dust and explored the two scenarios where the ionizing radiation is dominated by the thermal emission from a CNS or the non-thermal spectrum of a PWN, respectively. Both models are compatible with the observations, including the newly identified [\ion{Ca}{5}] and [\ion{Cl}{2}] lines. However, the narrow [\ion{Fe}{2}]~1.6440~$\mu$m line is severely underpredicted, suggesting either a more complex physical scenario, such as a very high Fe abundance and a wider distribution of dust grain sizes, or that the feature is not real. Its significance depends on the background from the broad [\ion{Si}{1}]\ 1.6459\ $\mu$m line and will need to be confirmed with future NIRSpec observations. Adding the spectra from different epochs will improve the S/N and allow for improved constraints on the [\ion{Fe}{2}] feature, as well as the other predicted lines in the NIRSpec range.

The PWN and CNS models primarily differ in the predictions for lines at short wavelengths, such as [\ion{S}{3}]~0.9533\ $\mu$m, [\ion{C}{1}] ~0.9827, 0.9853\ $\mu$m, and  [\ion{Si}{6}]~1.9646\ $\mu$m, which we propose are not detected due to dust scattering. This situation is expected to improve over time, as the optical depth will decrease as $\tau \propto t^{-2}$ due to the expansion of the ejecta. Continued monitoring of the emission from the center is therefore of great interest.

\begin{acknowledgments}

This work is based on observations made with the NASA/ESA/CSA James Webb Space Telescope. The data were obtained from the Mikulski Archive for Space Telescopes at the Space Telescope Science Institute, which is operated by the Association of Universities for Research in Astronomy, Inc., under NASA contract NAS 5-03127 for JWST. These observations are associated with program \#3131. The specific observations analyzed can be accessed via \dataset[DOI: 10.17909/tm6r-a511]{https://doi.org/10.17909/tm6r-a511}. Observations from programs \#1232 and \#2763 have also been analyzed. The DOIs for those data sets are provided in \cite{Larsson2023} and \cite{Jones2023}. 

This research is based in part on observations made with the NASA/ESA Hubble Space Telescope obtained from the Space Telescope Science Institute, which is operated by the Association of Universities for Research in Astronomy, Inc., under NASA contract NAS 5–26555. These observations are associated with program 16996. 

This paper makes use of the following ALMA data: ADS/JAO.ALMA\#2021.1.00707.S and ADS/JAO.ALMA\#2015.1.00631.S. ALMA is a partnership of ESO (representing its member states), NSF (USA) and NINS (Japan), together with NRC (Canada), NSTC and ASIAA (Taiwan), and KASI (Republic of Korea), in cooperation with the Republic of Chile. The Joint ALMA Observatory is operated by ESO, AUI/NRAO and NAOJ.

JL acknowledges support from the Knut \& Alice Wallenberg foundation. 
PJK acknowledges support from Research Ireland Pathway programme under Grant Number 21/PATH-S/9360.
BS acknowledges support from the JWST Cycle 2 grant JWST-GO-03131.002-A.
M.M. acknowledges support from the STFC Consolidated grant (ST/W000830/1).
CG is supported by a Villum Young Investigator grant (VIL25501) and a Villum Experiment grant (VIL69896) from VILLUM FONDEN.
RDG was supported, in part, by the United States Airforce.
NH and MM acknowledge that a portion of their research was carried out at the Jet Propulsion Laboratory, California Institute of Technology, under a contract with the National Aeronautics and Space Administration (80NM0018D0004). NH and MM acknowledge support through NASA/ JWST grant 80NSSC22K0025.
SR acknowledges support from SNF project number 212143. 

\end{acknowledgments}

\vspace{5mm}
\facilities{JWST (NIRSpec \& MRS), HST (WFC3), ALMA}

\software{astropy \citep{Astropy2022},
 Matplotlib \citep{Hunter2007}  
          }



\appendix

\section{Properties of the diffuse emission surrounding SN~1987A}
\label{sec:appendix:diffuse}

\begin{figure*}[t]
\centering
\includegraphics[width=\hsize]{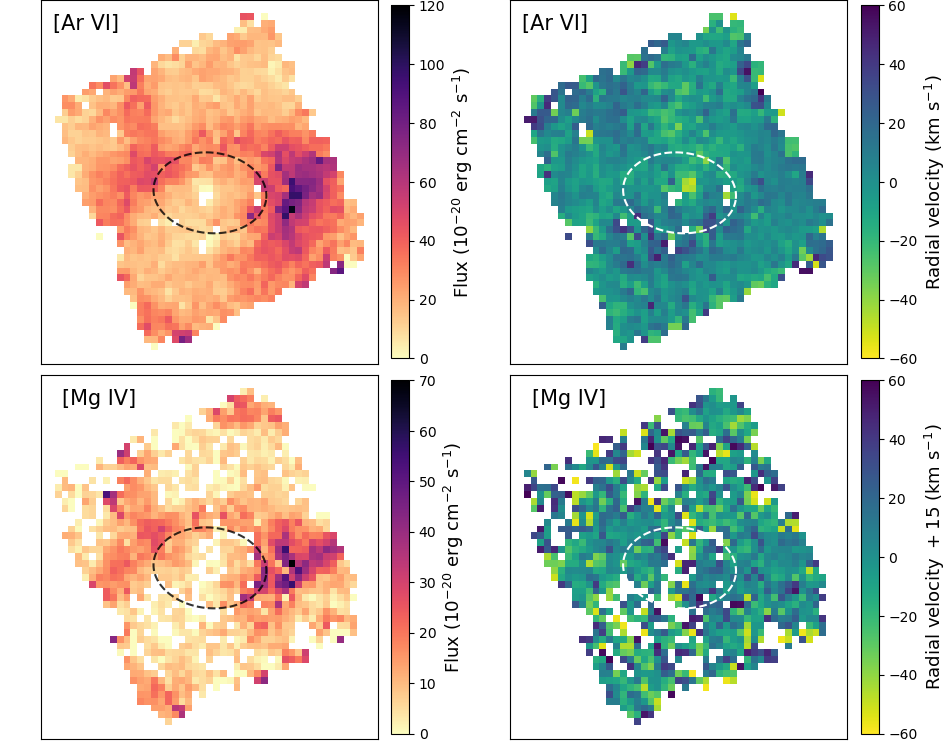}
\caption{Results of fitting the unresolved [\ion{Ar}{6}] and [\ion{Mg}{4}] lines with a Gaussian in each spaxel. The width of the Gaussian was kept fixed at the instrumental resolution. The left and right panels show the integrated fluxes and centroid velocities, respectively, with [\ion{Ar}{6}] results shown in the top panels and [\ion{Mg}{4}] results in the bottom panels. The dashed line in all the panels shows the position of the ER for reference. A constant of 15~\kms\ has been added to the [\ion{Mg}{4}] velocities, which results in a similar median velocity as the [\ion{Ar}{6}] line and is compatible with the uncertainties in the rest wavelengths of the two lines. Spaxels with missing values imply that the fit parameters could not be constrained. This is due to low S/N except for a few spaxels at the center for [\ion{Ar}{6}], where the central point source instead completely overwhelms the narrow diffuse component.}
\label{fig:ar6_mg4_gaussims}
\end{figure*}

Here we investigate the properties of the narrow [\ion{Ar}{6}] component at  $\sim 0$~\kms\ in more detail. To characterize the line, we fitted a Gaussian to the continuum-subtracted spectra in each spaxel in the whole FOV. Initial fits showed that the line was consistent with being unresolved, so the width of the Gaussian was fixed at the instrumental resolution, which corresponds to $\rm{FWHM}=96$~\kms\ at this wavelength. 
In the central region, we performed the fit in a narrow velocity interval of $\pm 170$~\kms\ and also added a straight line to the model to account for the extended red wing of the line from the central source (cf.~Figure~\ref{fig:ar6profiles}). The resulting fluxes and centroid velocities of the best-fit Gaussian lines are shown in the top row of Figure~\ref{fig:ar6_mg4_gaussims}. The flux does not trace the ER or ORs, though the bright region in the west is just outside the brightest part of the ER. The centroid velocity has a median value of $0.0$~\kms\ and does not show any significant spatial variations.    

Spatially extended emission from narrow lines with low radial velocities have previously been reported in the analysis of MRS data, including [\ion{S}{4}]~10.5105~$\mu$m, [\ion{Ne}{6}]~7.6524~$\mu$m, [\ion{Ne}{5}]~14.3217~$\mu$m, [\ion{S}{3}]~18.7130~$\mu$m, [\ion{Ne}{6}]~24.3175~$\mu$m, and [\ion{O}{4}]~25.8903~$\mu$m (\citealt{Jones2023}). With the improved spectral resolution in the new NIRSpec data, we are now able to identify additional such lines in the NIR, including [\ion{Mg}{4}]~4.4867~$\mu$m and [\ion{Ar}{6}]. The results of fitting the [\ion{Mg}{4}] line with a narrow Gaussian across the whole FOV are shown in the bottom row of Figure~\ref{fig:ar6_mg4_gaussims}. The S/N is too low for the fits to be constrained in all spaxels, but it is nevertheless clear that the line shows similar properties as the [\ion{Ar}{6}] line.\footnote{To the best of our knowledge, the most precise wavelength of the [\ion{Mg}{4}] line reported in the literature is $4.47668 \pm 0.00030\ \mu$m (\citealt{Feuchtgruber1997}). Using this wavelength results in a median radial velocity of $-15$~\kms\ for [\ion{Mg}{4}], compared to $0.0$~\kms\ for [\ion{Ar}{6}], for which the rest wavelength is $4.52922\pm 0.00015$ \citep{Casassus2000}. The offset of $-15$~\kms\ is within the wavelength uncertainties of the two lines.} 
We thus conclude that there are a number of narrow, high-ionization lines with low radial velocity in an extended region around SN~1987A. This is most likely local gas ionized by the UV/X-ray emission from the initial shock breakout and the ER \citep{Lundqvist1996,Lundqvist1999}.


\section{Upper limits on narrow lines from the central ejecta}
\label{sec:appendix:ul}

Here we present upper limits on narrow  Ar-like components from the central ejecta in other emission lines. We select lines that can constrain the photoionization and shock models, as well as potentially give information on dust scattering and absorption. For lines in the NIRSpec range, we use both the spectral and spatial properties of the [\ion{Ar}{6}] line when determining the limits, focusing on the peak of the line profile, where the signal from the central source compared to the surrounding ejecta is maximized. 

For each line of interest, we first resample the velocity (spectral) dimension of the cube to match that of the [\ion{Ar}{6}]  line, using linear interpolation. We then produce an image by integrating over the two spectral bins at the peak of the line, which corresponds to a velocity width of 90~\kms. We also create an average image of two spectral bins on each side of the peak (displaced by $\pm 90$ ~\kms\ from the peak), which we subtract as background. The same procedure is then carried out to create an equivalent image of the [\ion{Ar}{6}] line, using the version of the cube where the diffuse 0-\kms\ line has been removed (Section~\ref{sec:analysis:ar6}). 

For each line image, we add the [\ion{Ar}{6}] image multiplied by a constant, and determine the lowest value of the constant for which a central point source is detected in the resulting image. We use the \texttt{DAOStarFinder} tool from \texttt{photutils} \citep{Bradley2023} to carry out the source detection,  with the detection threshold set to the median + 3 standard deviations of the surrounding ejecta region in each image.  Finally, we use the properties of the [\ion{Ar}{6}] line to translate the constant that results in a detection to a luminosity for the narrow blueshifted component of the line. The resulting limits are presented in Table~\ref{tab:limits}. 

Before adding the renormalized [\ion{Ar}{6}] images to obtain the upper limits, we verified that no central point sources were detected for any of the lines in Table~\ref{tab:limits}. On the other hand, the central sources in the  [\ion{Fe}{2}] and [\ion{Ca}{5}] lines discussed in Sections~\ref{sec:analysis:fe} and \ref{sec:analysis:ca} are detected at 5 and $4 \sigma$ significance, respectively. The main systematic uncertainty affecting the limits in Table~\ref{tab:limits} is the background subtraction,  which is clearly approximate due to the complexity of the underlying spectra, which contain broad, asymmetric lines from the ejecta, as well as contributions from the northern outer ring and scattered light from the ER in some cases.

\begin{deluxetable}{lccrcc}[t]
\tablecaption{Upper limits on [\ion{Ar}{6}]-like components in other lines in the JWST/NIRSpec range.  \label{tab:limits}}
\tablecolumns{3}
\tablenum{3}
\tablewidth{0pt}
\tablehead{
\colhead{Line} &
\colhead{Wavelength} & 
\colhead{$L_{\rm lim}$\tablenotemark{a}} & \\
\colhead{} &
\colhead{($\mu$m)} &
\colhead{$(10^{29}\ {\rm erg\ s^{-1}})$} &
}
\startdata
$[$\ion{S}{3}$]$ & 0.9533 & 1.05 \\
$[$\ion{C}{1}$]$  & 0.9827  & 1.58 \\
$[$\ion{C}{1}$]$  & 0.9853  & 2.09 \\
$[$\ion{S}{2}$]$  & 1.0323\tablenotemark{b}  & 3.44 \\
$[$\ion{Si}{1}$]$  & 1.0994  & 3.04 \\
$[$\ion{Si}{1}$]$  & 1.6073  & 0.23 \\
$[$\ion{Si}{1}$]$  & 1.6459  & 5.70 \\
$[$\ion{Si}{6}$]$  & 1.9646  & 2.47 \\
$[$\ion{Ca}{4}$]$  & 3.2068  & 0.89 \\
\enddata
\tablenotetext{a}{The luminosities are 3$\sigma$ upper limits on narrow, blueshifted components with the same properties as observed for the [\ion{Ar}{6}] line.}
\tablenotetext{b}{Weighted wavelength of four lines.}
\end{deluxetable}

For lines in the MRS wavelength range,  we used the spectral properties of the [\ion{Ar}{2}] line to estimate upper limits from spectra extracted from the ``ejecta''  region \citep[see][Kavanagh et al., in prep.]{Jones2023}. Upper limits were determined using a Gaussian line profile centered at the blueshifted velocity of the [\ion{Ar}{2}] line with an amplitude three times the rms noise. The FWHM was either as reported for [\ion{Ar}{2}] in Kavanagh et al. (in prep.) or, if the line is unresolved, set to the instrument resolution \citep{Jones2023a}. Emission lines from the ER were masked when determining the rms noise. In the case of the [\ion{Fe}{2}]~25.9844~$\mu$m and $[$\ion{O}{4}$]$~25.8903~$\mu$m lines, we fitted a spline model using the \texttt{SplineExactKnotsFitter} \citep{Astropy2022} to remove the broad underlying Fe line complex and estimated the rms noise on the residuals. The resulting upper limits are reported in Table~\ref{tab:mrslimits}.

\begin{deluxetable}{lccrcc}[t]
\tablecaption{Upper limits on [\ion{Ar}{2}]-like components in other lines in the JWST/MIRI MRS range. \label{tab:mrslimits}}
\tablecolumns{3}
\tablenum{4}
\tablewidth{0pt}
\tablehead{
\colhead{Line} &
\colhead{Wavelength} & 
\colhead{$L_{\rm lim}$} & \\
\colhead{} &
\colhead{($\mu$m)} &
\colhead{$(10^{31}\ {\rm erg\ s^{-1}})$} &
}
\startdata
$[$\ion{Fe}{2}$]$ & 5.3402 & 1.07 \\
$[$\ion{Ar}{5}$]$  & 7.9016  & 0.25 \\
$[$\ion{Ar}{3}$]$  & 8.9914  & 0.17 \\
$[$\ion{Cl}{1}$]$  & 11.3334  & 0.16 \\
$[$\ion{Ar}{5}$]$  & 13.1022  & 0.09 \\
$[$\ion{Fe}{2}$]$  & 17.9360  & 0.22 \\
$[$\ion{Ar}{3}$]$  & 21.8291  & 0.35 \\
$[$\ion{O}{4}$]$  & 25.8903  & 1.73 \\
$[$\ion{Fe}{2}$]$  & 25.9884  & 1.25 \\
\enddata
\tablenotetext{a}{The luminosities are 3$\sigma$ upper limits on narrow, blueshifted components with the same properties as observed for the [\ion{Ar}{2}] line. Vacuum wavelengths are from \citet{vanHoof2018}.}
\end{deluxetable}


\bibliography{sn87a_co_jwst_refs.bib}{}
\bibliographystyle{aasjournal}

\end{document}